\documentclass[prc,aps,preprintnumbers,amsmath,amssymb]{revtex4}
\usepackage{graphicx}

\def\J{J/\Psi}

 \def\gsim{\mathrel{\rlap{\lower4pt\hbox{\hskip1pt$\sim$}}
 \raise1pt\hbox{$>$}}}

 \newcommand\la{\langle}
 \newcommand\ra{\rangle}
 \newcommand\beq{\begin{equation}}
 \newcommand\noi{\noindent}
 \newcommand\eeq{\end{equation}}
 \newcommand\beqn{\begin{eqnarray}}
 \newcommand\eeqn{\end{eqnarray}}

\def\fm{\,\mbox{fm}}
\def\GeV{\,\mbox{GeV}}
\def\TeV{\,\mbox{TeV}}
\def\lsim{\mathrel{\rlap{\lower4pt\hbox{\hskip1pt$\sim$}}
    \raise1pt\hbox{$<$}}}         %less than or approx. symbol
\def\gsim{\mathrel{\rlap{\lower4pt\hbox{\hskip1pt$\sim$}}
    \raise1pt\hbox{$>$}}}         %greater than or approx. symbol

\def\fm{\,\mbox{fm}}
\def\GeV{\,\mbox{GeV}}

\def\beq{\begin{equation}}
\def\eeq{\end{equation}}

\def\beqy{\begin{eqnarray}}
\def\eeqy{\end{eqnarray}}
\def\bm{\begin{minipage}}
\def\em{\end{minipage}}
\def\bc{\begin{center}}
\def\ec{\end{center}}

\begin{document}
%\date{today}

\title{\bf High-\boldmath$p_T$ paradigms revisited}

\author{B. Z. Kopeliovich}

\affiliation{Departamento de F\'{\i}sica, 
Universidad T\'ecnica Federico Santa Mar\'{\i}a; and
\\
Instituto de Estudios Avanzados en Ciencias e Ingenier\'{\i}a; and\\
Centro Cient\'ifico-Tecnol\'ogico de Valpara\'iso;\\
Casilla 110-V, Valpara\'iso, Chile\\
and\\
Institut f\"ur Theoretische Physik der Universit\"at,\\
Philosophenweg 19, 69120 Heidelberg, Germany}

\author{J. Nemchik}
      \affiliation{ Czech Techn. University in Prague, FNSPE,
B\v rehov\' a 7,11519 Prague, Czech Republic, and\\
Institute of Experimental Physics SAS, 
Watsonova 47,04001 Ko\v sice, Slovakia}
\vspace*{1cm}
\begin{abstract}
\noindent
We present an attempt at a critical overview of the current status of modeling for high-$p_T$ processes in nuclei. The paper covers several topics including coherence phenomena, in particular gluon shadowing and CGC; nuclear effects related to the restrictions imposed by energy conservation at large $x_L$ and $x_T$ ; space-time development of hadronization of highly virtual light and heavy partons and the related time scales; and the role of early production and subsequent attenuation of
pre-hadrons in a dense medium. We identify several intriguing problems in the current paradigms for high-$p_T$ processes and propose solutions for some of them.
\end{abstract}

%\date{\today}


\maketitle

%\newpage

\vspace{0.5cm}
\tableofcontents

%\newpage

\section{Introduction}

During the last decade of RHIC operation the physics of high-$p_T$ reactions have attracted lot of attention, because these processes can serve as a probe for the early stage of the development of the quark-gluon matter created in heavy ion collisions. Intensive experimental and  theoretical studies led to essential progress in our understanding in this field, especially for high-$p_T$ processes on nuclei. Yet there is plenty of room for improvements and many experimental puzzles are waiting for solutions. This paper presents a brief critical overview of several key paradigms related to high-$p_T$ physics. In particular, we challenge some of the current standard approaches, which have been settled by voting, rather than by real arguing. We also propose solutions for some of experimentally observed effects.

In Sections~2-4 we start with the issue of the parton distribution functions (PDF) in nuclei, which are desperately needed for calculation of hard reactions at high energies. We present some of the theoretical tools for calculation of the gluon shadowing. Reviewing several popular gluon nuclear 
PDFs (nPDF), which  resulted from global data analyses, we found some of them to be either ad hoc, or incorrect.  The $p_T$ dependent nPDFs are
subject to the phenomenon of color glass condensate (CGC) controlled by the saturation scale. We demonstrate that the saturation scale in nuclei can be directly accessed by measuring $p_T$ broadening in
different reactions. Even more nuclear modifications affect the gluon distribution in the case of heavy ion collisions. We found a novel effect of mutual boosting for the saturation scales in the colliding nuclei,
which can substantially enhance the effect of saturation in $AA$ compared with $pA$ collisions.

In section~5 we study the restrictions imposed by energy conservation on hard reactions at large $x_L$ and $x_T$, in the vicinity of the kinematic bounds. In view of the current controversy in interpretation of the suppression of high-$p_T$ hadrons produced at forward rapidities in $d$-$Au$ collisions at RHIC we propose a few tests, sensitive to the source of suppression, whether it is the deficit of energy, or coherence effects. Data available for these tests seem to favor the former mechanism.

Section~6 presents a critical analysis of the current models for jet quenching observed at RHIC. We propose several tests for the popular scenario of energy loss, which is based on the unjustified assumption that hadronization is lasting a long time, and production of leading hadrons always occurs outside of the medium. We find that available data disfavor such a space-time structure of  hadronization. Moreover,  theoretically the production range of leading hadrons should be expected to be rather short because of the intensive vacuum gluon radiation ignited by the high-$p_T$ scattering. 

In Section~7 we present a model based on our calculations for the  pre-hadron production length. 
We demonstrate that the produced pre-hadron dipole quickly expands its size, especially if it contains an open heavy flavor.  This leads to a strong absorption in a dense medium created in heavy ion collisions. In the limit of short mean free-path of such dipoles the nuclear suppression factor $R_{AA}$ for high-$p_T$ hadrons can be predicted in a parameter free way. RHIC data for light and heavy flavored hadrons at high $p_T$  seem to support existence of such a regime.

The main observations are summarized in section~8.

\section{Saturation (CGC, gluon shadowing)}\label{cgc}
\begin{figure}[htb]
\bc
\includegraphics[width=40mm]{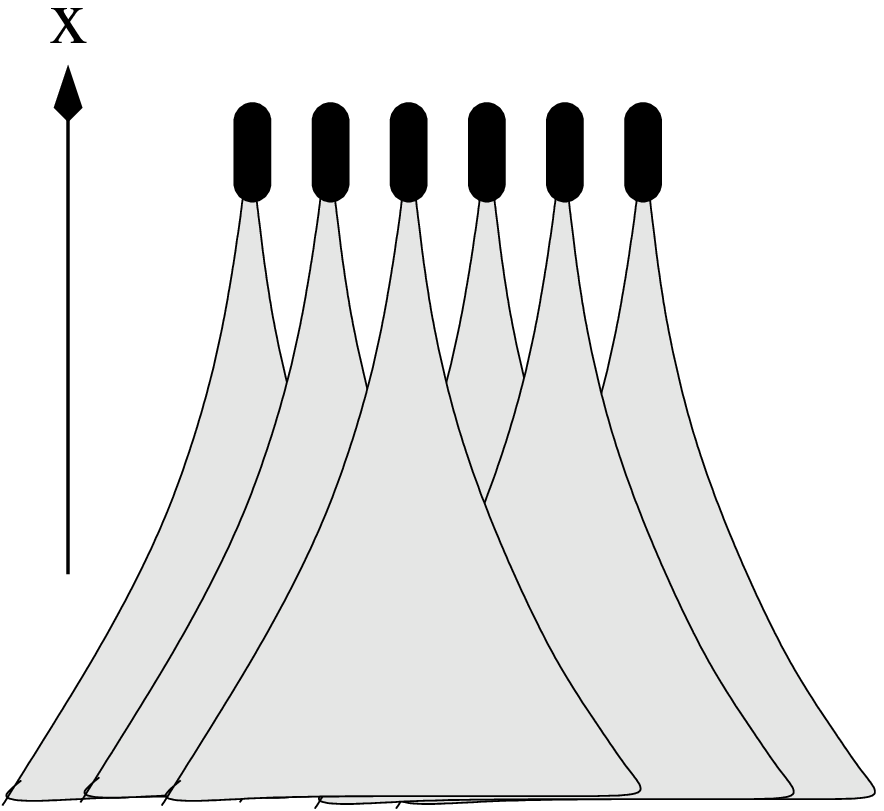}
\hspace{2cm}
 \includegraphics[width=8.5cm]{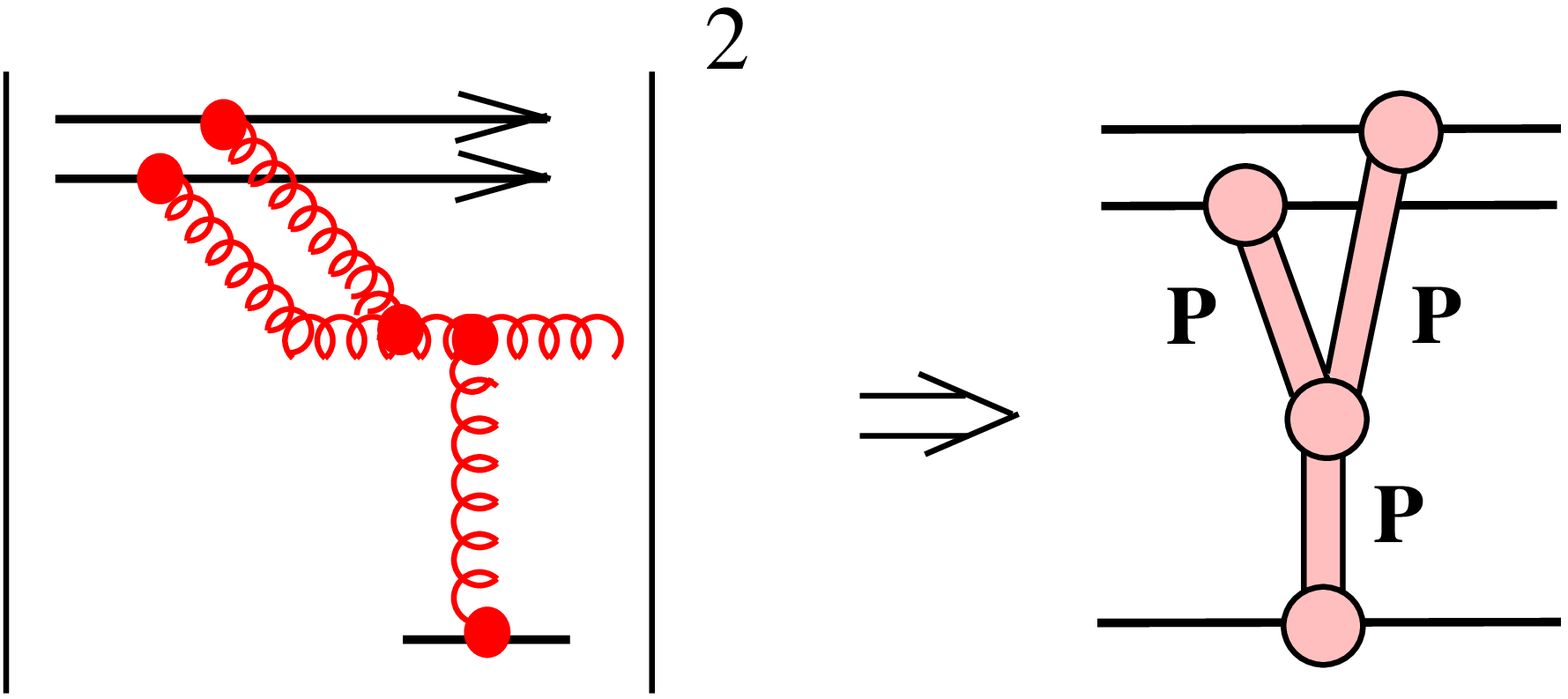}
 \caption{ \label{fusion}{\it Left:} Bjorken $x$ (vertical axis) dependence of gluon clouds spread over longitudinal coordinate (horizontal axis) in a high-energy nucleus. The Lorentz factor decreases downwards small $x$, and the clouds start overlapping.
 {\it Right:} Fusion and interaction with the target of gluons originating from different bound nucleons.
 The amplitude squared gives the triple-Pomeron graph.}
 \ec
 \end{figure}

\noi
Bound nucleons are well separated in the nuclear rest frame. They still do not overlap in the nucleus Lorentz boosted to the infinite-momentum frame. Indeed, although the Lorentz boosted nucleus looks like a pancake, not only the nucleon spacing, but also the nucleons themselves are subject to Lorentz contraction. However, partons carrying a small fraction $x$ of the nucleon momentum are contracted $x$ times less. As a result the parton clouds originated from different nucleons overlap in the longitudinal direction at small $x\ll1$ \cite{kancheli}, as is illustrated in Fig.~\ref{fusion}.

This leads to a dense packing of the radiated gluons in the phase space. However,
according to the Landau-Pomeranchuk (LP) principle \cite{lp} multiple interactions do not generate multiple radiation of gluon spectra, if the coherence time of radiation is large, $l_c\gg R_A$.
The radiation process does not resolve between single and multiple interactions, only the total momentum transfer matters.
In other words, when the interaction becomes sufficiently strong, it starts screening itself. In particular, radiation of soft gluons, which has large cross section, can be strongly shadowed. In terms of the Fock state decomposition, a gluon in a given Fock state can be radiated only once.

Therefore, the spectrum radiated with small transverse momenta, $k_T^2\ll\la k_T^2\ra$ from multiple interactions must saturate when the phase space of radiated gluons is densely packed. Only at sufficiently large transverse momentum of gluons, $k_T\gsim\la k_T\ra$,
where the phase space becomes dilute, multiple interactions
start contributing to the multiplicity of gluons, increasing the range of $k_T$. Eventually, one arrives at
the Bethe-Heitler regime of radiation when each of the multiple interactions equally contributes to the radiation spectrum.
The transverse gluon momentum characterizing the transition scale between the two regimes is called {\it saturation momentum} and is defined below. 
This effect is known as the  color glass condensate (CGC) \cite{mv,cgc1,cgc2}.

\subsection{CGC: how to measure the saturation scale}

The rise of total cross sections with energy, discovered back in 1973, was the first manifestation of
the  increasing population of partons at small $x$. As the parton density increases,
the inverse process of parton fusion becomes important, and eventually the parton density is expected \cite{glr} to saturate. 

In the nuclear rest frame, the same phenomenon looks like Glauber shadowing and color filtering for a dipole (quark-antiquark, or glue-glue) propagating through  nuclear matter \cite{al}.
The partial elastic dipole-nucleus amplitude at impact parameter $b$ reads \cite{zkl}
\beq
f_{\bar qq}^A(b)=1-e^{-{1\over2}\sigma_{\bar qq}^N(r_T,E)\,T_A(b)},
\label{100}
\eeq
where $r_T$ and $E$ are the transverse separation and energy of the dipole respectively; $T_A(b)=\int_{-\infty}^{\infty} dz\,\rho_A(b,z)$ is the nuclear thickness function, integral of nuclear density along the trajectory of the projectile at impact parameter $b$. 
We assume here that the Lorentz dilated length (coherence length) of the dipole size fluctuations is much longer than the nucleus.
The $\bar qq$ dipole cross section on a nucleon should vanish at small transverse separation
$\sigma_{\bar qq}(r_T\to0,E)\propto r_T^2$ \cite{zkl}.  The energy dependence is discussed below.
For large $ T_A(b)$ only the small-$r_T$ part of the cross section contributes, so one can use the $r_T^2$ approximation, 
\beq
\left.\sigma_{\bar qq}^N(r_T,E)\right|_{r_T\to0}\approx C_q(E,r_T)\,r_T^2,
\label{120}
\eeq
where $C_q(E,r_T)$ is logarithmically divergent at small $r_T$ \cite{zkl}. In this limit the factor $C_q$
can be related to the gluon distribution \cite{fs},
\beq
C_q(E,r_T)=\frac{\pi^2}{3}\,\alpha_s(1/r_T^2)\,xG(x,1/r_T^2),
\label{130}
\eeq
where $1/x=2m_NE\,r_T^2 $.

The quark saturation momentum $Q_{qA}$  is usually introduced as,
\beq
f_{\bar qq}^A(b)=1-\exp\left[-\frac{r_T^2\,Q_{qA}^2(b,E)}{4}\right],
\label{140}
\eeq
so comparing with  Eqs.~(\ref{100})
and (\ref{120}) one gets,
\beq
Q_{qA}^2(b,E)=2C_q(E,r_T=1/Q_{qA})T_A(b).
\label{160}
\eeq
Here we fixed the dipole separation at the typical value $r_T\sim1/Q_{qA}$ relying on the weak, logarithmic, $r_T$ dependence of $C(E,r_T)$.

\begin{figure}[htb]
\bc
 \includegraphics[width=6.8cm]{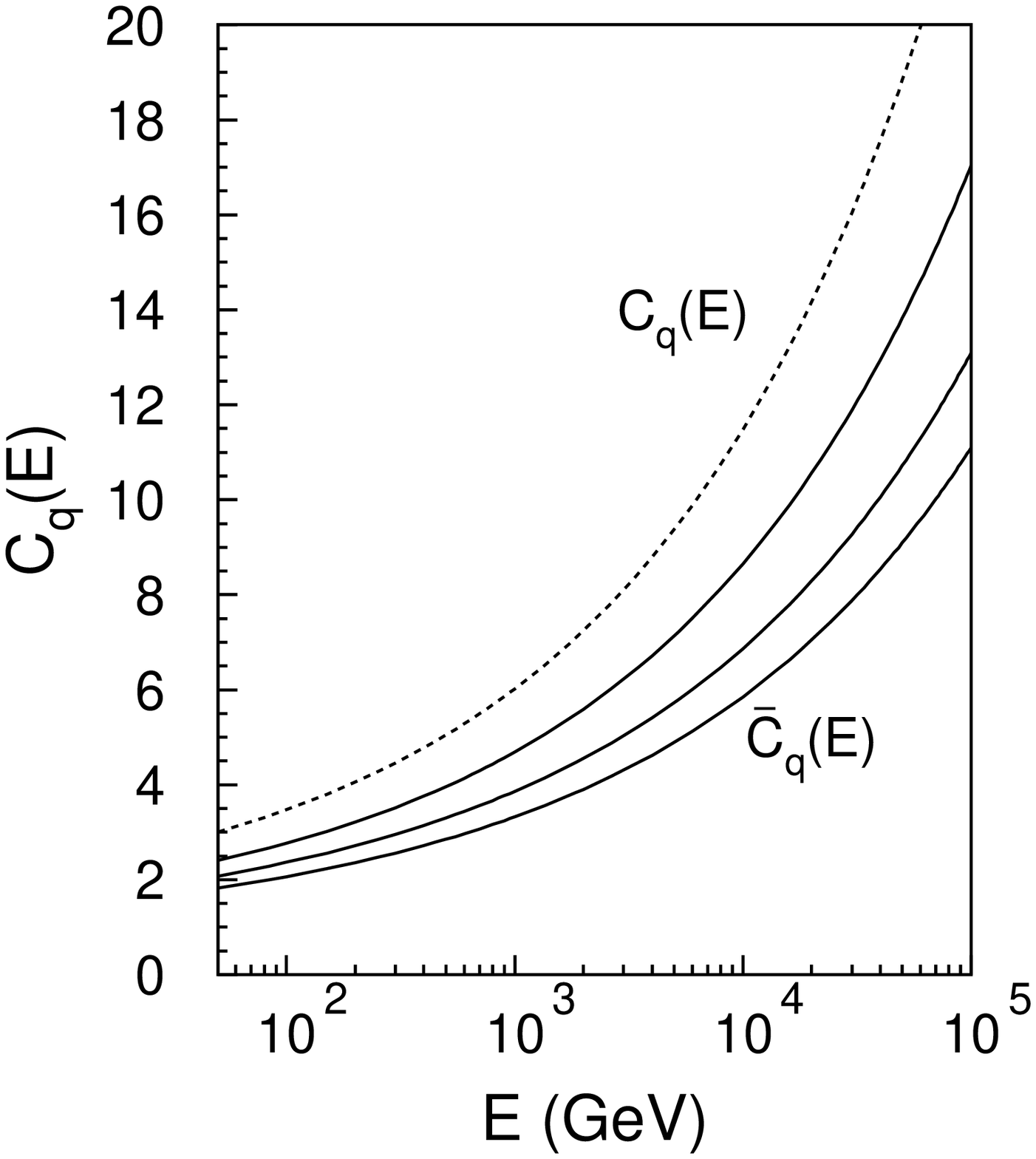}\hspace{15mm}
 \includegraphics[width=6.5cm]{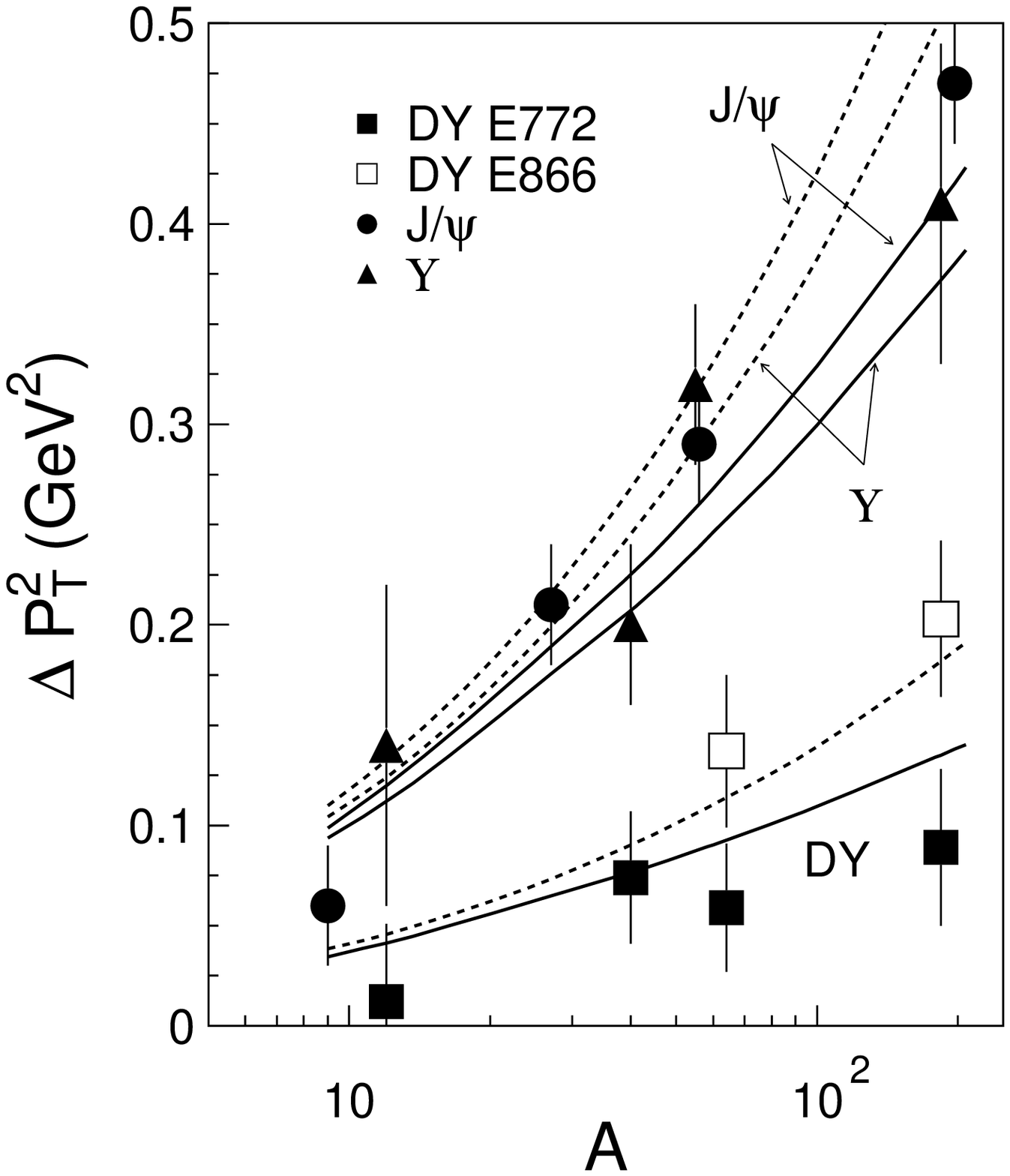}
 \caption{ \label{broad-fig}{\it Left:}
 Factor $C(E)$ defined in (\ref{120}), which controls the saturation scale in nuclei through 
 Eqs.~(\ref{180})-(\ref{190}). The dashed curve shows the dependence on quark energy in the leading  
 order approximation.The solid curves show the modified broadening factor $\widetilde C_q(E)-R_g(E)
 \,C_q(E)$ damed by gluon shadowing, which depends on nuclear thickness $T_A$ propagated by the 
 quark. The curves from bottom to top correspond to $T_A=1.5,\ 1.0,\ 0.5\fm^{-2}$.
{\it Right:} Broadening in 
 Drell-Yan reaction on different nuclei as measured in the E772  (closed squares) \cite{e772}
 and E866 (open squares) \cite{e866-pt} experiments, respectively.  Broadening for $J/\Psi$ and 
 $\Upsilon$ \cite{e772,mmp} is shown by circles and triangles, respectively.
 The dashed and solid curves correspond to the predictions without and with the corrections for gluon       
 shadowing, respectively.}
 \ec
 \end{figure}

The energy dependence of the factor $C(E)$ fitted to DIS and photoproduction data is depicted in the left panel of Fig.~\ref{broad-fig}. It steeply rises with energy up to very high values. We remind that because $Q_{qA}$ increase with energy, the mean value of $\la r_T\ra$, which we fixed above, should be reconsidered.

The same factor $C(E)$, as in Eq.~(\ref{160}), controls broadening of the transverse momentum of a 
single parton propagating through a nucleus \cite{dhk,jkt},
\beq
\Delta p_T^2=T_A(b)
\left.\vec\nabla_{r_T}^2\sigma_{\bar qq}^N(r_T)\right|_{r_T=0}=
2T_A(b)\,C_q(E,r_T=0).
\label{180}
\eeq
Thus, we arrive at a divergent result, since $C(E,r_T)\propto \ln(1/r_T)$ at $r_T\to0$ \cite{zkl}.
This is not a surprise, as the mean transverse momentum squared $\la p_T^2\ra$ is ultraviolet divergent.
Moreover, this divergency is not canceled in broadening $\Delta p_T^2=\la p_T^2\ra_A-
\la p_T^2\ra_N$ \cite{mikkel-raufeisen,kovner-wied}.
To settle the problem, one should fix $r_T$ at a characteristic value like in (\ref{160}) at $r_T^2\sim 1/\Delta p_T^2$.

Thus, broadening and the saturation momentum are equal:
\beq
Q_{qA}^2(b,E)=\Delta p_T^2(b,E),
\label{190}
\eeq
so one has  a direct access to the saturation scale by measuring broadening.

The experiments at HERA provided detailed information about the dipole cross section $\sigma_{\bar qq}(r_T,x)$
as function of $x$ and the dipole size. Therefore the factor $C(E,r_T)$ in (\ref{120}) and broadening can be well predicted.  The results are depicted by the bottom dashed curve in Fig.~\ref{broad-fig} (right) in comparison with broadening in Drell-Yan reaction measured in the E772 and E866 experiments \cite{e772,e866-pt}.
Notice that this is broadening or the saturation scale for quarks. For gluons it is expected (perturbatively) to be $9/4$ times larger. That should be compared with broadening of heavy quarkonia.
Comparison of the predicted broadening shown in Fig.~\ref{broad-fig} by the two dashed curves, with data for $J/\Psi$ and $\Upsilon$ \cite{e772,mmp} also demonstrates good agreement.

Notice that  to calculate nuclear broadening for heavy quarkonium production one does not need to know its mechanism  provided that the coherence length of quarkonium production is short, i.e., 
\beq
l_c=\frac{s x_1}{m_N M_{\bar QQ}^2}\ll R_A.
\label{330}
\eeq
 The data for broadening of $\Upsilon$ depicted in Fig.~\ref{broad-fig} satisfy well this condition, only the data for $J/\Psi$ production are somewhat out of this kinematic domain.

\section{Gluon shadowing}

As was explained above and is illustrated in Fig.~\ref{fusion}, the overlap of gluons originated from different nucleons leads to their fusion and to a reduction of the gluon density at small $x$. This is how gluon shadowing looks like in the infinite momentum frame of the nucleus. In the nuclear rest frame the same phenomenon is interpreted quite differently. It is related to higher Fock components in the incoming hadron, containing gluons. Multiple interactions of those gluons give rise to gluon shadowing. This interpretation has the advantage of being more intuitive and directly linked to the optical analogy of shadowing.

\subsection{Theoretical expectations}

Unfortunately, no satisfactory theoretical description of gluon shadowing, which would  work in all kinematic regimes, has been developed so far. Each of the existing approaches has limitations, which makes a direct comparison with data difficult.

The most rigorous quantum-mechanical treatment of gluon shadowing is based on the path-integral technique \cite{kst2,krtj}. It includes both the phase shifts between different production points and
a proper treatment of the dipole attenuation. Therefore this approach should correctly reproduce the onset of shadowing at medium-small $x$ where the magnitude of shadowing is small. Actually, shadowing is expected to remain weak even at smaller $x$ because of the specific nonperturbative effects suppressing gluon radiation \cite{kst2}.  Indeed, high-statistics data for single diffraction, $pp\to pX$, show that gluon radiation, which controls the large invariant mass tail, $d\sigma_{sd}/dM_X^2\propto1/M_X^2$, is an order of magnitude smaller that one could expect in pQCD. This problem has been known in the Regge phenomenology as smallness of the triple-Pomeron coupling \cite{kklp}. The only way to suppress gluon radiation is to reduce the mean quark-glue, or glue-glue separation. In order to explain data for diffraction one has to assume this separation to be as small as $r_0\sim0.3\fm$ \cite{kst2}. Indeed, lattice calculations lead to a glue-glue correlation radius of this order of magnitude \cite{pisa}. There are many other experimental evidences supporting the existence of such a semihard scale in hadrons \cite{spots}. 

\begin{figure}[htb]
\begin{center}
\includegraphics[width=8cm]{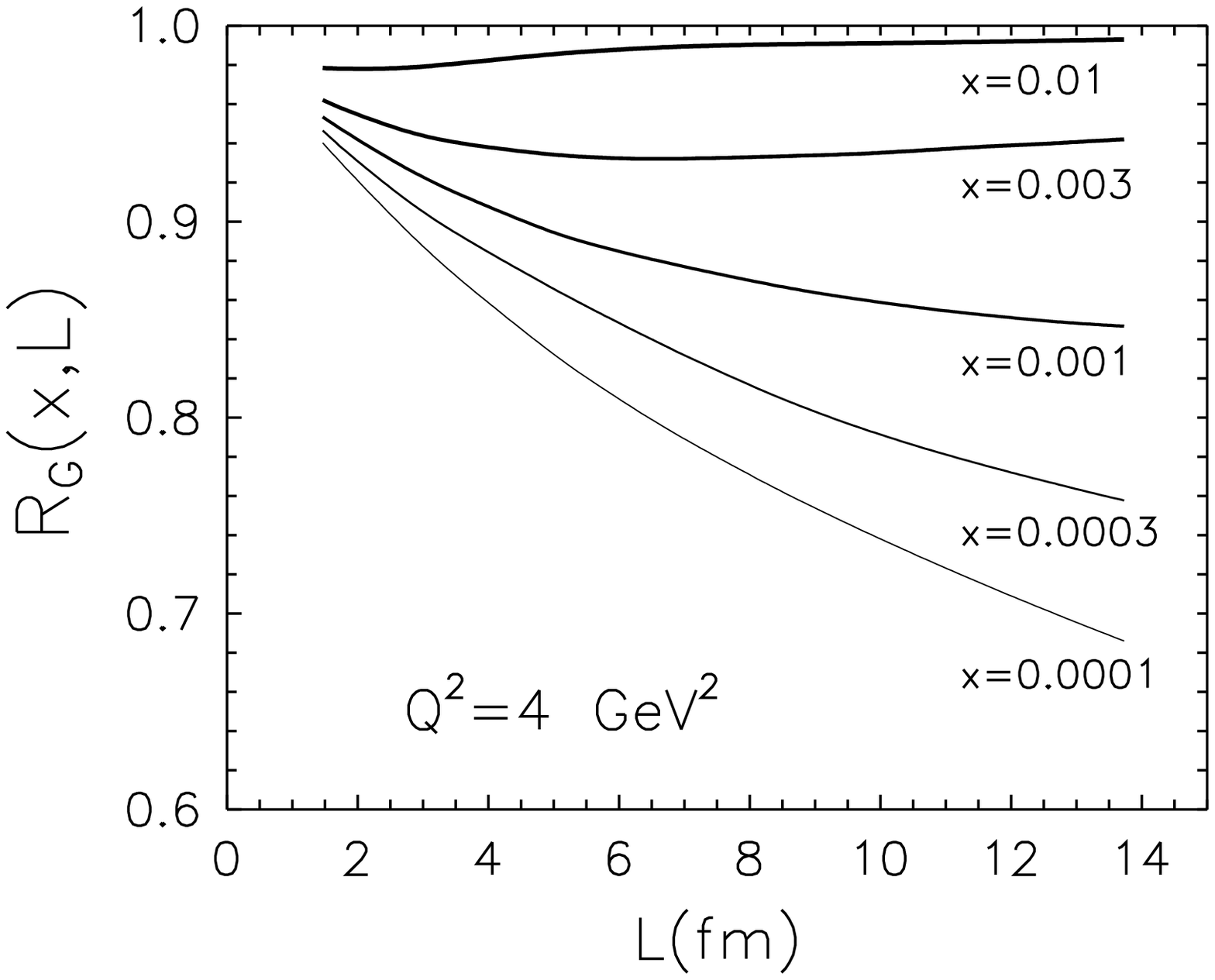}
\hspace{15mm}
\includegraphics[width=5.5cm]{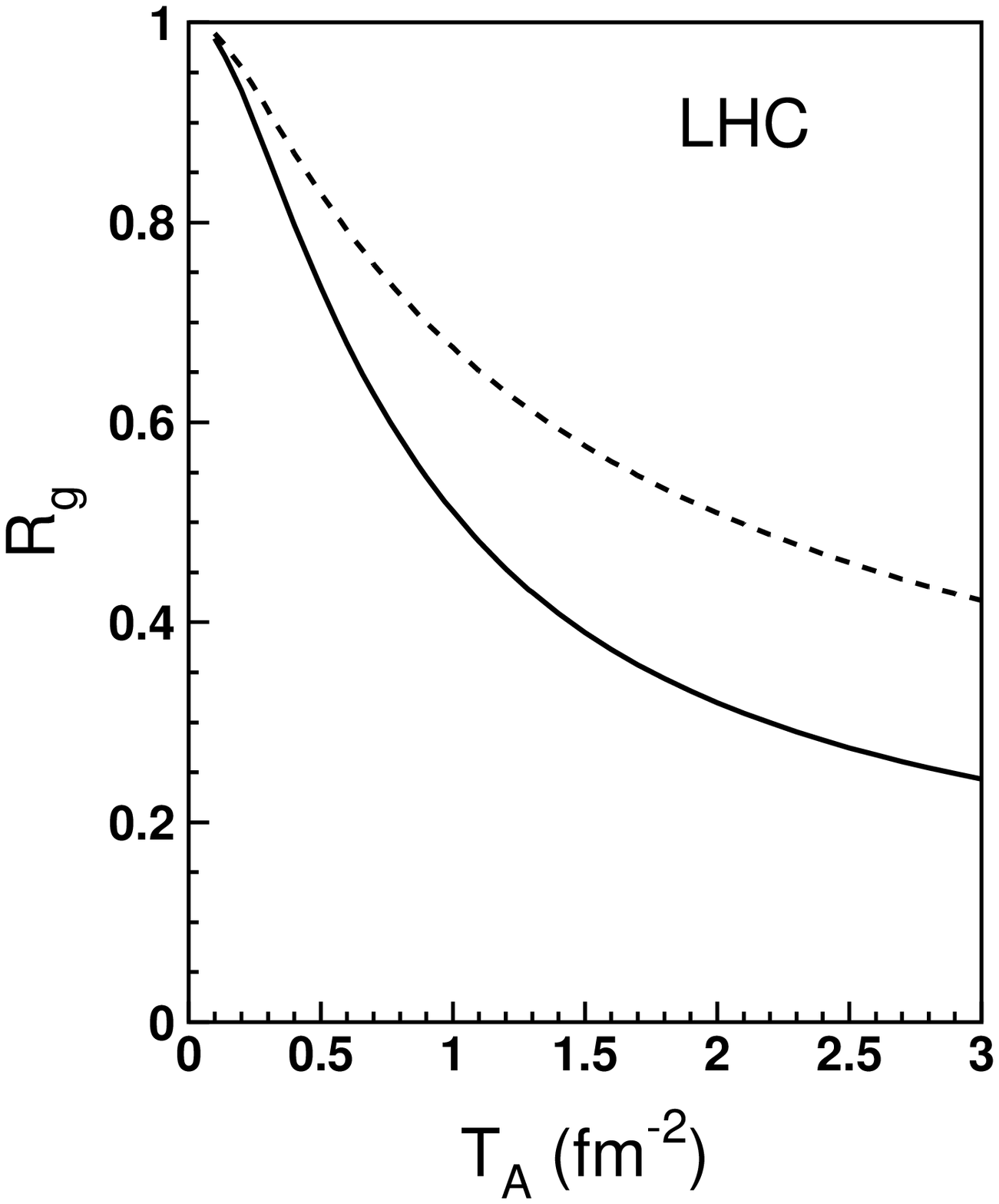}
\caption{   \label{glue-shad}{\it Left:} 
The nuclear ratio $R_g$ for gluons as function of path length in nuclear matter, calculated in \cite{jan} at 
$Q^2=4GeV^2$ for several fixed values of $x$.
{\it Right:} $R_g$ as function of 
nuclear thickness $T_A\approx 0.16\fm^{-3}\times L$, calculated  at the semihard scale $Q_0^2=1.7GeV^2$ and $x=10^{-4}$. 
The dashed curve presents gluon shadowing corresponding to
hadron-nucleus collisions. The solid curve includes the boosting effects specific for central nucleus-nucleus collisions ($T_A=T_B$). }
 \end{center}
\end{figure}

Thus, the magnitude of gluon shadowing evaluated in \cite{kst2} is rather small, as  is confirmed by  calculations performed in \cite{jan} and depicted in the left panel of Fig.~\ref{glue-shad}.

As we mentioned above, this is the most accurate method, which is valid in all regimes of gluon radiation, from incoherent to fully coherent. Nevertheless,  this is still
the lowest order calculation, which might be a reasonable approximation only for light nuclei, or for the onset of shadowing. Contribution of higher Fock components is still a challenge. This problem has been solved so far only in the unrealistic limit of long coherence lengths for all radiated gluons, in the form known as the Balitsky-Kovchegov equation (BK) \cite{b,k}.
Numerical solution of this equation is quite complicated and includes lot of modeling \cite{wiedemann}.
A much simpler equation, which only employs a modeled shape of the saturated gluon distribution, was derived in \cite{saturation}. It leads to a gluon distribution in nuclei, which satisfies the unitarity bound \cite{bound}, and the results are quite similar to the numerical solutions \cite{wiedemann} of the BK equation. The new equation reads \cite{saturation},
\beq
R_g=1-
\frac{R_g^2\,n_0^2\,n_{eff}}{(1+R_g\,n_0)^2(1+n_{eff})}
\label{400}
\eeq
where 
\beqn
n_0(E,b)&=& \frac{9\,C(E)}{2\,Q_{qN}^2(E)}\,T_A(b);
\nonumber\\
n_{eff}(E,b)&=&{9\over4}\,C(E)\,r_0^2\,T_A(b).
\label{420}
\eeqn
The energy dependent factor $C(E)$ was introduced above in (\ref{120})-(\ref{130}). The mean size of a gluonic dipole, $r_0\approx 0.3\fm$, as was already mentioned, is dictated by data \cite{kst2,spots}. 
We rely on the saturated shape of the dipole-nucleon cross section with the saturation scale $Q_{qN}(E) = 0.19\GeV\times(E/1GeV)^{0.14}$, fitted to DIS data \cite{kst2,saturation}. 
The solution of equation (\ref{400}) at $x=10^{-4}$ and $Q_0^2=1.7\GeV^2$ \cite{nontrivial} is depicted by dashed curve in the right panel of Fig.~\ref{glue-shad}.
Compared to the results depicted in the left panel we see that Eq.~(\ref{400}) predicts a somewhat stronger shadowing than the path-integral method ($Q^2$ dependence is rather weak).
This is not a surprise, the left figure was calculated accounting for phase shifts, which are not yet small even at $x=10^{-4}$. Indeed the $x$-dependence of shadowing shown in the left panel does not seem to saturate at this value of $x$. This means that in the saturated regime with very long coherence lengths for all gluons, which was assumed for the equation (\ref{400}), shadowing should be stronger.

Notice that Eq.~(\ref{400}) does not contain any hard scale, but only a semi-hard one controlled by $r_0$ \cite{spots}. Therefore its solution should be treated as the starting gluon distribution
at the semi-hard scale $Q_0^2\approx 4/r_0^2$. Then shadowing should be DGLAP evolved up to an appropriate hard scale.

\subsection{Gluon shadowing from data}\label{shadowing}

Since gluons dominate at small $x$, the nuclear modification of the gluon distribution is desperately needed, if one wants to understand nuclear effects in hard reactions at high energies. There is wide spread believe in the community that gluon distributions in nuclei are known from available global analyses of data.
Unfortunately, we should admit that most of the nuclear gluon PDFs, which are currently used in calculations and data analyses, are either an educated guess having little too do with data, or just plain wrong.

As an example, data for $J/\Psi$ production from the PHENIX experiment were analyzed \cite{phenix-psi} using two popular gluon PDFs, which according to the authors of \cite{phenix-psi} come from data:
"The modified  nuclear PDFs from EKS and NDSG are 
constrained from other experimental measurements such as 
deep inelastic scattering from various nuclear targets and the 
resulting $F_2(A)$ structure functions." Well, let us check whether such a claim is really justified.

\begin{figure}[htb]
\begin{center}
\includegraphics[width=7cm]{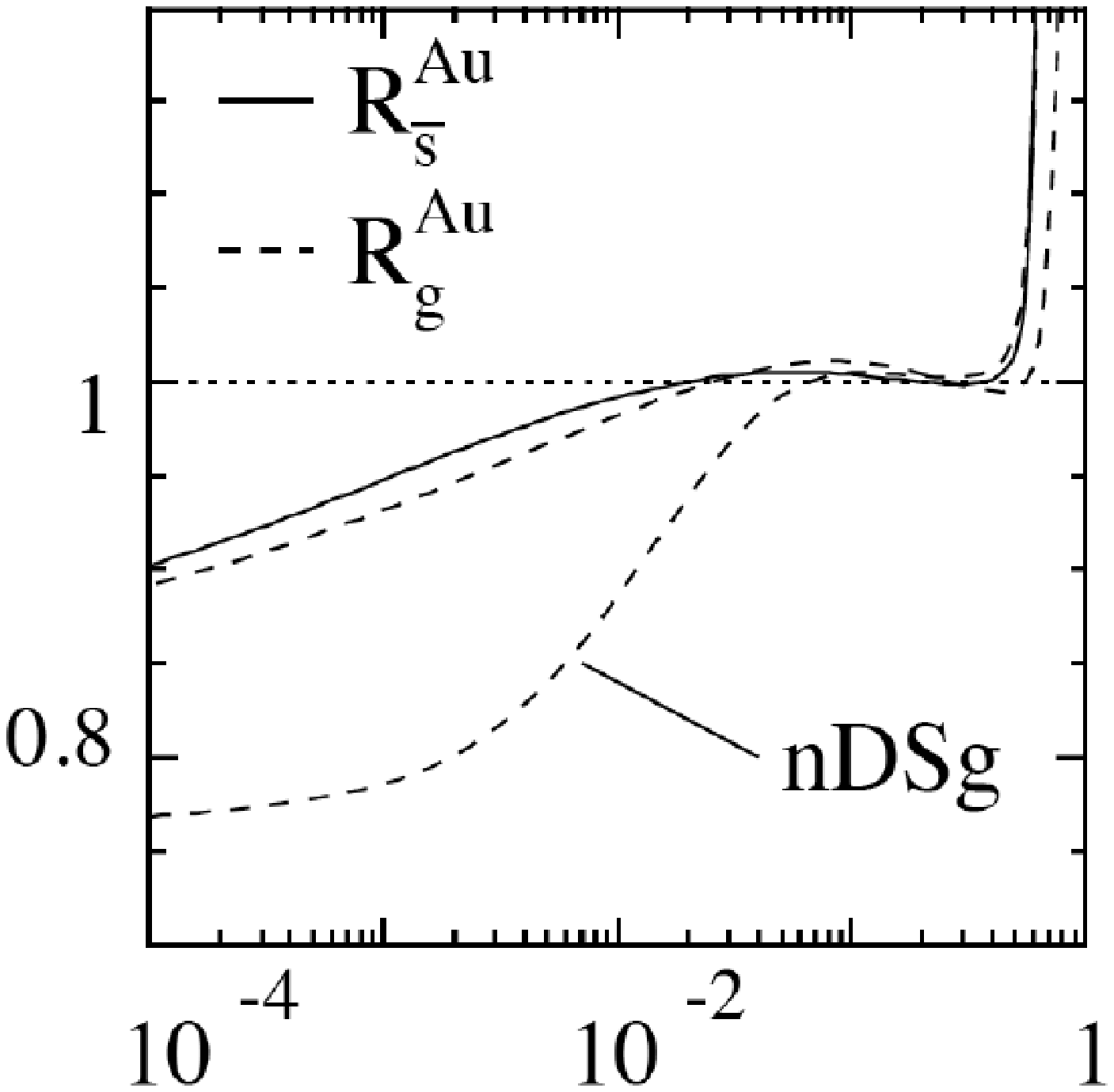}
\hspace{15mm}
\includegraphics[width=6.4cm]{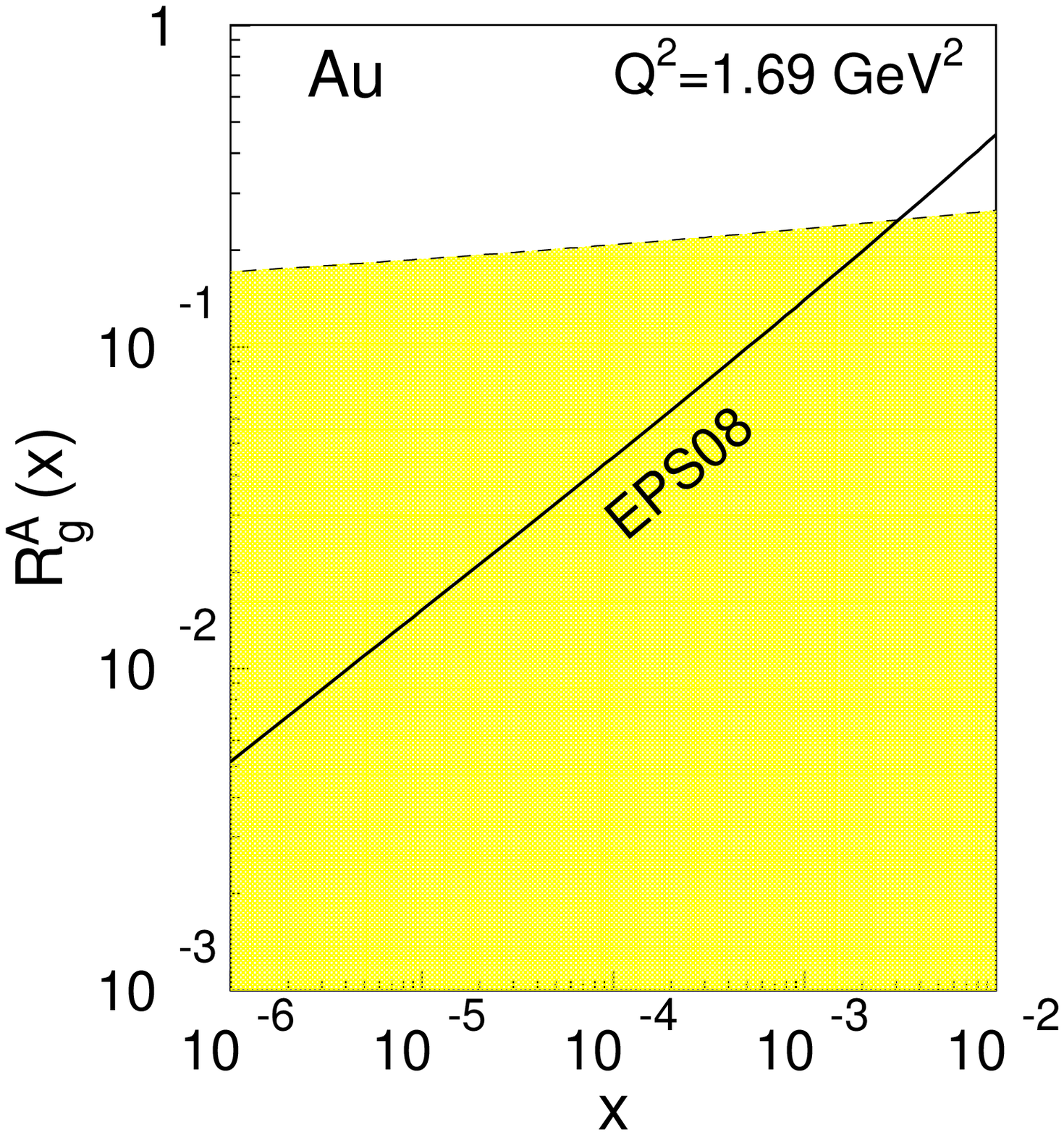}
\caption{  \label{Rg}{\it Left:} 
Nuclear ratios at $Q^2=10\GeV^2$ as function of $x$ fitted in \cite{DS} to DIS data.
The upper dashed curve presents the results of the fit for the gluon ratio $R_g$.
The bottom dashed curve (nDSg) is an ad hoc trial function (see text).
{\it Right:} The gluon ratio $R_g$ provided by the EPS08 code at $Q^2=1.69\GeV^2$.
The shaded area is forbidden by the unitarity bound \cite{bound} shown by dashed curve.}
 \end{center}
\end{figure}

First,  the nDSg plotted in the left panel of Fig.~\ref{Rg}, which presents the result of the global analysis of DIS data performed in \cite{DS}.
The gluon ratio $R_g$ which comes from the data analysis is depicted by the upper dashed curve,
while the bottom dashed curve, labelled by nDSg, is an ad hoc curve, which actually contradicts the DIS data. This is what the authors of \cite{DS} say about it: "We provide the result in set called 
nDSg, constrained to satisfy $R^{Au}_g = 0.75$ at $x_N = 0.001$ 
and $Q^2 = 5$. The $\chi^2$ value of this analysis is around 
550, considerably larger than the unconstrained fit, and 
should be considered only as a mean to study variations 
on, mainly, the gluon nuclear distribution." 
Strangely, the authors of \cite{phenix-psi} ignored the real solution for $R_g$, but picked up the fake curve, which provides a several times overestimated gluon shadowing.

The magnitude of gluon shadowing in EKS98 is similar to nDSg, and does not come from data either.
The gluon shadowing at small $x$ also was fixed at an ad hoc large value, $R_g=R_{F_2}$, what resulted in an overestimated gluon shadowing and antishadowing.
This is what the authors of \cite{eks98} say: "We point out, however, that even though we obtain a very good agreement with the NMC data [3] and with the analysis of Ref. [5], we can confirm our initial assumption of gluon shadowing at small values of x only on fairly qualitative grounds (stability of 
the evolution), rather than through a direct comparison with the data." 

The recent new global analyses \cite{eps08} tried to enhance the input from data by including
BRAHMS results for hadron production at forward rapidities \cite{brahms}. This was a risky addition, because
the mechanisms of suppression are still under debate (see Sect.~\ref{forward-sect}), and no consensus has been reached so far. This was probably the reason why this fit led to such astonishing results. The gluon shadowing
was found incredibly strong, stronger than is allowed by the unitarity bound \cite{bound}, as is demonstrated in the right panel of Fig.~\ref{Rg}. Apparently, the problem is caused by
misinterpretation of the suppression observed in \cite{brahms} relating it to gluon shadowing.

Indeed, the later analysis EPS09 \cite{eps09}, which wisely excluded the BRAHMS data, led to the PDFs, which failed to reproduce the suppression observed in the BRAHMS experiment at $\eta=3.2$.
It disagrees even more with STAR data at $\eta=4$ \cite{star-forward}. 
However, in the new analysis EPS09  the BRAHMS data  was replaced
by a new source of information about gluon nPDF, the Cronin effect in neutral pion production at the mid rapidity at RHIC, which created a new problem. The mechanism of the Cronin enhancement was quite misinterpreted in \cite{eps09}. This effect is a manifestation of the CGC phenomenon, which is a modification of the transverse momentum distribution of partons in nuclei increasing their mean transverse momentum up to the saturation scale. This also can be understood as broadening of partons propagating through a nucleus, as we discussed in Sect.~\ref{cgc}.

Such a modification of the $p_T$-distribution in nuclei has been always considered as a mechanism for the Cronin effect. This mechanism explains well data
from fixed target experiments \cite{cronin-prl,cronin-wang} and correctly predicted \cite{cronin-prl} the Cronin effect for pions measured at RHIC \cite{cronin-phenix}.
These data were interpreted in \cite{eps09} in the collinear approximation, but the Cronin exhancement was attributed to the nuclear modifications in nPDFs.
This mechanism can be disproved comparing it with data from fixed target experiments, where the Cronin enhancement is huge and cannot be explained by nPDFs.
So we conclude that gluon nPDF in EPS09 are not trustable, since the main source of information about nuclear modification of the gluon distribution is based on the
incorrect dynamics for the Cronin effect.

We should also comment on the global analysis HKN07 \cite{hkn07}, which was performed at the NLO level.
Among different kinds of modeling in the current theoretical estimates of gluon shadowing, some are more debatable, some less. In particular, our knowledge of the coherence time of gluon radiation, which controls the onset of gluon shadowing, is pretty solid. Shadowing is possible only if the coherence time becomes longer than the mean internucleon spacing, about $2\fm$.
It is clear that a $|\bar qqg\ra$ fluctuation of a virtual photon is heavier than a $|\bar qq\ra$ one,
therefore the coherence time for gluon shadowing must be shorter than for quark shadowing.
This is a very solid,  non-debatable statement dictated by kinematics. Thus the onset of gluon shadowing must occur at smaller $x$ than for quarks. The latter is well known both from theory \cite{krt2} and data \cite{nmc} to occur at $x\sim 0.07$. 

Such a restriction should be used as a physical input when the initial $x$-dependences of the PDFs at the starting scale are shaped. This apparently has not been done in the analysis \cite{hkn07}, which led to an unphysical $x$-dependence of gluon shadowing, which starts at $x=0.2$ and is  rather strong already at $x=0.1$, where it is excluded by the above arguments.

\section{Boosting the saturation scale in AA collisions}

Due to broadening a nuclear target probes the parton distribution in the beam hadron with a higher resolution. Therefore, the effective scale $Q^2$ for the beam PDF drifts to a higher value $Q^2+Q_{sA}^2$. At first glance this seems to contradict casuality, indeed, how can the primordial parton distribution in the hadron depend on the interaction which happens later? However, there is nothing wrong. The interaction performs a special selection of Fock states in the incoming hadron. The same phenomenon happens when one is measuring the proton parton distribution in DIS. The proton PDF "knows" in advance about the virtuality of the photon which it is going to interact with.

The shift in the scale also can be interpreted as a manifestation of the Landau-Pomeranchuk principle \cite{lp}: at long coherence times gluon radiation (which causes the DGLAP evolution) does not depend on the details of multiple interactions, but correlates only with the total momentum transfer, $\vec q+\Delta\vec p_T$, which after squaring and averaging over angles results in $Q^2+\Delta p_T^2$.

As far as the PDF of the projectile proton has a harder scale in $pA$ collisions than in $pp$, the ratio of parton distributions should fall below one at forward and rise above one at backward rapidities. This may look like a breakdown of $k_T$-factorization, however, it is a higher twist effect. 

\begin{figure}[htb]
\bc
 \includegraphics[width=6cm]{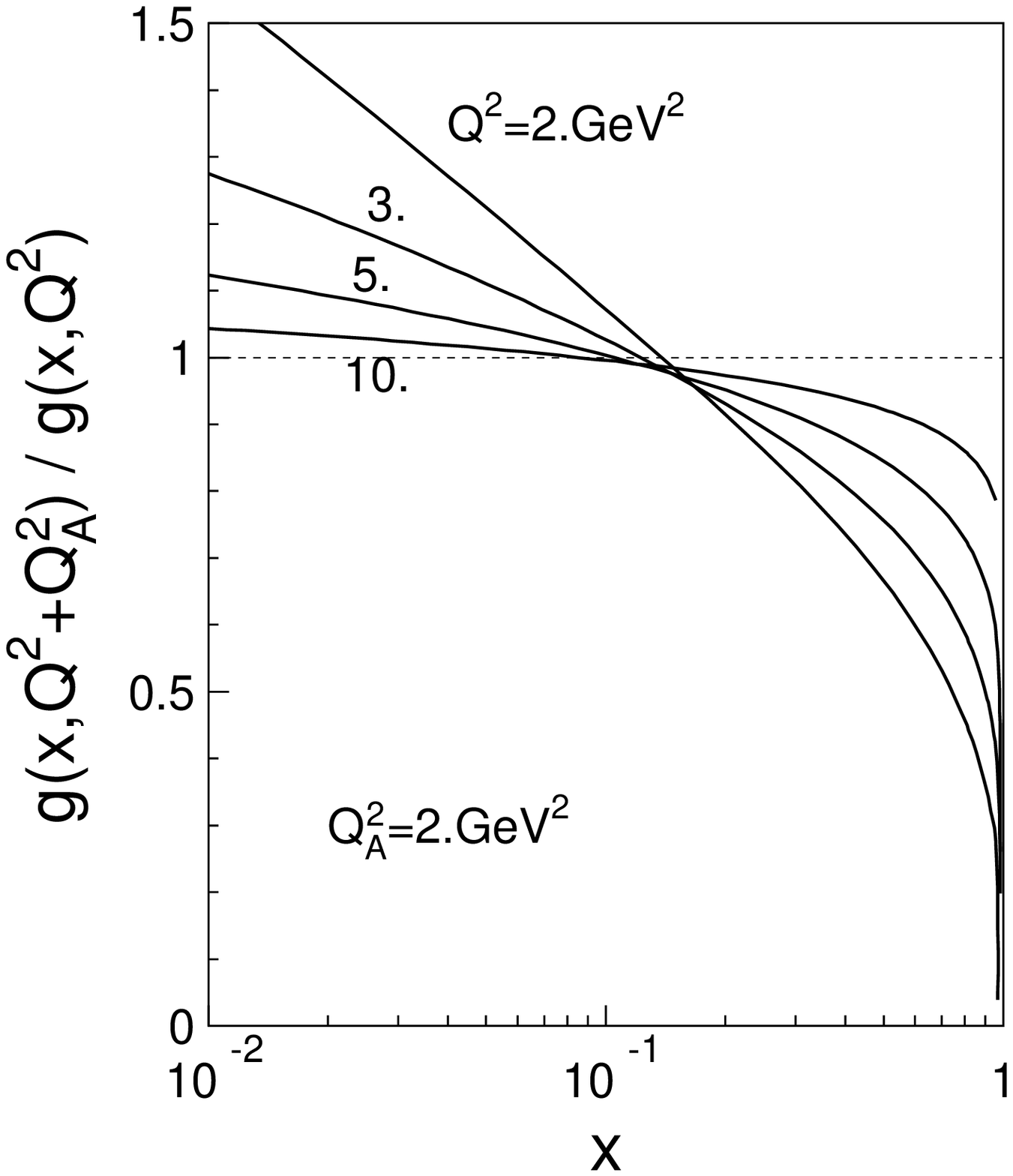} \hspace{25mm}
 \includegraphics[width=6cm]{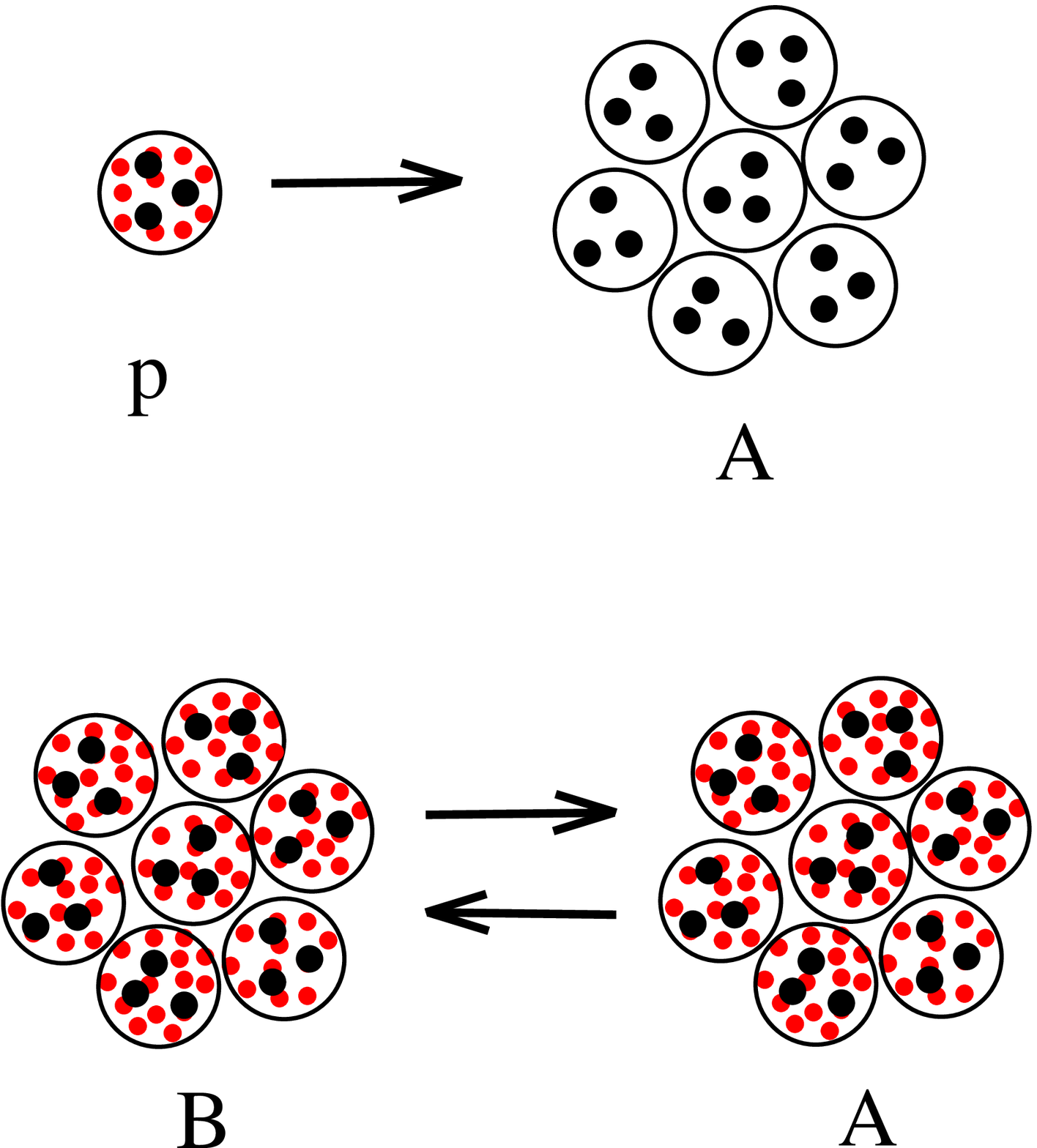}
 \caption{\label{rescaling} {\it Left panel:} Nucleus to proton ratio of the gluon distribution functions in a reaction characterized by the saturation momentum $Q_{sA}^2=2\GeV^2$ and  hard scale $Q^2=2,\ 3,\ 5,\ 10\GeV^2$.   
{\it Right panel (top):}  $pA$ collision in which the
colliding proton is excited by multiple interactions up to the
saturated scale $Q_{sA}^2$, which leads to an increased multiplicity
of soft gluons in the incoming proton. {\it Right panel (bottom):} nuclear
collision in which participating nucleons on both sides are boosted
to the saturation scales, $Q_{sA}^2$ in the nucleus $B$, and
$Q_{sB}^2$ in the nucleus $A$. As a result, the low-$x$ gluon
population is enriched in both nuclei.}
\ec

 \end{figure}

Examples of $pA$ to $pp$ ratios $R_A(x,Q^2)$ calculated with MSTW2008 \cite{mstw} are shown in Fig.~\ref{rescaling} for gluon distributions in a hard
reaction (high-$p_T$, heavy flavor production, etc.). 
We see that the shift in the hard scale caused by saturation in the
nucleus leads to a sizable suppression in the projectile parton
distribution at large $x\to1$ and enhancement at small $x\ll1$. We
also observe that the magnitude of nuclear modification quickly decreases with $Q^2$ confirming that this is a high twist effect.

Notice that in $pA$ collisions the modification of the PDFs of the beam and target are not symmetric. Namely, the scale of the PDF of the beam proton gets a shift, $Q^2\Rightarrow Q_{eff}^2=Q^2+Q_{sA}^2$,
while the PDFs of the bound nucleons, which do not undergo multiple interactions, remain the same as in $pp$ collisions. This is illustrated pictorially in the top part of the right panel in Fig.~\ref{rescaling}.

The situation changes in the case of a nucleus-nucleus collision: the bound
nucleons in both nuclei participate in multiple interactions,
therefore the scales of PDFs of all of them are modified. However,
this modification goes beyond the simple shift $Q^2\Rightarrow
Q^2+Q_{sA}^2$. Indeed, in an $AB$ nuclear collision not only the two
nucleons (one from $A$ and one from $B$) participating in the hard
reaction undergo multiple interactions, but also many other
nucleons, the so called participants, experience multiple soft
interactions. For this reason their parton distributions are boosted
from the soft scale $\mu^2$ up to the saturation scale
$\mu^2\Rightarrow \mu^2+Q_{sA(B)}^2$, which is usually much larger.
Thus, the participant nucleons on both sides are boosted to a higher
scale and get softer PDFs, with larger parton multiplicities at
small $x$. This is illustrated on the cartoon in the bottom part of the right panel in Fig.~\ref{rescaling}.

\subsection{Reciprocity equations}

Thus, a participating nucleon simultaneously plays the roles of a
beam and of a target. Its PDF is boosted to a
higher scale due to multiple interactions it undergoes in another
nucleus. As a target such a nucleon, being boosted to a higher scale
$Q_0^2\Rightarrow Q_0^2+Q_{sB}^2$,  increases broadening of
partons from another nucleus, since the factor  $C_q(E,Q^2)$
Eq.(\ref{130}) rises. This leads to a mutual enhancement of the
saturation scales in both nuclei. Indeed,  multiple rescatterings of
nucleons from the nucleus $A$ on the  boosted nucleons in $B$
proceed with a larger cross section, so broadening, i.e. the
saturation scale in $B$ increases, $Q_{sB}^2\Rightarrow \tilde
Q_{sB}^2>Q_{sB}^2$. For this reason, the nucleon PDFs in $A$ get
boosted more. Then the partons from $B$ experience even stronger
multiple interactions with such double-boosted nucleons in $A$. This
results in an additional boost of the saturation scale in $A$, then, as a result, in $B$, and so on. 

\begin{figure}[htb]
\begin{center}
 \includegraphics[width=12cm]{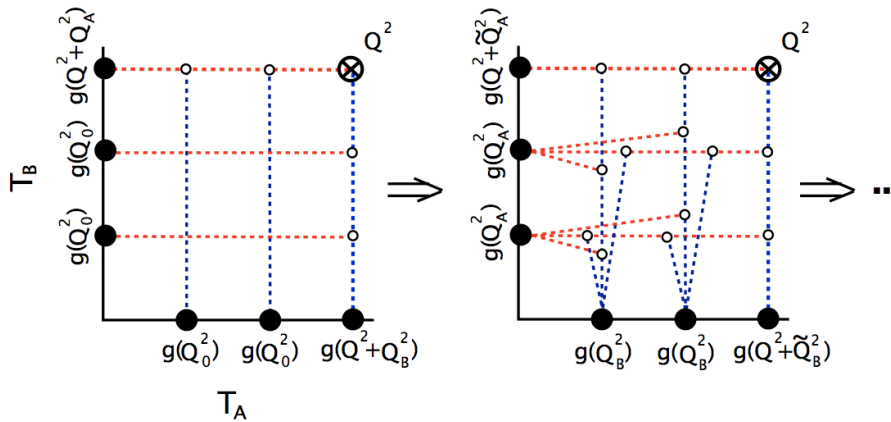}
 \end{center}
\caption{\label{boosting-fig} Collision of two one-dimensional rows of nucleons $T_A$ and $T_B$ displayed on horizontal and vertical axes. 
Multiple interactions of colliding gluons propagated through both nuclei,
including additional multiple scatterings of gluons, which carry out the interactions, are shown as is described in text. }
 \end{figure}

Such a multi-iteration mutual boosting of the saturation scales is illustrated
pictorially in Fig.~\ref{boosting-fig}, where two rows of nucleons, $T_A$ and $T_B$ are displayed on horizontal and vertical axes. 

To proceed further we present the function $C(E,Q^2)$ in the form \cite{frs,bartels},
\beq
C(E,Q^2)=\frac{\pi^2}{3}\,\alpha_s(Q^2+Q_0^2)\,xg_N(x,Q^2+Q_0^2),
\label{410}
\eeq
where $g_N(x,Q^2)$ is the gluon distribution function in the nucleon, and 
$x=2m_NE/s$ is the fractional momentum of a gluon experiencing broadening.
The scale shifting parameter $Q_0$ is introduced in order to have the correct infrared behavior: 
in the limit $Q^2\to0$ Eq.~(\ref{410}) has to reproduce the known function $C(E)$ depicted in the left panel of Fig.~\ref{broad-fig}. This condition fixes the parameter $Q^2\approx 1.7\GeV^2$, which is nearly independent of energy.

Following Eq.~(\ref{410}) the modified gluon saturation scales $\tilde Q_s^2$ in the collision of two rows of nucleons  $T_A$ and $T_B$ can be found solving the reciprocity equations \cite{boosting},
 \beqn
\tilde Q_{sB}^2(x_B)&=&\frac{3\pi^2}{2}\,\alpha_s(\tilde Q_{sA}^2+Q_0^2)\,
x_B g_N(x_B,\tilde Q_{sA}^2+Q_0^2)\,T_B;
\nonumber\\
\tilde Q_{sA}^2(x_A)&=&\frac{3\pi^2}{2}\,\alpha_s(\tilde Q_{sB}^2+Q_0^2)\,
x_Ag_N(x_A,\tilde Q_{sB}^2+Q_0^2)\,T_A.
\label{430}
\eeqn
These equations take into account the modification
of the properties of bound nucleons in each of the colliding nuclei
due to multiple interactions in another nucleus and the following
increase of the scale. 

\begin{figure}[htb]
\begin{center}
 \includegraphics[width=6.5cm]{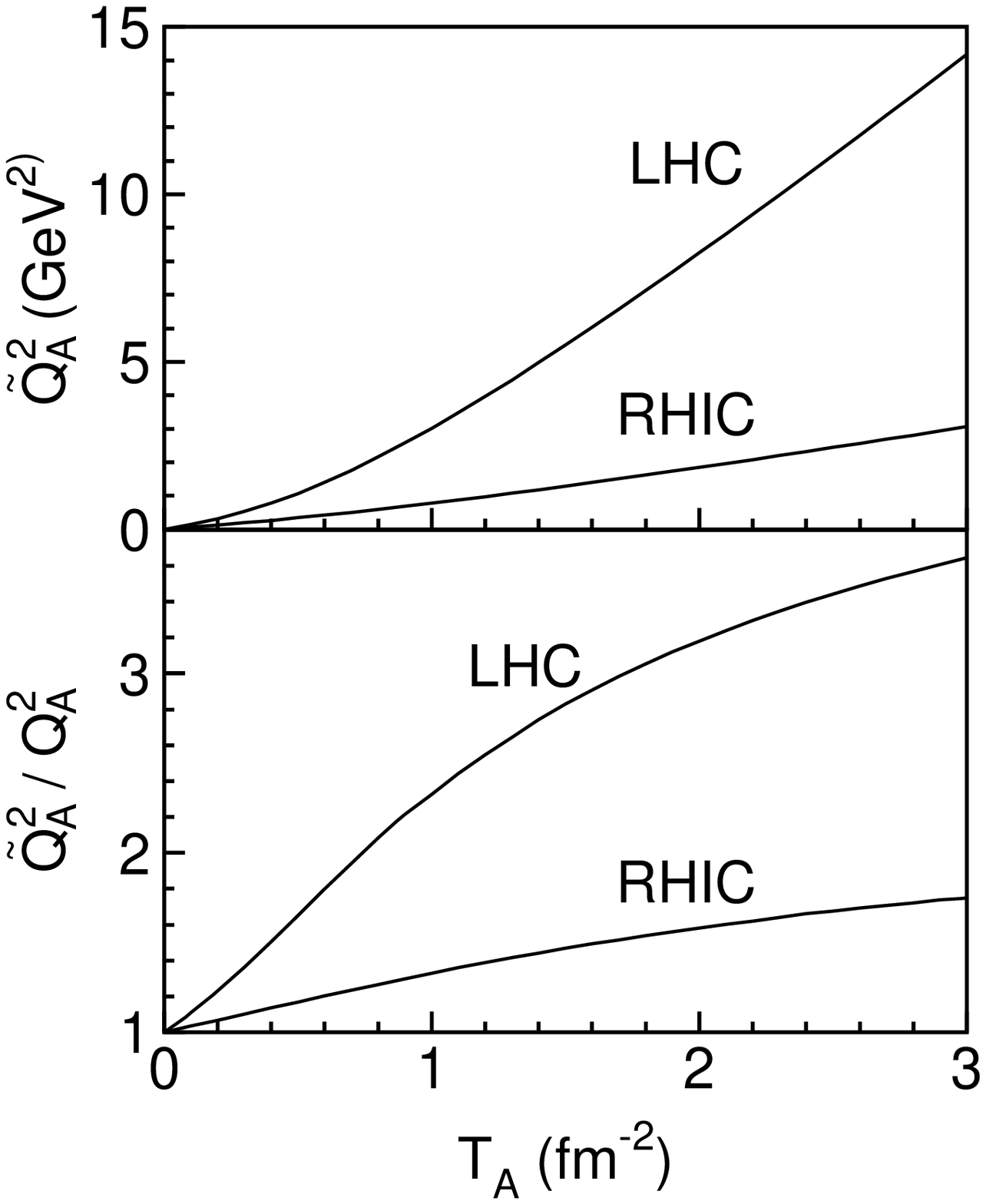}\hspace{1.5cm}
 \includegraphics[width=6.5cm]{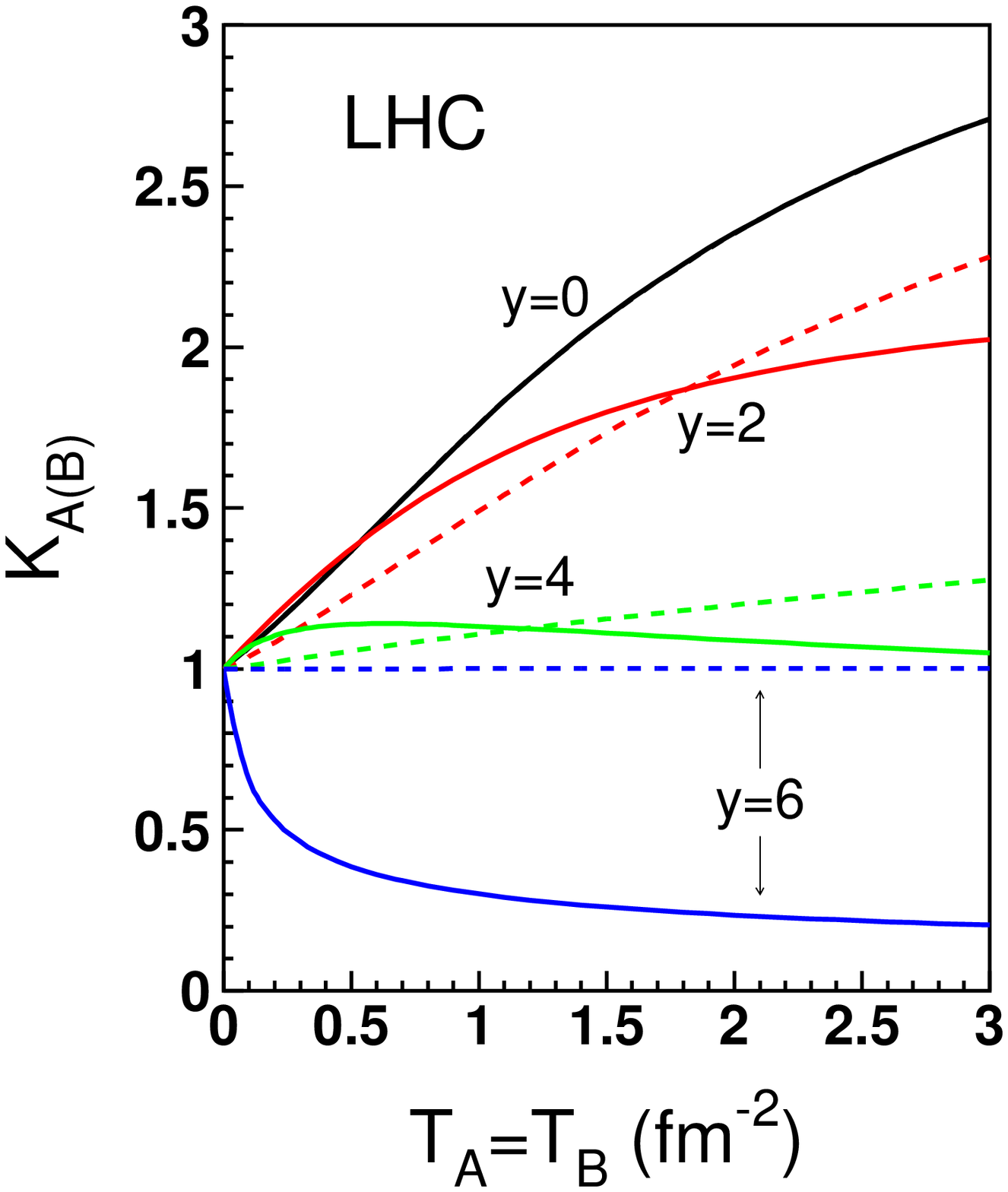}
 \end{center}
 \caption{ \label{results} {\it Left panel (top):}  the boosted values of the saturation momentum squared $\tilde Q_{sA}^2$ calculated at the energies of RHIC and LHC as function of nuclear thickness.
 {\it Left panel (bottom):} the boosting factor $K_A=\tilde Q_{sA}^2/Q_{sA}^2$ as function of $T_A$
 at the energies of RHIC and LHC.
{\it Right panel:} The boosting factors $K_A(T_A)$ 
 (solid curves) and $K_B(T_A)$ (dashed curves) as function of nuclear thickness at $\sqrt{s}=5.5\TeV$. Each pair of curves is marked by the rapidity for which it is calculated. 
 }
 \end{figure}

Solutions of these equations \cite{boosting} for $x_A=x_B=Q_0^2/s$ for central collision of two identical nuclei at the energies of RHIC and LHC  is plotted in the left panel of Fig.~\ref{results} as function of $T_A$, for the modified saturation scale (top) and for the ratio $K_A=\tilde Q_{sA}/Q_{sA}$. 
 For $g(x,Q^2)$ we use the LO gluon distributions of the recent analysis MSTW2008 \cite{mstw}. 
We see that the saturation scale of heavy nuclei may be as large as
about $10\GeV^2$ at the LHC.
In the right panel of Fig.~\ref{results} we plotted the ratios $K_A(T_A)$ and $K_B(T_A)$ with $x_A$ and $x_B$ calculated for charmonium production at different rapidities
in the central collision of identical  nuclei, $A=B$ at $\sqrt{s}=5500\GeV$ \cite{nontrivial}.
The saturation scales in $A$ (solid curves) and $B$ (dashed curves) are different, because the sign of rapidity is defined relative to the "beam"  nucleus $A$.

\subsection{Boosted gluon shadowing}

As we have already mentioned, in terms of the Fock decomposition gluon shadowing corresponds to multiple interactions of higher Fock states, containing gluons. One may wonder of gluon shadowing in nuclear collisions factorize?
\beq
R_g^{AB}(\vec b,\vec\tau)=R_g^A(\tau)\,R_g^B(\vec b-\vec\tau).
\label{450}
\eeq
The answer is yes. The lifetime of gluonic fluctuation produced by a nucleon in the nucleus $A$ may be sufficiently long only relative to the nucleus $B$, but is very short relative to the parent nucleus $A$. Therefore no gluonic fluctuations undergo double color filtering. In terms of Gribov inelastic shadowing this means that the diffractive excitation of the nucleons of $A$ propagate through $B$ independently of the excitations of $B$ propagating through $A$.

One has to rely on a fully developed theoretical model for gluon shadowing to predict its $b$-dependence.
The effect of boosted saturation scale leads to a modification of the factor $C(E)$
which is different for nuclei $A$ and $B$,
\beq
C(E)\,\Rightarrow\, \tilde C_{A}(E,T_A)=K_{A}(T_A)\,C(E).
\label{440}
\eeq
Solving equation (\ref{400}) with such a boosted saturation scale one arrives at a modified nuclear ratio for gluons in nuclei $A$ and $B$,
$\tilde R^{A(B)}_g(E,b)$, for which the factorized relation (\ref{430}) can be used \cite{nontrivial},
\beq
\tilde R_g^{AB}(\vec b,\vec\tau)=\tilde R_g^A(\tau)\,\tilde R_g^B(\vec b-\vec\tau).
\label{460}
\eeq

A numerical example for the boosting effect on gluon shadowing is depicted by the solid curve in the right panel of Fig.~\ref{glue-shad}.
 We see that the boosting effect considerably enhances gluon shadowing.

\section{Forward rapidities: challenging the suppression mechanisms}\label{forward-sect}

\subsection{Saturation effects}

The effects of saturation and shadowing are enhanced at small $x$, i.e. at higher energies. If the energy of collisions is fixed or restricted, one still can go to smaller $x$ relying on the relation
\beq
x_1\,x_2=\frac{k_T^2}{s},
\label{480}
\eeq
where $x_{1,2}$ are not the values of Bjorken $x$ for colliding partons, but are the fractional light-cone momenta of the produced high-$k_T$ parton relative to 
the beam and target respectively. Thus, one can reach smaller $x_2$ at fixed $s$, simply increasing
$x_1\to1$, i.e. moving to forward rapidities. Then the coherence phenomena (gluon shadowing, CGC) are expected to set up and suppress the particle production rate. 
However, as we multiply emphasized, these effects are rather weak, especially for the kinematics of RHIC experiments where the values of $x_2\gsim 10^{-3}$, which can be accessed at forward  rapidities, are not sufficiently small for gluon shadowing.

\begin{figure}[htb]
\begin{center}
 \includegraphics[width=8.5cm]{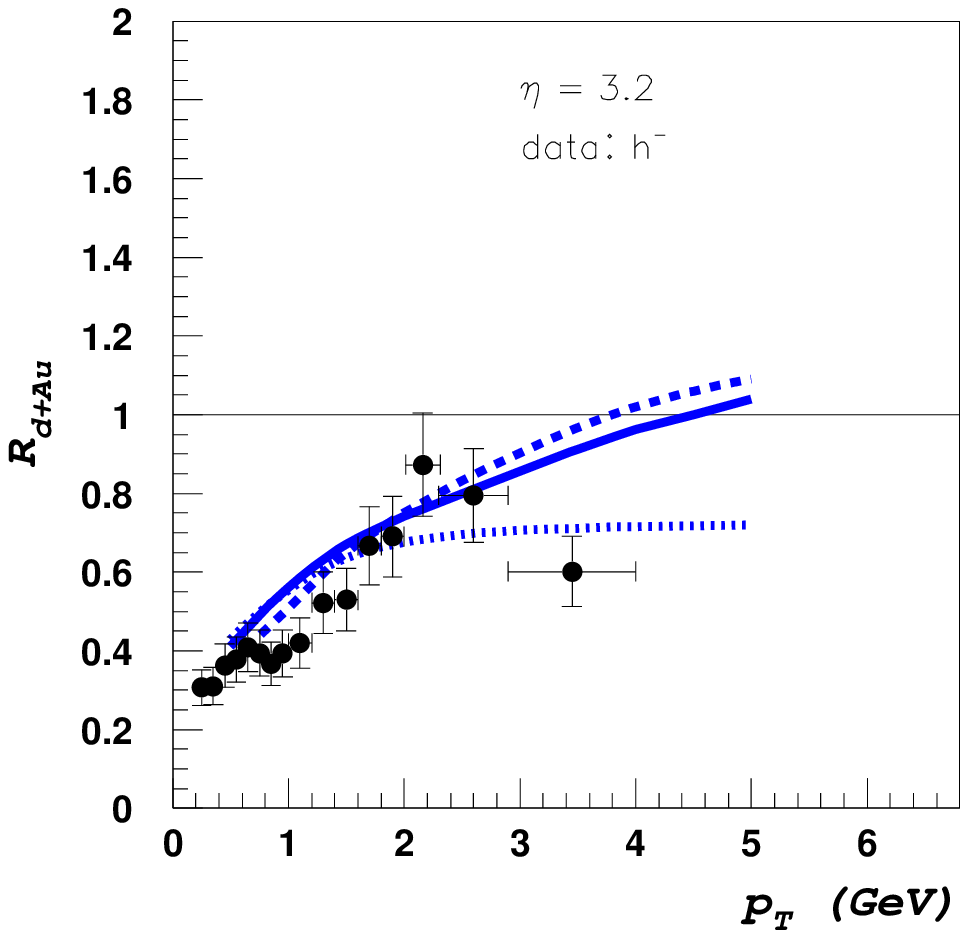}%\hspace{-5mm}
 \includegraphics[width=6.5cm]{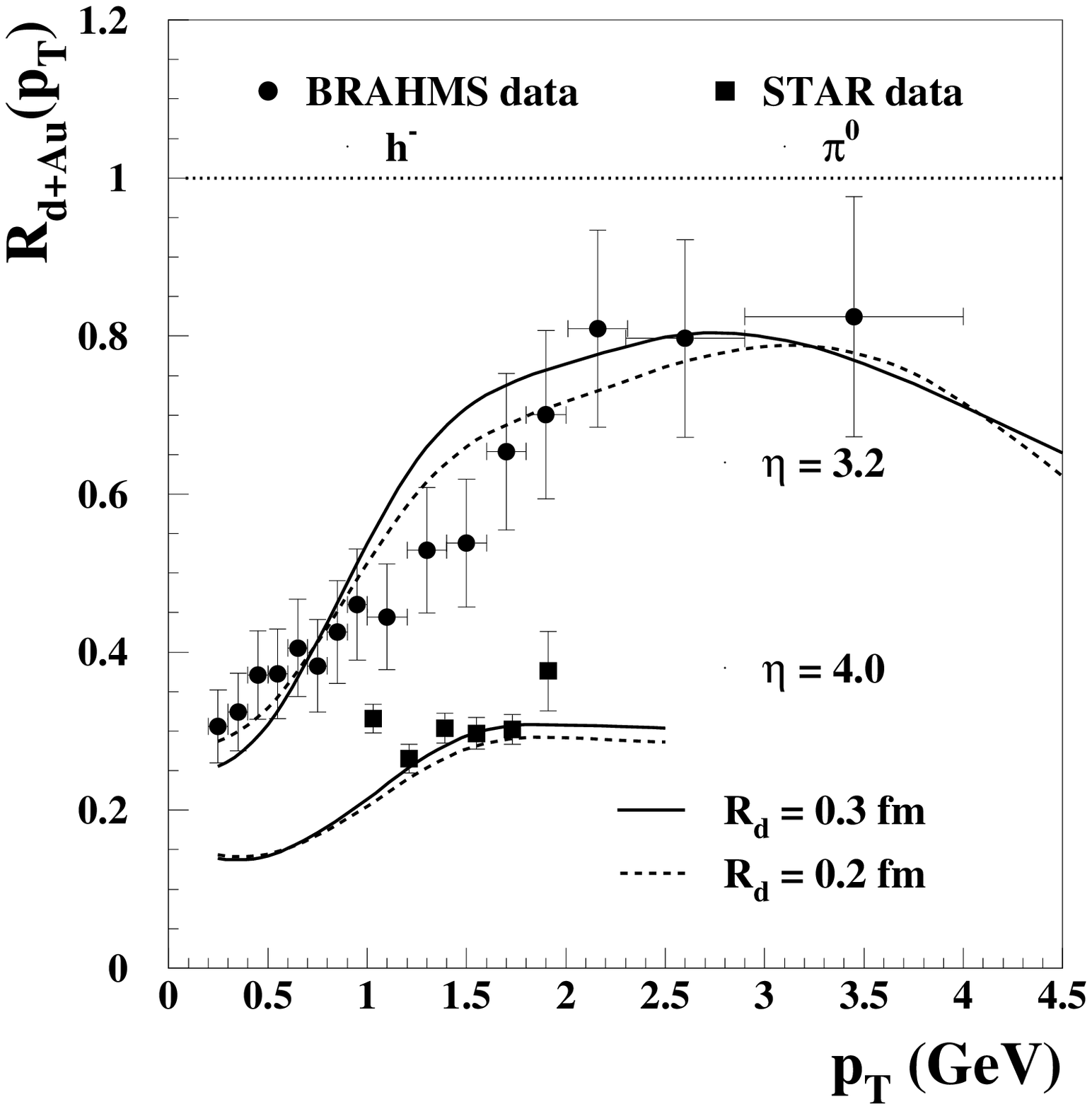}
\end{center}
 \caption{ \label{kkt-fig} {\it Left panel:}  Nuclear modification factor $R_{dAu}$ of negative particles 
at $\eta=3.2$.  Data is from \cite{brahms}.
The solid line represents the main result of \cite{kkt} for the $h^-$ contribution. The meaning of other curves is explained in \cite{kkt}.
{\it Right panel:} Nuclear ratio, $d-Au$ to $pp$, for negative particles as function of $p_T$ at pseudorapidity $\eta = 3.2$ \cite{brahms}, and data for neutral pion
production at $\eta = 4$ vs \cite{star-forward} are depicted by round points and squares respectively.
Solid and dashed curves correspond to calculations with the diquark size $0.3\fm$
and $0.2\fm$ respectively \cite{knpjs,npps-08}.  
 }
 \end{figure}

Nevertheless, when a suppression was indeed observed in the BRAHMS experiment \cite{brahms}, it was quite tempting to explain data by the effects of CGC. The model proposed in \cite{kkt} contained few parameters fitted to the BRAHMS data. The results are depicted by the thick curve in Fig.~\ref{kkt-fig}.
%\vspace*{-10mm}
 Notice that latest analysis \cite{albacete} also fits the absolute value of the ratio allowing different $K$-factors for nuclear and proton targets.
Although agreement with data is reasonable, when data to be explained are fitted, the missed sources of suppression can be easily absorbed  into the fit. One of them, the deficit of energy, is presented in the next section.

\subsection{ \label{e-loss}Deficit of energy at forward rapidities}

Multiple interactions of the projectile hadron and its debris propagating through the nucleus should cause a dissipation of energy. This intuitive expectation is supported by consideration of the Fock state decomposition. The projectile hadron can be expanded over different states which are the fluctuations of this hadron.  In the limit of infinite momentum frame those fluctuations live forever. One can probe the Fock state expansion by interaction with a target. The interaction modifies the weights of the Fock states, some interact stronger, some weaker. An example is the light-cone wave function of a transversely polarized photon \cite{krt2}. In vacuum it is overwhelmed by $\bar qq$ Fock states with vanishingly small separation (this is why the normalization of the wave function is ultraviolet divergent). However, those small size fluctuations have a vanishingly small interaction cross section, and the photoabsorption cross section turns out to be finite.

In each Fock component the hadron momentum is shared by the constituents, and the momentum distribution depends on their multiplicity: the more constituents are involved, the smaller is the mean energy per a constituent parton, i.e. the softer is the fractional energy distribution of a leading parton.
So on a nuclear target the projectile parton distribution  falls at {$ x\to1$} steeper than on a proton. This is similar to the rescaling of the PDFs by the saturation scale we observed earlier (see Fig.~\ref{rescaling}, left panel), but a much stronger effect. 

In the case of a hard reaction on a nucleus, this softening of the projectile parton momentum distribution can be viewed as an effective loss of energy of the leading parton in the nuclear medium, because the initial state multiple interactions enhance the weight factors for higher Fock states in the projectile hadron. Those components with large number of constituents have a tough energy sharing, so the mean energy of the leading parton decreases compared to lower Fock states, which dominate the hard reaction on a proton target. Such a reduction of the mean energy of the leading parton can be treated as an effective energy loss, which is proportional to the initial hadron energy.
Indeed, the partons responsible for multiple Glauber collisions carry substantial fractions of the initial energy (formally, they should be included into the hadron-multipomeron vertex, rather than into a ladder \cite{agk}). Thus, the effective loss of energy is proportional to the initial energy.

There is an important difference between this effect and energy loss of a single parton propagating through a medium and experiencing induced gluon radiation.
In this case the mean fractional energy carried the radiated gluons vanishes with initial energy $E$ as $\Delta E/E\propto1/E$ \cite{feri,bh,bdmps}. This energy loss is independent of the parton energy.

As far as multiple collisions suppress the cross section of leading particle production (heavy dilepton, high-$p_T$ hadron, heavy quarkonium, etc.), we assume that every collision brings in a suppression
factor $S(\xi)$ \cite{knpjs}, where
\beq
\xi=\sqrt{x_L^2+x_T^2};
\label{510}
\eeq
\beqn
x_L&=& \frac{2p_L}{\sqrt{s}};
\nonumber\\
x_T&=& \frac{2p_T}{\sqrt{s}};
\label{520}
\eeqn
and $p_{L,T}$ are the longitudinal and transverse components of the momentum of the produced particle in the c.m. frame. Notice that $x_L$ coincides with the Feynman variable, $x_L=x_F$.

This factor $S(\xi)$ should cause a strong (for heavy nuclei) suppression at $\xi\to1$, but some enhancement at small $\xi$ due to the feed down from higher $\xi$. This is because energy conservation does not lead to disappearance (absorption) of particles, but only their re-distribution in $\xi$.
 
 Since at $\xi\to1$ the kinematics of an inelastic collision corresponds to no particle produced within 
the rapidity interval $\Delta y\sim-\ln(1-\xi)$, the suppression factor $S(\xi)$ can be also treated as survival probability of a large rapidity gap, analogous to the Sudakov suppression for no gluon radiation. 
Assuming the Poisson distribution for the radiated gluons and using the gluon production rate estimated in \cite{gb}, the probability of such a rapidity gap was found \cite{knpjs} to be approximately,
\beq
S(\xi)\approx1-\xi.
\label{540}
\eeq
This also goes along with the results of the dual parton model \cite{kaidalov, capella}.

With such a suppression factor and applying the AGK cutting rules \cite{agk} with the Glauber weight factors, one achieves a parameter-free description of data depicted in the right panel of Fig.~\ref{kkt-fig}.
With no adjustment the model also well describes the STAR data at $\eta=4$ \cite{star-forward}.

\begin{figure}[htb]
\begin{center}
  \includegraphics[width=6cm]{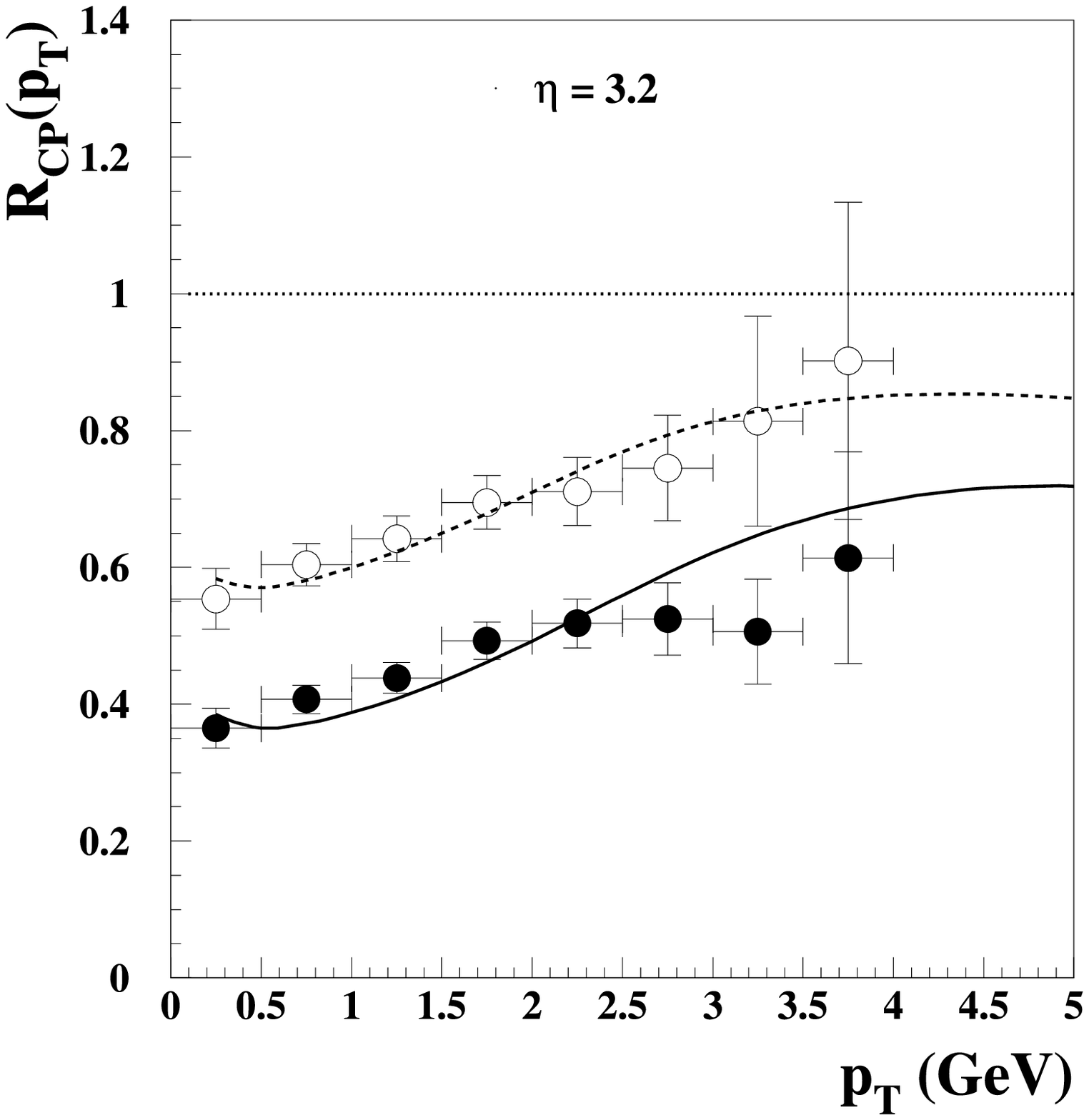}\hspace{2cm}
\includegraphics[width=6cm]{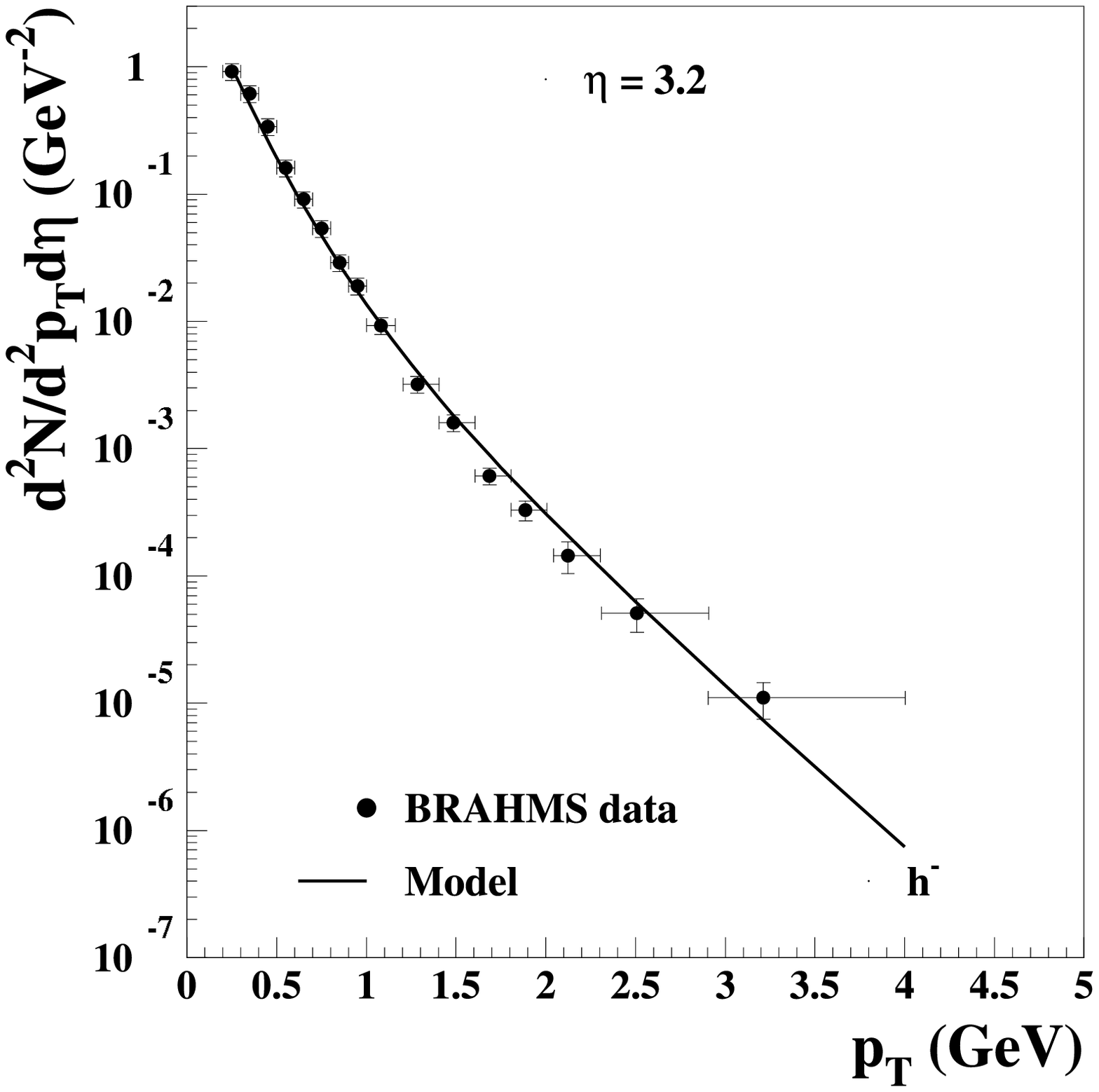}
 \caption{ \label{forward-fig}{\it Left panel:}  Ratio of negative particle production in central (0-20\%) and
semi-central (30-50\%) to peripheral (60-80\%) $d$-$Au$ collisions, shown by
closed and open points respectively. The results of 
calculations  \cite{knpjs} are plotted by solid and dashed curves.
{\it Right panel:} Number of negative hadrons versus $p_T$ produced in $pp$ collisions at
$\sqrt{s}=200\GeV$ and pseudorapidity $\eta=3.2$. The results of the dipole model
plotted by solid curve, are compared with BRAHMS
data \cite{brahms}.
}
\end{center}
\end{figure}

As an additional test, one can also look at the $p_T$ spectra at different centralities \cite{brahms}.
The model agrees well with this data also, as is demonstrated in the left panel of Fig.~\ref{forward-fig}.
 
 The last, but not least check is comparison with the $p_T$ distribution of pions produced in $pp$ collisions at this rapidity. The dipole model used in the calculations performed in \cite{knpjs}
 describes well both the absolute value and $p_T$ dependence of data \cite{brahms} presented in the right panel of Fig.~\ref{forward-fig}.
 
 Having two competing models, which are able to explain data at forward rapidities, one should
 look for specific reactions and kinematic domains where the models can be disentangled.
 Several tests are proposed below.
 
\subsubsection{Test \#1: Down to smaller energies, other processes}

As we mentioned above, the main motivation for moving to forward rapidities is getting access to smaller Bjorken $x$. If smallness of $x$ indeed is the reason for the observed suppression, this effect should
disappear at lower energies, because $x$ rises as $\propto 1/\sqrt{s}$. Having no other mechanism
contributing to the suppression observed by BRAHMS, but only CGC, one should expect no suppression at forward rapidities at lower energies.

On the other hand, the suppression caused by energy deficit scales in Feynman $x_F$ and should exist at any energy. Thus lowering the collision energy would be a sensitive test for the models.

\begin{figure}[htb]
\begin{center}
 \includegraphics[width=6cm]{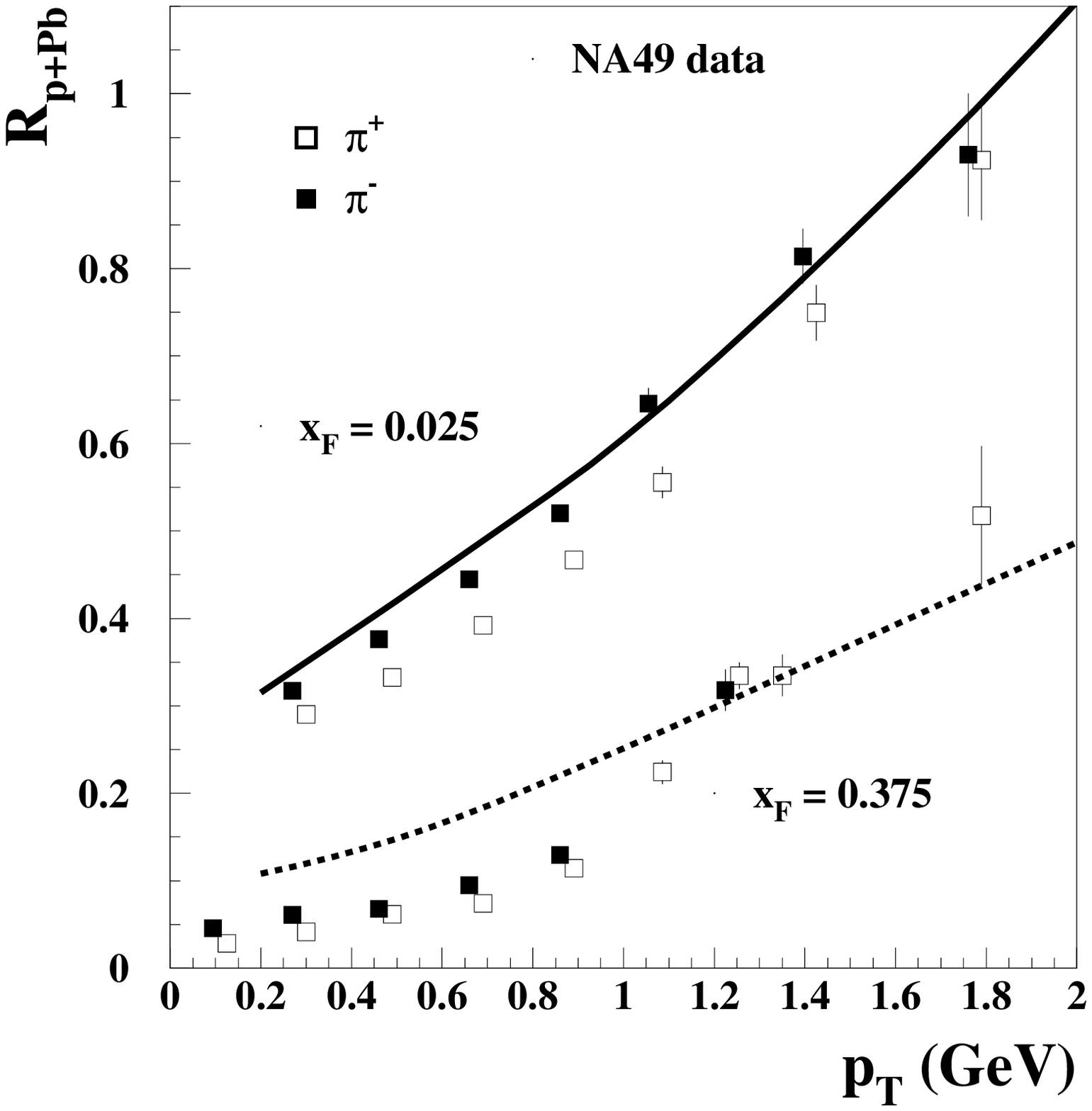}\hspace{2cm}
 \includegraphics[width=6.7cm]{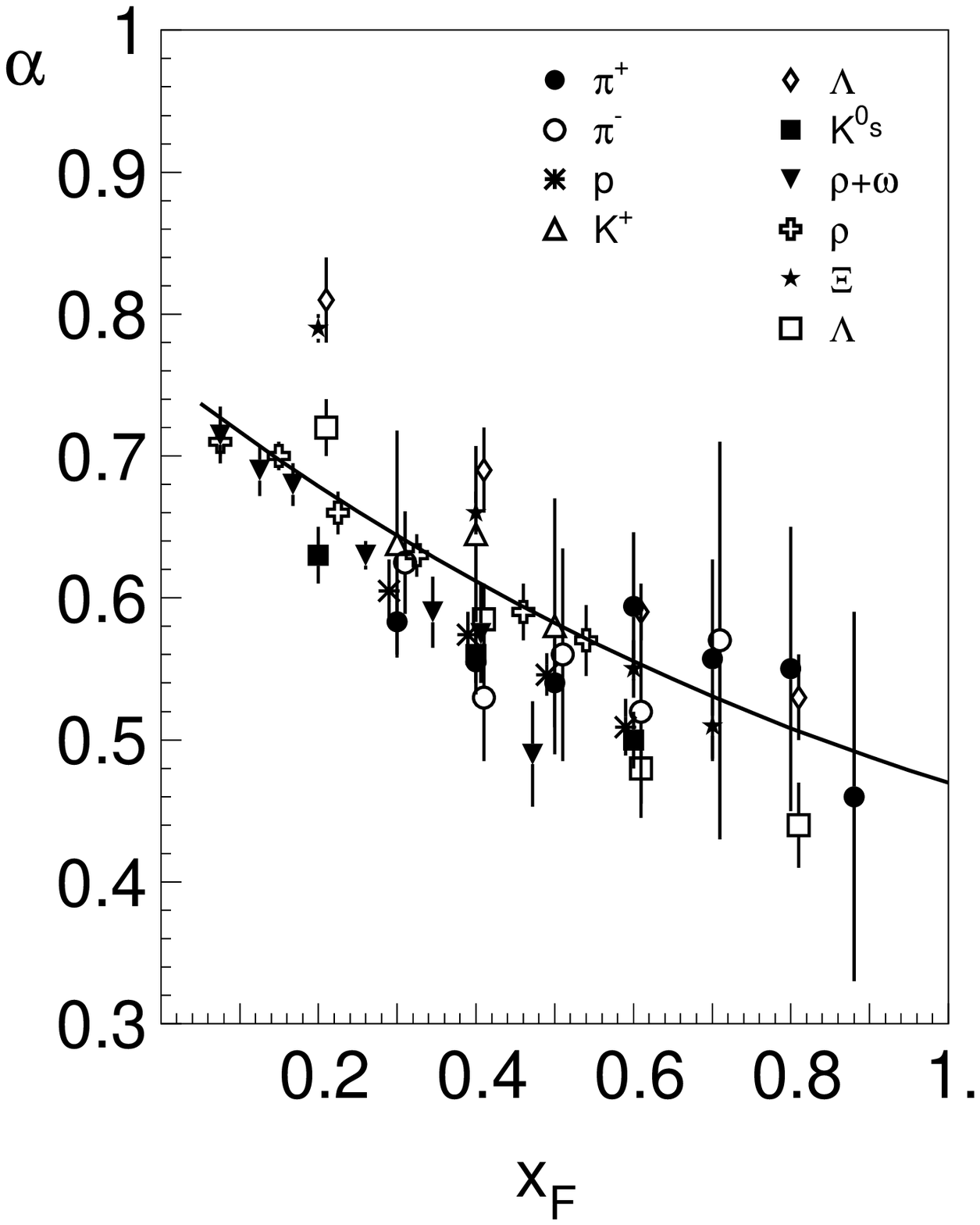}
 \end{center}
 \caption{ \label{fixed-target} {\it Left panel:}  The nuclear ratio for pions produced in proton-lead collisions at $E_{lab}=158\GeV$ as function of $p_T$ for two values of Feynman $x_F=0.025$ and $0.375$ \cite{na49}. The curves present the results of the parameter free calculations \cite{knpjs}
described in Sect.~\ref{e-loss}.
{\it Right panel:} The exponent for the $A^\alpha$ dependence fitted to data on total yield of different species of hadrons produced in $pA$ collisions at different energies as function of $x_F$.
The curve  present the results of the parameter free description of \cite{knpjs} (see Sect.~\ref{e-loss}).
 }
 \end{figure}

The NA49 experiment at SPS has performed measurements \cite{na49} similar to BRAHMS, but at much lower energy, where the value of $x_2$ is two orders of magnitude larger than in the BRAHMS data. The results depicted in the left panel of Fig.~\ref{fixed-target} show that the effect of suppression at forward rapidities is still there. 

Moreover, it turns out that any reaction measured so far at forward rapidities exposes the same effect of an increased suppression. As an example, we present a collection of data for nuclear dependence (the exponent $\alpha$ fitted to data in the form of $A^\alpha$) of the total yields of different species of hadrons in $pA$ collisions as function of $x_F$, in the right panel of Fig.~\ref{fixed-target}. The curve shows the result of the parameter-free calculation in \cite{knpjs}
within the model described in Sect.~\ref{e-loss}.
 
Other reactions, like Drell-Yan process of heavy dilepton production, or charmonium production at forward rapidities also demonstrated a stronger suppression at large $x_F$ (see the data and references in \cite{knpjs}). All these data have been taken at low energies of fixed target experiments, where no coherence effects can be expected. Apparently, another mechanism, which causes suppression at forward rapidities, is at work. It should be also important for the interpretation of RHIC data.

\subsubsection{Test \#2: Towards large $x_T$}\label{xT}

One can approach the kinematic limit $\xi\to1$ increasing either {$ x_F$}, or {$ x_T$}.
In both cases the constraints imposed by energy conservation cause nuclear suppression. 
Therefore, we expect a suppression  at large $x_T$ to be rather similar to what is observed for forward rapidities corresponding to large $x_L$.
This would be a rigorous test for the mechanisms of suppression, because
no coherence is possible at large $x_T$.

\begin{figure}[htb]
\begin{center}
\includegraphics[width=6.5cm]{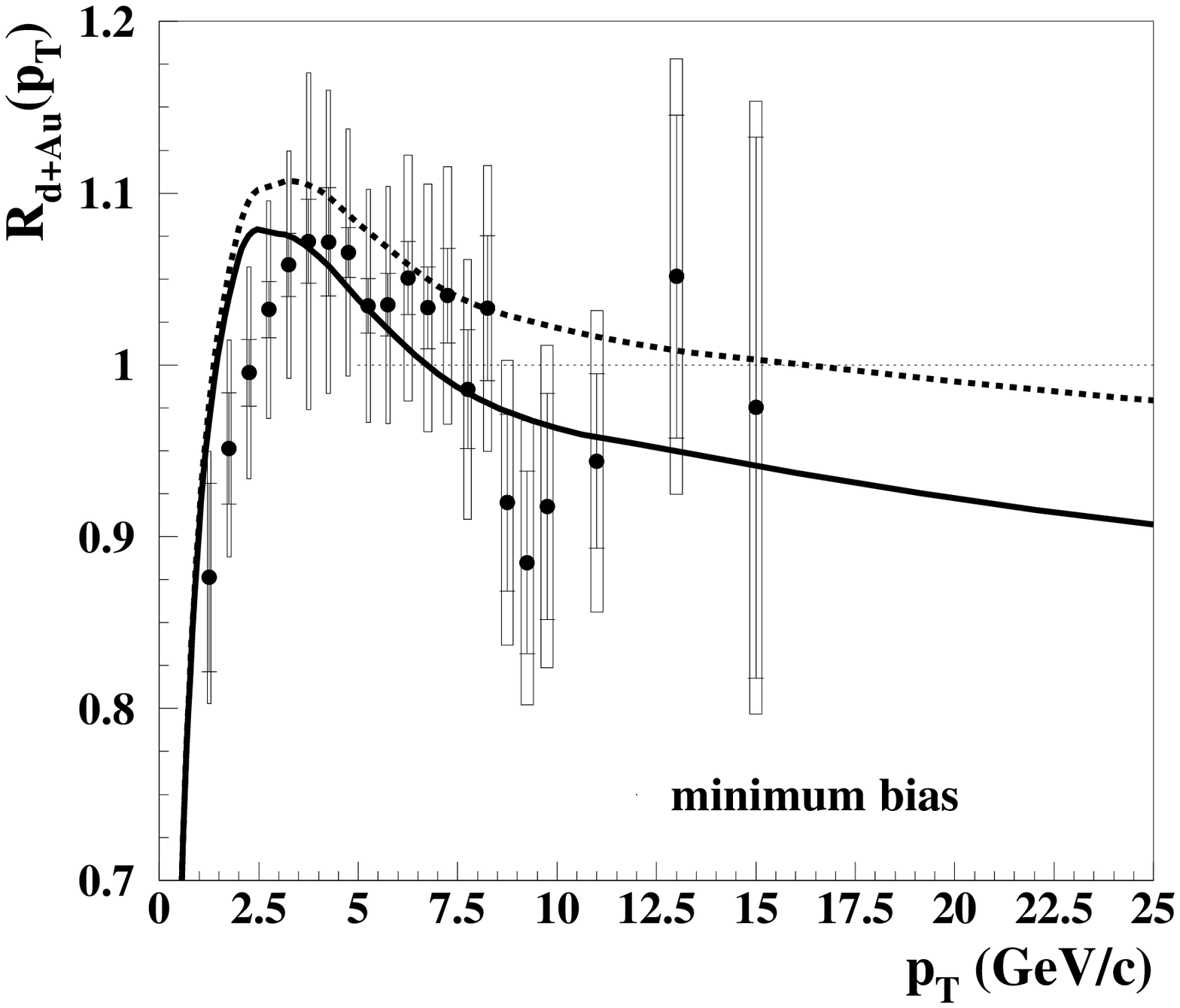}\hspace{15mm}
\includegraphics[width=6.5cm]{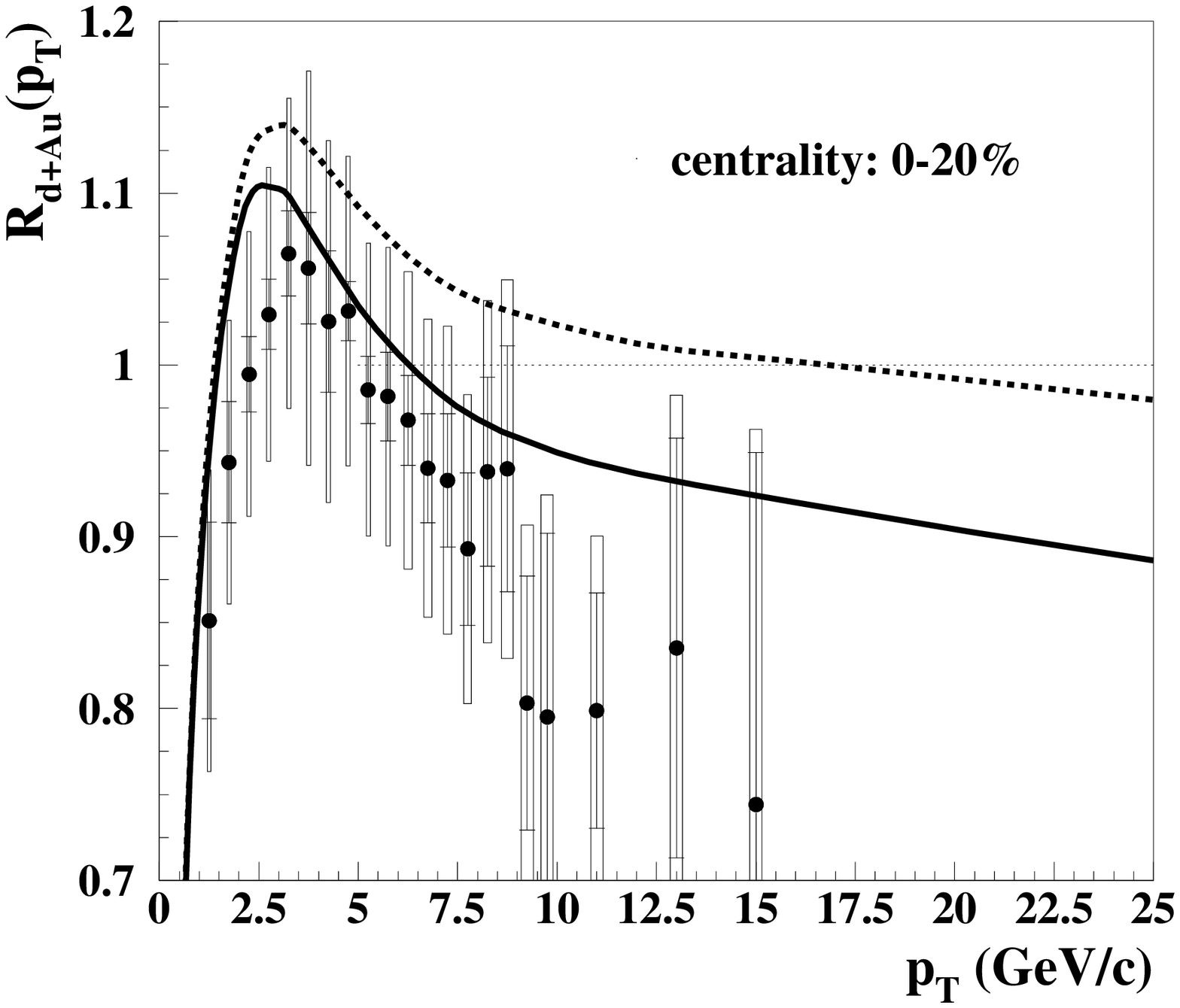}
\end{center} \caption{\label{cronin} Nuclear attenuation factor $R_{dAu}(p_T)$ as function of
$p_T$ of  $\pi^0$ mesons produced in $d$-$Au$ collisions
at $\sqrt{s}=200\GeV$ and $\eta = 0$. The solid and dashed curves
represent the model predictions calculated with and without corrections Eq.~(\ref{540}) for energy conservation.  The data \cite{cronin-phenix} and calculations correspond to either minimum bias (left), or central (right) $d$-$Au$ collisions.
 }
 \end{figure}

The nuclear ratio $R_{pA}(p_T)$ is known to expose nuclear enhancement at medium high $p_T$, the effect \cite{james.cronin} named after James Cronin. 
The rather small magnitude $\sim10\%$ of the Cronin enhancement for the {$ dA$} to {$ pp$} ratio, predicted in \cite{cronin-prl} was confirmed by the later measurements of neutral pion production in the PHENIX experiment \cite{cronin-phenix}, as is depicted by the dotted curves in Fig.~\ref{cronin}.
The expectation based on QCD factorization that the ratio must approach $R_{dAu}(p_T)\to1$ at large {$ p_T$}, is confirmed by the behavior of the dotted curves, with a small correction for the isotopic effects. 

However, energy conservation becomes an issue with rising $x_T$, and eventually causes a considerable suppression, as is shown by the solid curves in Fig.~\ref{cronin}.
The same suppression factor Eq.~(\ref{540}) was used in this calculations, as at forward rapidities.
The data do not contradict the predicted suppression at large $p_T$, even support it,
especially in central collisions. Of course the observed suppression has nothing to do with the  coherence effects.

Notice, that the suppression at large $p_T$ caused by energy deficit also contributes to the $p_T$ dependence of hadron suppression in $AA$ collisions. The observed flat $p_T$ dependence of $R_{AA}(p_T)$ would be rising, if the suppression caused by energy conservation were switched off.

\begin{figure}[htb]
\begin{center}
\includegraphics[width=7cm]{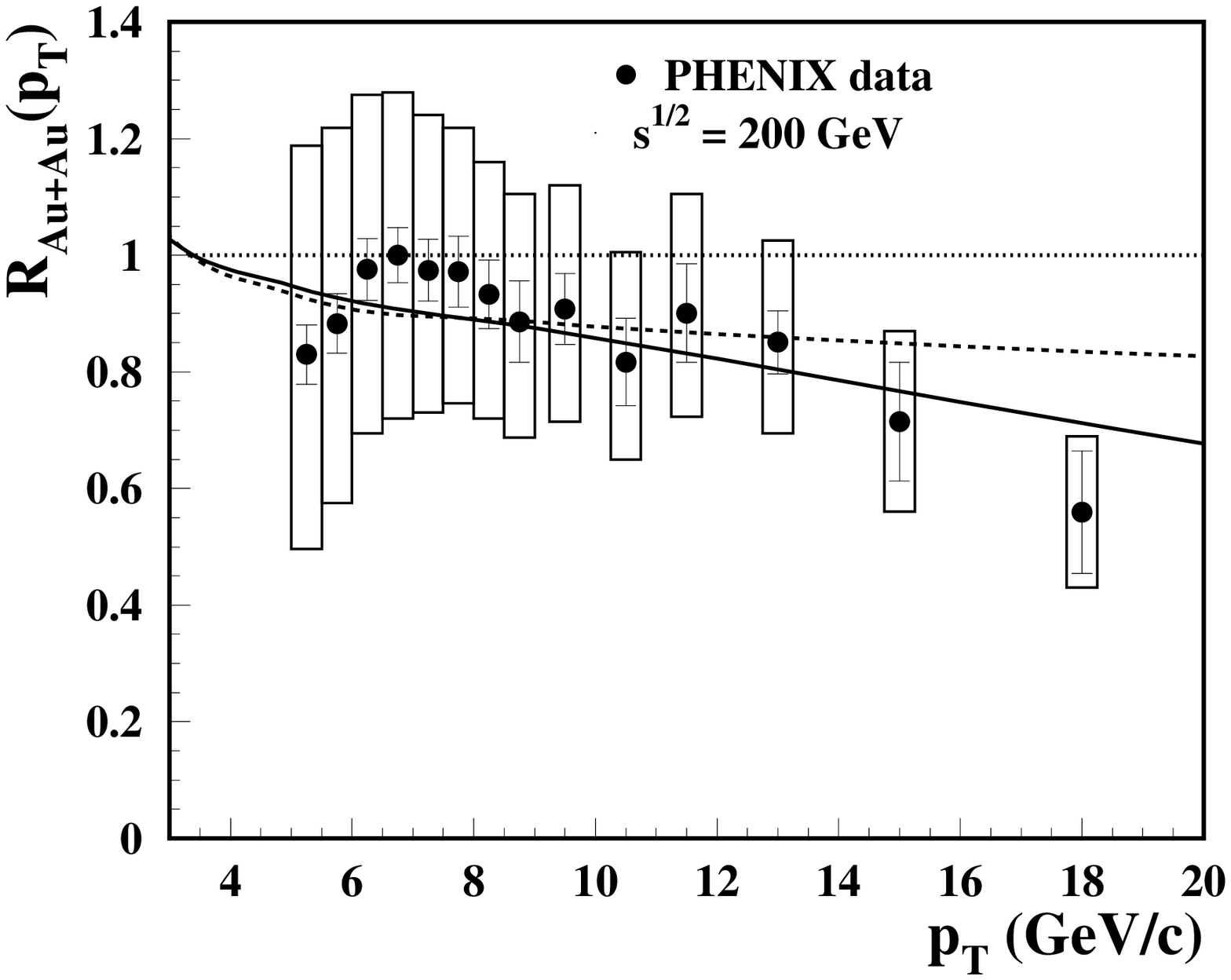}\hspace{15mm}
\includegraphics[width=7cm]{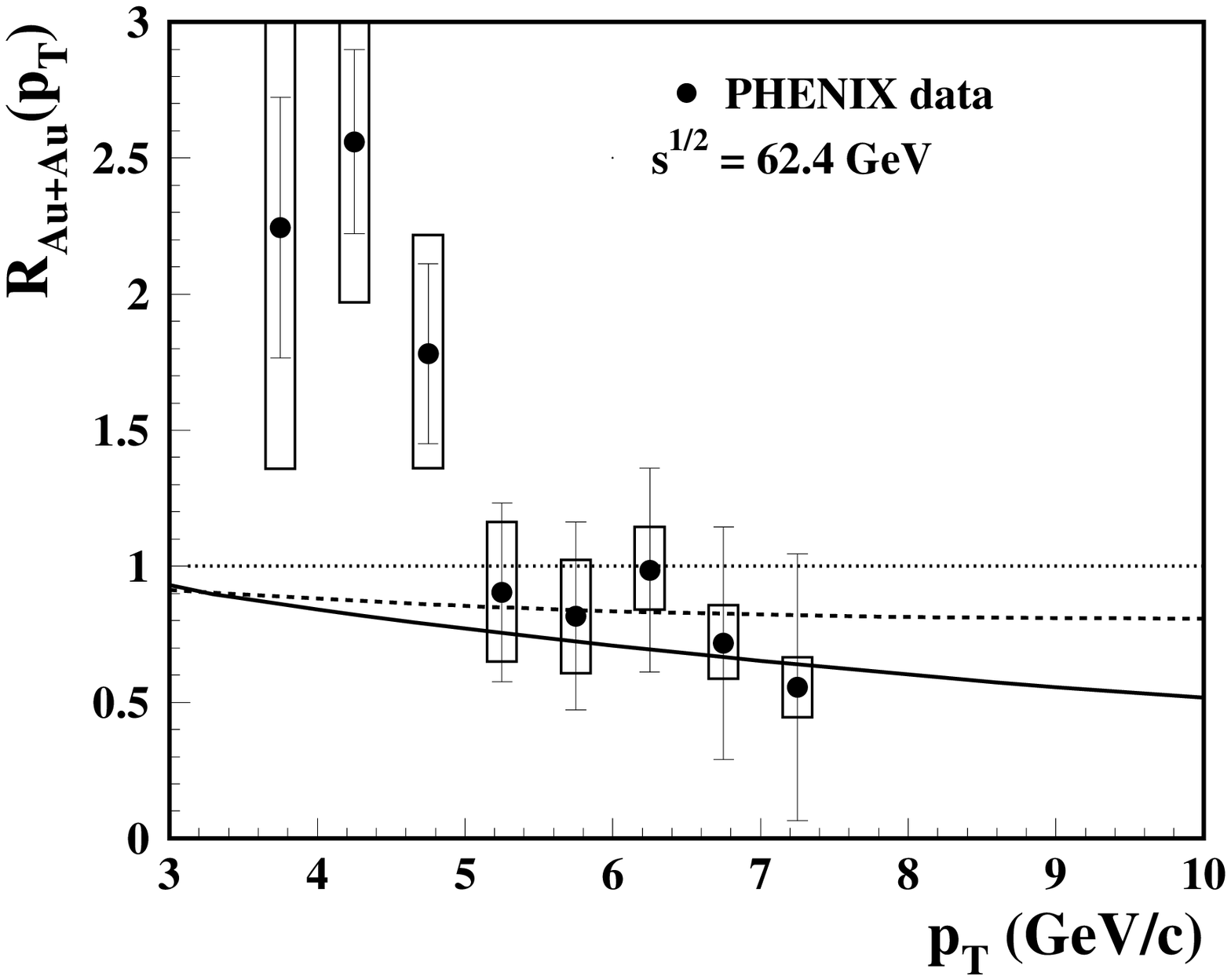}
\end{center} \caption{\label{gammas} Ratio of cross sections of direct photon production in gold-gold to proton-proton collisions at $\sqrt{s}=200\GeV$ (left panel) and at $\sqrt{s}=62\GeV$ (right panel). Solid and dashed curves 
correspond to calculations done with and without the corrections for energy deficit, the factor Eq.~(\ref{540}) suppressing multiple interactions, 
respectively. The data are from PHENIX experiment at RHIC \cite{phenix-gamma}. }
 \end{figure}

A similar test can be performed with direct photons produced with high $p_T$. 
This is even a cleaner probe, than hadron production, since photons have no final state interactions, and no convolution with the jet fragmentation function is required.
 Production of direct photons with large {$ p_T$} approaching the kinematic limit also is subject to the energy sharing problem, so is universally suppressed  by multiple interactions. Our predictions for high-$p_T$ photons at the energies $\sqrt{s}=200$ and $62\GeV$ are depicted in Fig.~\ref{gammas}
 in comparison with data from PHENIX experiment \cite{phenix-gamma}.
The energy range of RHIC is in between the two  regimes of photon radiation, coherent and incoherent. To avoid the technically complicated interpolation between the two regimes, we have done calculations for both regimes and found the difference to be quite small for $p_T>3\GeV$. The solid and dashed curve are calculated for incoherent photon radiation, either including or excluding the corrections for energy limitations, respectively. Corrections for isotopic effects are included into these calculations.

While nothing certain can be concluded from comparison with data at $\sqrt{s}=62\GeV$,
data  at $\sqrt{s}=200\GeV$ provide some evidence for the predicted suppression.
In fact, the observed strong suppression of direct photons has been considered as a puzzle, since no energy loss or absorption in the produced hot medium is expected for photons. We explain suppression for particle production at large $x_L$ and large $x_T$ by the same mechanism. Of course the CGC scenario for forward rapidities \cite{kkt}
would have no effect on hadrons or photons produced with very large $p_T$.

\subsubsection{Test \#3: The magnitude of gluon shadowing}

This test has been already done within the global fit \cite{eps08} aimed at extraction of PDFs from data.
The BRAHMS data \cite{brahms} were included in the analysis, assuming that the observed suppression is related to gluon shadowing. The results depicted in the right panel of Fig.~\ref{Rg}.
strongly violate the unitarity bound \cite{bound}.
Apparently, the source of the trouble was the misinterpretation of the BRAHMS data.
Moreover, even that strong shadowing was not sufficient to explain the STAR data at higher rapidity
$\eta=4$ \cite{star-forward}.

\section{Critical examination of the energy loss scenario}

A parton propagating through a medium experiences multiple interactions, which induce gluon radiation taking away a part of the parton energy. Therefore, one can expect that the production
leading hadrons, which carry the main fraction of the jet energy will be suppressed compared to the same reaction in $pp$ collisions \cite{miklos}.
Indeed a significant suppression, usually called jet quenching, was found at RHIC for all species of hadrons produced with large $p_T$ in central collisions of heavy ions \cite{phenix-pt,star-pt}.

However, the energy loss scenario is based on the unjustified assumption that gluon radiation continues throughout the dense medium created in the collision, and the leading hadron
(or a colorless state, called pre-hadron, which does not have any certain mass) is always produced outside the medium. 

This is quite a debatable issue, since there are solid theoretical and experimental arguments favoring the alternative scenario:
the pre-hadron frequently is produced inside the medium and strongly attenuates. Indeed,
it has been realized long time ago \cite{k-nied} that the production length for leading hadrons, may be rather short,
\beq
l_p\sim\frac{E}{|dE/dl|}(1-z_h).
\label{600}
\eeq
Here $E$ is the energy of the jet, $dE/dl$ is the rate of vacuum energy energy loss, which is usually much larger than the medium induced one, especially for highly virtual partons.
Here the suppression mechanism related to the deficit of energy is again at work. If the highly virtual parton intensively radiates, its energy steeply decreases (see next section) and can fall  below the minimal energy required for production of the leading hadron of energy $E_h=z_h\,E$. Only creation of a colorless pre-hadron, which does not radiate gluons any more, can stop the dissipation of energy. 

In the string model the rate of energy loss $-dE/dl=\kappa\approx 1\GeV/\fm$ is the string tension, known from the slope of the Regge trajectories, as well as from lattice calculations. The predicted $z_h$-dependence of $l_p$  \cite{k-nied} was confirmed by the study of time development of fragmentation within the Lund model \cite{bg}.

\subsection{Time dependence of vacuum radiation}\label{t-dep}

The color field of a quark originated from a hard reaction
(high-$p_T$, DIS, $e^+e^-$, etc.) is stripped off, i.e. such a quark is
lacking a color field up to transverse frequencies $q\lsim Q$.
Therefore the quark originated from such a hard process starts regenerating its field by radiating gluons, i.e.,
forming a jet. This can be described by means of an expansion
of the initial "bare" quark over the Fock states containing a physical
quark and different number of physical gluons with different
momenta. Originally this is a coherent
wave packet equivalent to a single bare quark $|q\ra$. However,
different components have different invariant masses and they start
gaining relative phase shifts as function of time. As a result,
the wave packet is losing coherence and gluons are radiated in
accordance with their coherence times.

This process lasts a long time proportional to the jet energy
($E\approx p_T$), since the radiation time (or length) depends on the gluon
energy and transverse momentum $k$ (relative to the jet axis),
 \beq
l_c=\frac{2E}{M_{qg}^2-m_q^2}= \frac{2Ex(1-x)}{k^2+x^2\,m_q^2}.
\label{610}
 \eeq Here $x$ is the
fractional light-cone momentum of the radiated gluon; $m_q$ is the quark mass;
$M_{qg}^2=m_q^2/(1-x)+k^2/x(1-x)$ is the invariant mass squared of
the quark and radiated gluon.

One can trace how much energy is radiated over the path length $L$ by the gluons which have lost coherence during this time interval \cite{knp,within,trieste,jet-lag,light-heavy},
 \beq
\Delta E(L) =
E\int\limits_{\Lambda^2}^{Q^2}
dk^2\int\limits_0^1 dx\,x\,
\frac{dn_g}{dx\,dk^2}
\Theta(L-l_c),
\label{620}
 \eeq
 where $Q\sim p_T$ is the initial quark virtuality; the infra-red cutoff is fixed at $\Lambda=0.2\GeV$.
 The radiation spectrum reads
 \beq
\frac{dn_g}{dx\,dk^2} =
\frac{2\alpha_s(k^2)}{3\pi\,x}\,
\frac{k^2[1+(1-x)^2]}{[k^2+x^2m_q^2]^2},
\label{640}
 \eeq
 where $\alpha_s(k^2)$ is the running QCD coupling, which is regularized at low scale by replacement $k^2\Rightarrow k^2+k_0^2$ with $k_0^2=0.5\GeV^2$.
 
In the case of heavy quark the $k$-distribution Eq.~(\ref{640}) peaks at $k^2\approx x^2\,m_q^2$, corresponding to the polar angle (in the small angle approximation) $\theta=k/xE=m_q/E$. This is known as the dead cone effect \cite{dead-cone,yura}.

The step function in Eq.~(\ref{620}) creates another dead cone: since the quark is lacking a gluon field, no
gluon can be radiated unless its transverse momentum is sufficiently
high, 
\beq k^2>\frac{2Ex(1-x)}{L}-x^2m_q^2. 
\label{660} 
\eeq 
This bound is relaxing with the rise of $L$ and reaches the magnitude
$k^2\sim x^2m_q^2$ characterizing the heavy quark dead cone at 
\beq
L_q= \frac{E(1-x)}{xm_q^2}. 
\label{680} 
\eeq 
We see that $L_q$ for
beauty is an order of magnitude shorter than for charm, but linearly
rises with the jet energy.

The radiation of such a "naked" quark has own dead cone controlled by its virtuality {$ Q^2\gg m_q^2$}.
This cone is much wider than the one related to the quark mass. There is no mass dependence of the radiation until the quark virtuality cools down to {$ Q^2\Rightarrow Q^2(L)\sim m_q^2$}. Therefore,
the results of \cite{yura} for a reduced energy loss of heavy quarks should be applied with a precaution. At the early stage of hadronization,
when {$ Q^2(L)\gg m_q^2$}, all quarks radiate equally.

The characteristic length $L_q$ may be rather long, since gluons are radiated mainly with small $x$. For instance, for a charm quark with $E=p_T=10\GeV$ the sensitivity to the quark mass is restored at $L\gsim1/x\,\fm$. Only at longer distances, $L\gsim L_q$, the dead cone related to the heavy quark mass sets up, and the heavy and light quarks start radiating differently.

\begin{figure}[htb]
\bc
 \includegraphics[height=7.5cm]{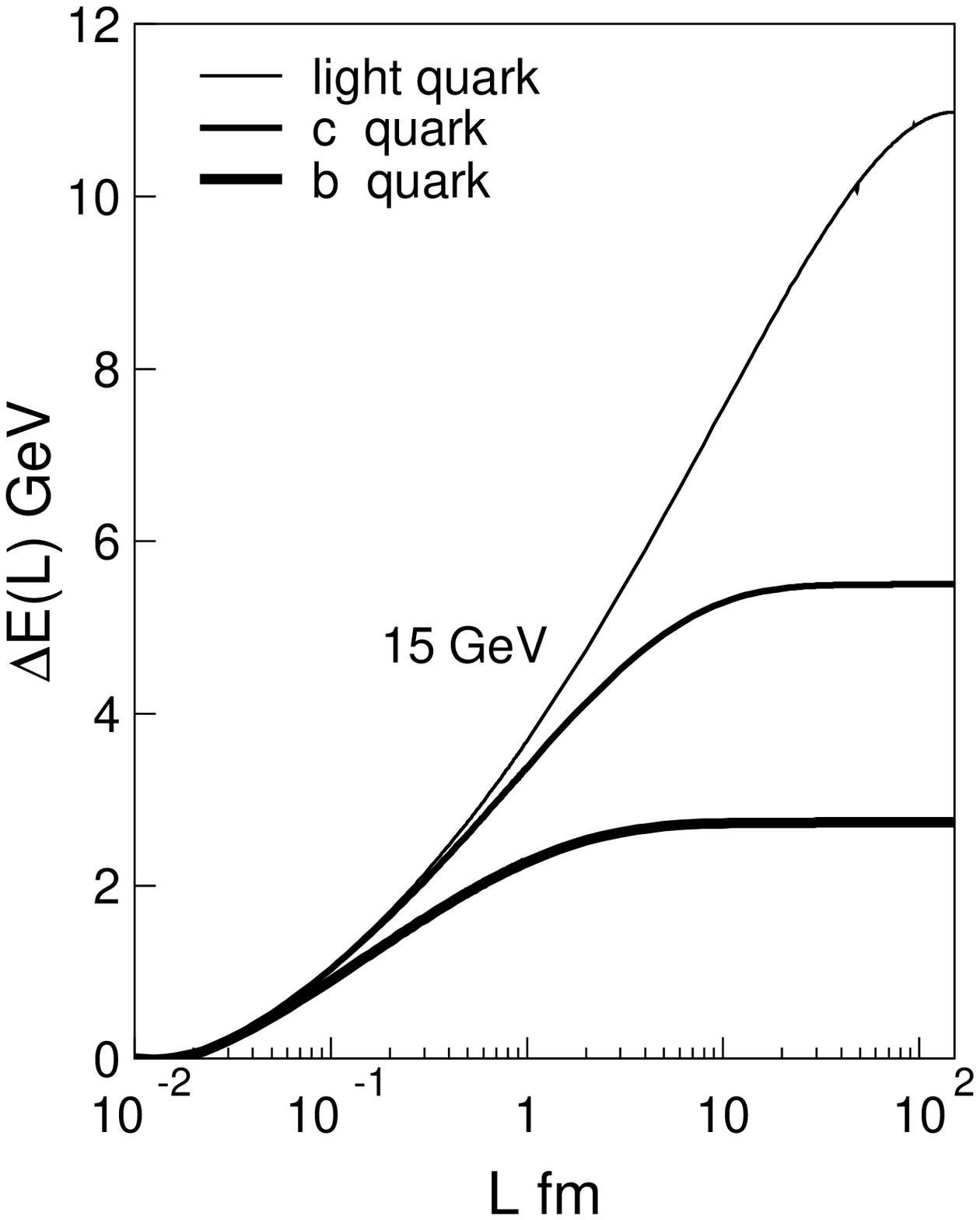}\hspace{20mm}
 \includegraphics[height=7.5cm]{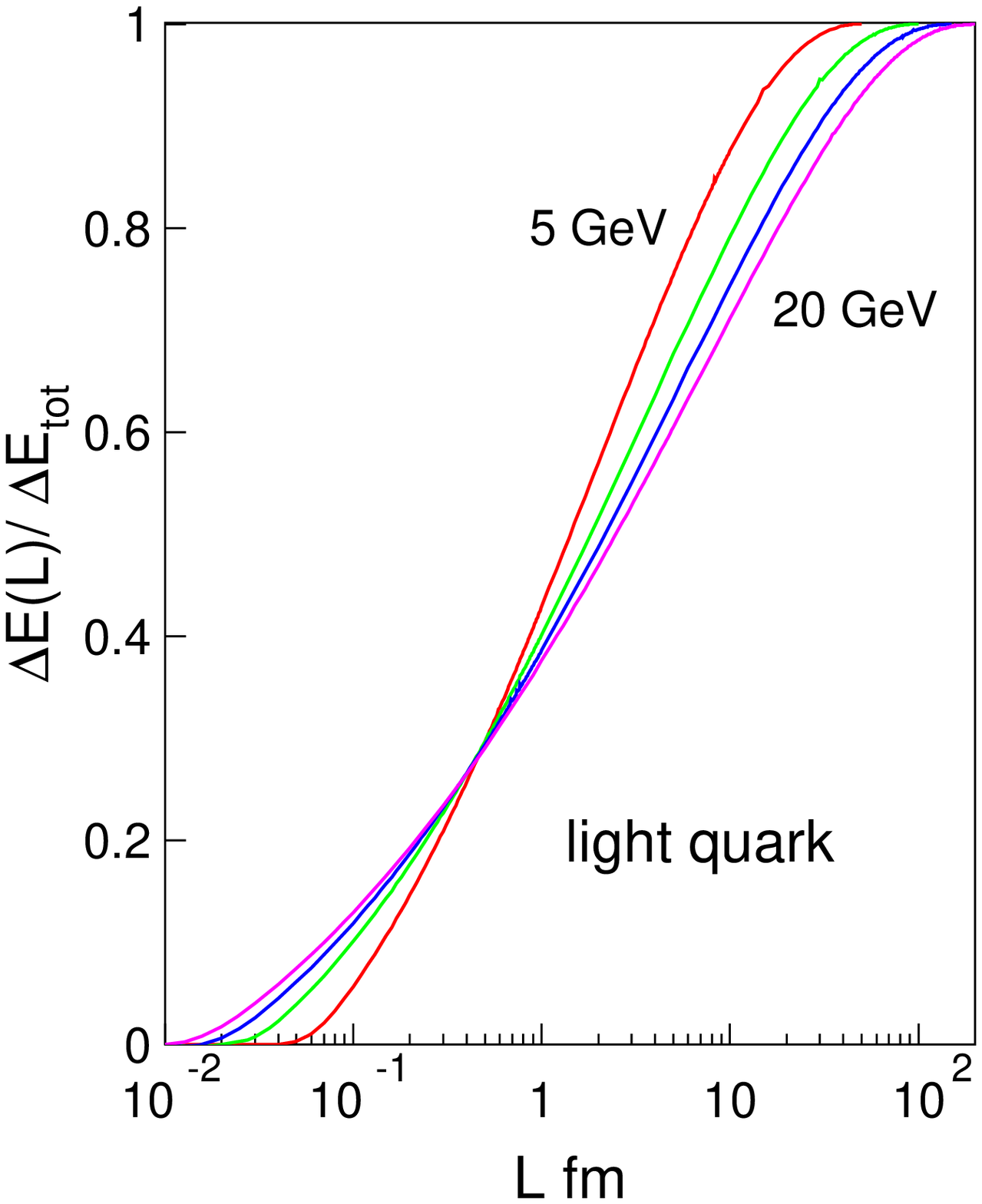}
\caption{\label{l-dep} {\it Left panel:} vacuum energy loss by light ($m_q=0$), charm ($m_c=1.5\GeV$) and bottom ($m_b=4.5\GeV$) quarks with $E=15\GeV$ as function of path length. {\it Right panel:} Fractional energy loss in vacuum as function of distance covered by a light quark. The jet energies, $E=p_T$, corresponding to the curves from top to bottom,
are $E=5,\ 10,\ 15,$ and $20\GeV$. }\ec
 \end{figure}

The numerical results demonstrating this behavior are depicted in the left panel of
Fig.~\ref{l-dep}.
One can see that a substantial difference between radiation of energy by the charm and light quarks
onsets at rather long distances, above $10\fm$, while  within several fermi the difference is insignificant. The $b$-quark radiation is suppressed already at rather short distances.
Moreover, it completely regenerates the color field already at a distance of the order of $1\fm$
and does not radiate any more. Of course, this $b$-quark still may have a medium induced radiation,
which is very weak according to \cite{yura}. Notice that the interference between vacuum and induced radiations, which was found in \cite{urs} to be important, is absent because they occur on different time scales.

\subsection{Production length}\label{lc}

The fractional energy loss in vacuum by a light quark is depicted in the right panel of Fig.~\ref{l-dep} as function of path length. Right after the hard scattering the quark is highly virtual and the radiation is very  intensive. As a result, the quark virtuality is cooling down and the rate of energy loss decreases. So the quark dissipates most of energy at the early stage of hadronization,  during the first {$ 1\fm$} about {$ 40\%$} of the total radiated energy. A heavy quark does it even faster \cite{light-heavy}.
This means that a hadron with large fractional momentum {$ z_h$} should be produced at a short time scale, otherwise too much energy will be radiated, and the rest will be insufficient for production of the hadron. This is what the equation (\ref{600}) says. Notice that in a high-$p_T$ hadron production the large values of {$ z_h$} are favored by the steeply falling transverse momentum spectrum of quarks convoluted with the fragmentations function.

\begin{figure}[htb]
\bc
 \includegraphics[height=7cm]{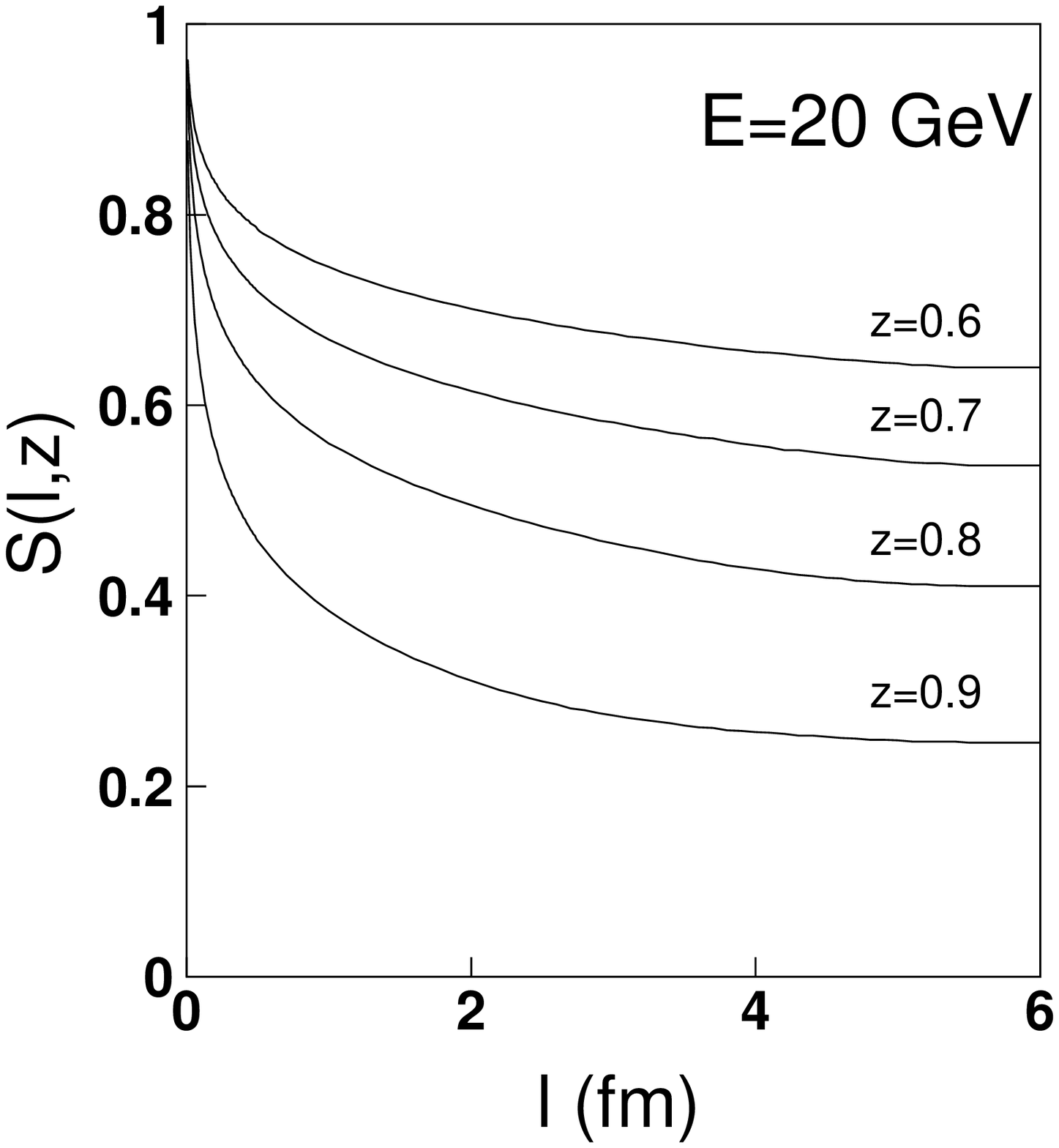}\hspace{15mm}
 \includegraphics[height=7.1cm]{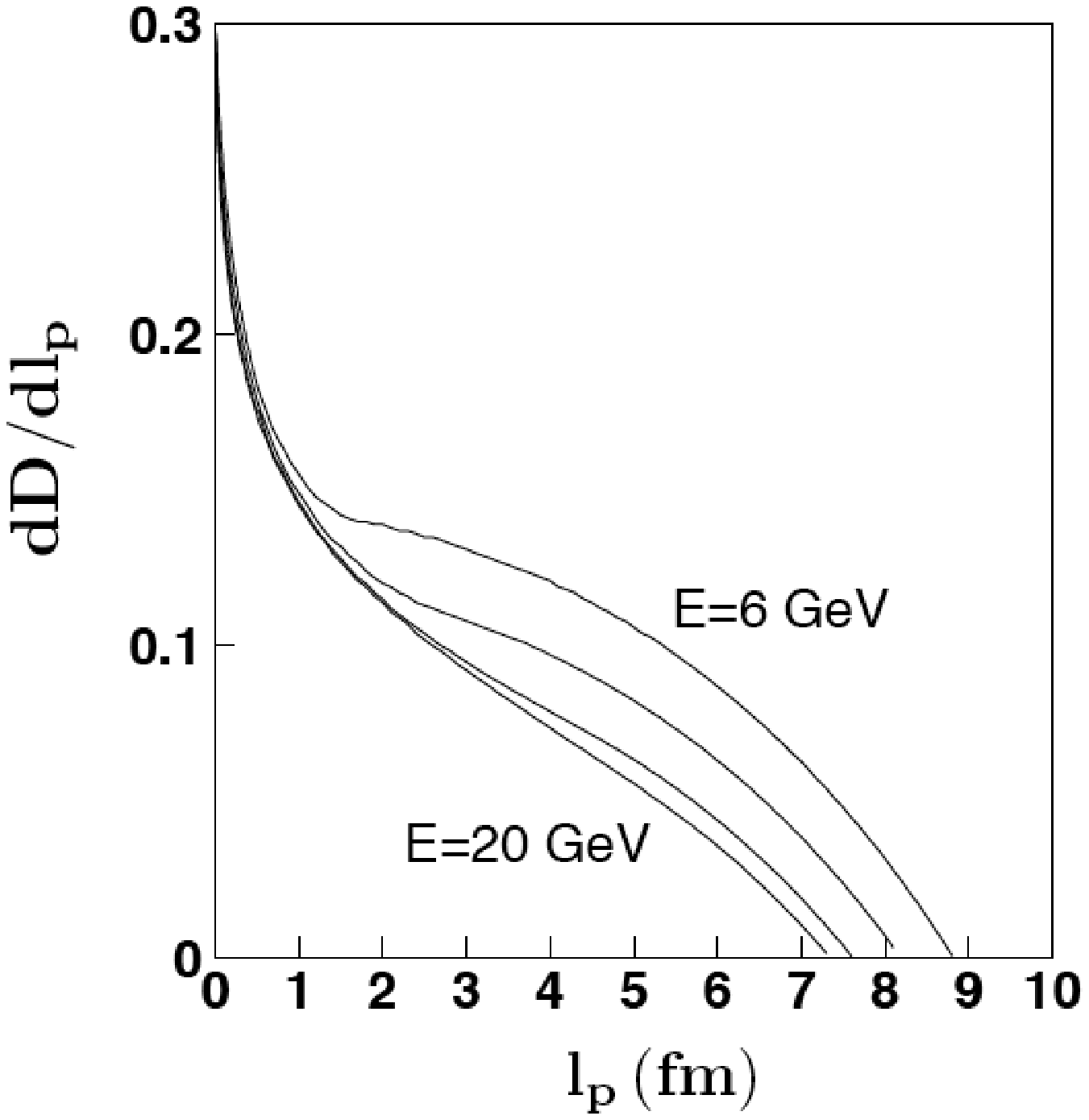}
\caption{ \label{dd-lp} {\it Left:} Sudakov suppression caused by a ban for radiation of gluons with
fractional energy higher than $1-z_h$. Calculations are done for a jet with
$E=20\GeV$.
{\it Right:}  The production length distribution $\partial D(z)/\partial l_p$
(in arbitrary units) for jet energies $6,\ 10,\ 16,\ 20\GeV$ and
$z_h=0.7$. }
 \ec
 \end{figure}

In fact energy restrictions make the dissipation of energy  for a given $z_h$ somewhat slower than it follows from Eq.~(\ref{620}) and is depicted in Fig.~\ref{l-dep}, because radiation of gluons with fractional momenta larger than $1-z_h$ is forbidden by energy conservation \cite{knp,jet-lag}. On the other hand, a ban for radiation of gluons with energy
$\omega>(1-z_h)E$ in (\ref{620}),  leads to the
Sudakov type suppression factor,
 \beq
S(L,z_h)=\exp\left[-\la n_g(L,z_h)\ra\right]\,,
\label{720}
 \eeq
 where $\la n_g(L,z_h)\ra$ is the mean number of nonradiated gluons during propagation over the distance $L$,
 \beq
 \la n_g(L,z_h)\ra=\int\limits_{1/Q}^{l_{max}} dl
\int\limits_{(2El)^{-1}}^{1}
d\alpha\,\frac{dn_g}{dld\alpha}\,
\Theta\left(\alpha+\frac{1-\alpha}{2lE}-1+z_h\right)
\,.
\label{730}
 \eeq
Here $\alpha$ is the fractional light-cone momentum of a radiated gluon;  $l_{max}={\rm min}\{L,\ E/2\lambda^2\}$, and $\lambda$ is the soft cutoff for transverse momenta of gluons, fixed in \cite{jet-lag} at $\lambda=0.7\GeV$. The step function in (\ref{730}) takes care of energy conservation.
An example of a Sudakov factor calculated at $E=Q=20\GeV$
for different values of $z$  is shown in the left panel of Fig.~\ref{dd-lp}.

Combination of these two effects, vacuum energy loss and Sudakov suppression, leads to a rather short production length, which slightly varies with jet energy and 
virtuality. 
The resulting production length distribution for jets with maximal virtuality $Q=p_T/z_h=E$ is depicted in the right panel of Fig.~\ref{dd-lp} for different jet energies as was calculated in \cite{jet-lag}.
We see that the mean production length is rather short, few fermi, and is slowly decreasing with energy. We took into consideration so far only vacuum energy loss. Apparently, adding medium induced energy loss will only enhance the energy deficit and make the production length even shorter.

The production length distribution calculated for light quarks is also valid for charm quarks, which
have a similar vacuum radiation during first several fermi. However, a bottom quark, according to Fig.~\ref{l-dep}, dissipates considerably less energy and its vacuum radiation ceases at the distance of about $1\fm$, because the quark completely restores its color field.  Of course according to confinement a colored quark even with a restored field cannot propagate freely. It keeps losing energy via nonperturbative mechanisms \cite{jet-lag}, like in the mentioned above string (flux tube) model. We can make a rough estimate of the production length
for hadronization of a bottom quark, relying on Eq.~\ref{600} and assuming the $p_T$ distribution
of the produced quarks $d\sigma/dp_T^2\propto p_T^{-n}$,
\beq
\la l_p\ra = n\,\frac{E}{\kappa}\int\limits_0^1 dz\,(1-z)\,z^{n-1}=
\frac{E}{(n+1)\kappa}.
\label{750}
\eeq
This is quite a short distance, e.g. for $E=15\GeV$ and $n=6$, it is only $l_p\approx2\fm$.

Notice that one should not mix up the production time with the time scale evaluated in \cite{vitev}, which  is just the well known coherence time (e.g. compare  Eq.~(2) in \cite{jet-lag} with Eq.~(3) in \cite{vitev} containing some misprints) , which is the time interval between the hadron production points where the production amplitudes  interfere. 
This is not the time of duration of hadronization which we are interested in.
If hadronization were lasting as long as the coherence time, energy conservation would be broken.
Besides, a pre-hadron does not have any certain mass, since according to the uncertainty relation it takes time, called formation time, to resolve between the ground and excited states, which have certain masses. Therefore, one cannot evaluate the production time of a pre-hadron relying on the
mass of the hadron.

Thus, theoretically we see no justification for a long production length, while any reasonable evaluation leads to a rather short duration time for hadronization which ends up producing a leading hadron. Nevertheless, it would be more convincing to test the models comparing with data.

\subsection{Within or without?}

\begin{figure}[htb]
\bc
 \includegraphics[height=6cm]{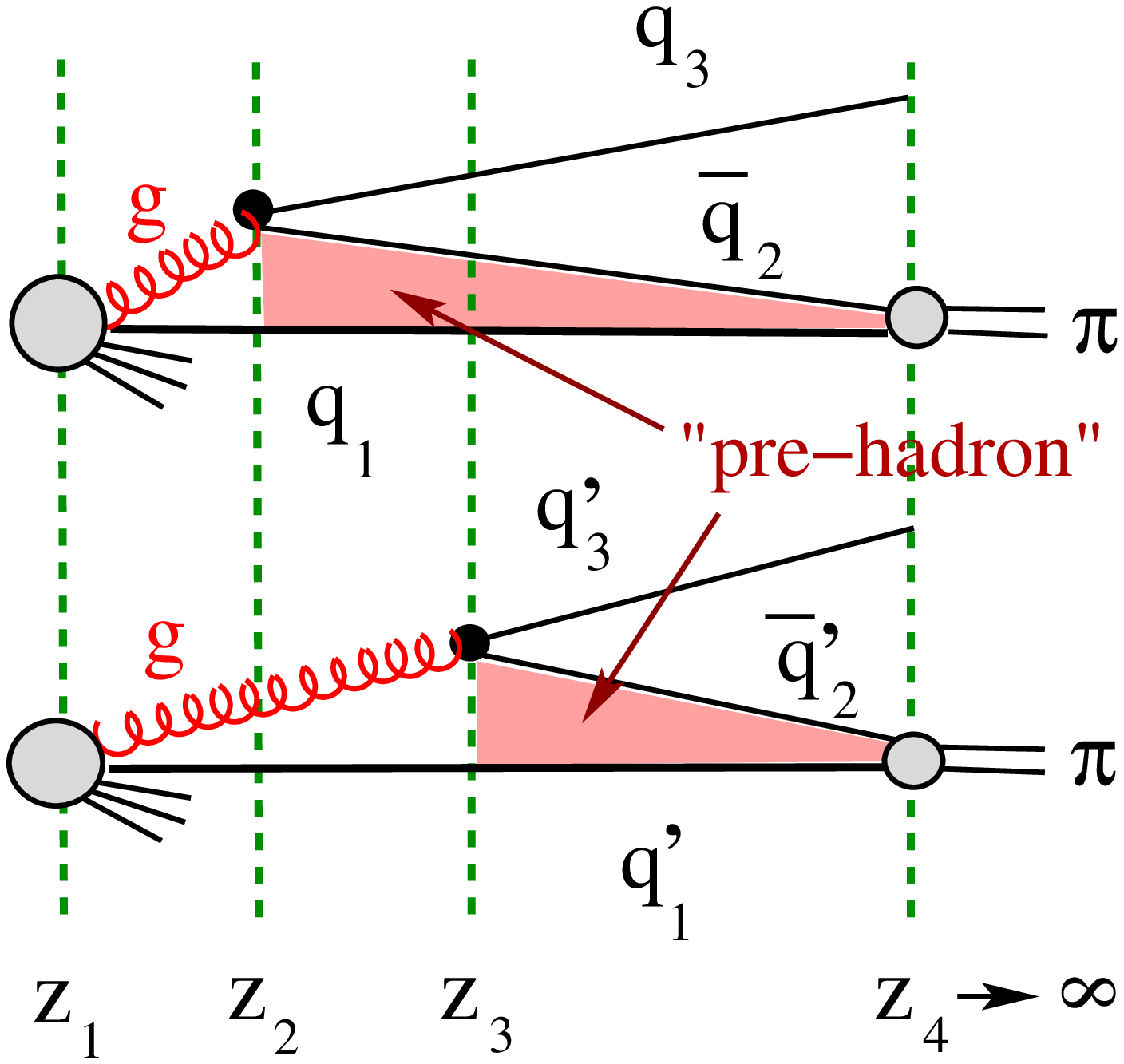}\hspace{1cm}
 \includegraphics[height=6.5cm]{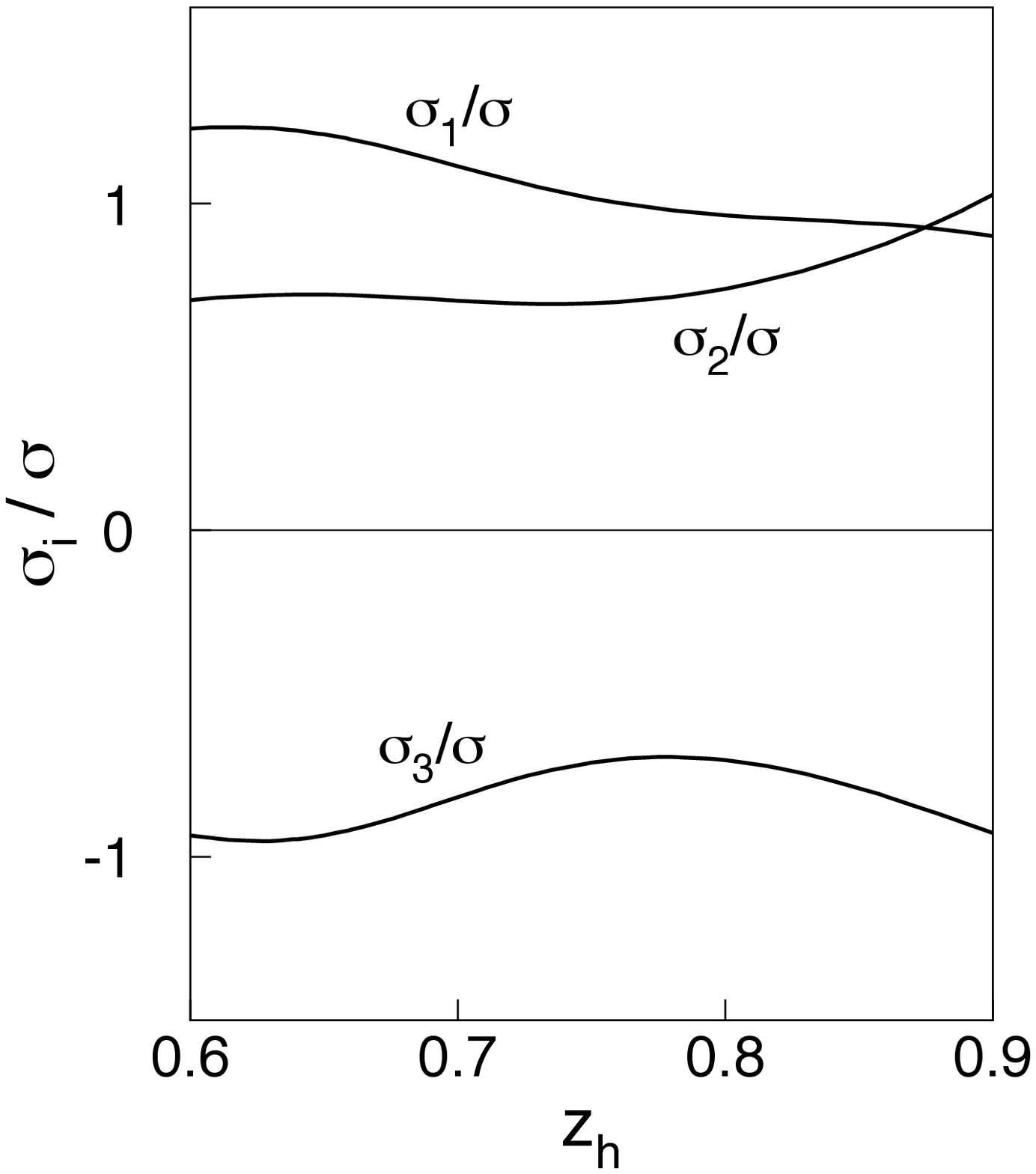}
\caption{ \label{interference} {\it Left:} Graphical representation for the interfering direct and conjugated
 amplitudes for production of a pre-hadron at points with coordinates $z_2$ and $z_3$ respectively. {\it Right:}  Fractional cross sections $\sigma_i/(\sigma_1+\sigma_2+\sigma_3)$ as function of $z_h$ calculated for lead at $E= 10\GeV$. $\sigma_{1,2,3}$ correspond to both coordinates $z_{2,3}$ of pre-hadron production being either outside or inside the nucleus, or to their interference respectively \cite{quant}.}
 \ec

 \end{figure}

Strictly speaking this very question is not well defined. Indeed, in quantum mechanics
one cannot always say with certainty whether the pre-hadron was produced within or without the medium.
The corresponding amplitudes interfere like is illustrated in the left panel of Fig.~\ref{interference}.
 We employ here the Berger model for perturbative fragmentation \cite{berger,pert,jet-lag}.
The  hard process occurs incoherently
 on a nucleon with longitudinal coordinate $z_1$. The radiated gluons decay into
 $\bar q_2 q_3$ and $\bar q_2^\prime q_3^\prime$ coherently in the two amplitudes.
 The colorless pre-hadrons (dipoles) $\bar q_2 q_1$ and $\bar q_2^{\,\prime} q_1^\prime$
 created at $z_2$ and $z_3$ respectively, are projected to the pion wave function in each
 of the two amplitudes.
When both points $z_2$ and $z_3$ are located outside, or inside the nucleus, the corresponding parts of the cross section are labeled as $\sigma_1$ and $\sigma_2$ respectively.
The interference term in the cross section, $\sigma_3$, corresponding to simultaneous production inside and outside, was found 100\% important \cite{quant}. An example of relative contribution of all three term is shown in the right panel of Fig.~\ref{interference}.

Unfortunately, the phenomenology based on such a quantum-mechanical description has not been sufficiently developed so far. Therefore, in what follows we treat the space-time development of in-medium hadronization in a semi-classical way,  and confront with data the two options: (i) depending on the process and kinematics the production time may be shorter or longer than the path length in the medium, correspondingly the hadronization ends inside or outside the medium; (ii) according to the basic assumption of the energy loss scenario  the pre-hadron is always produced outside the medium.

\subsubsection{Test \#1: Nuclear effects in SIDIS}

\begin{figure}[htb]
\bc
 \includegraphics[height=6cm]{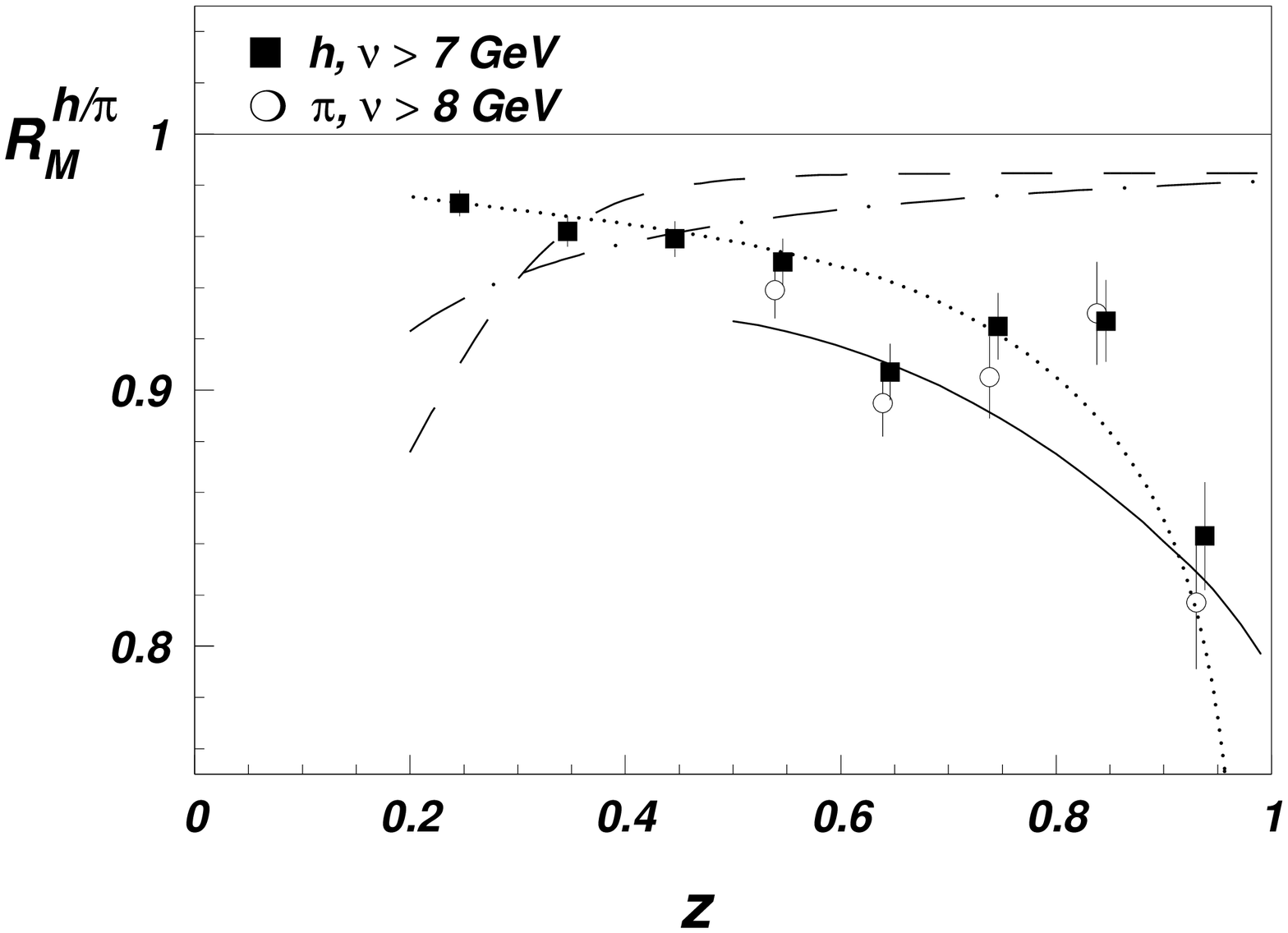}\hspace{10mm}
\vspace*{-8mm}
 \includegraphics[height=6.1cm]{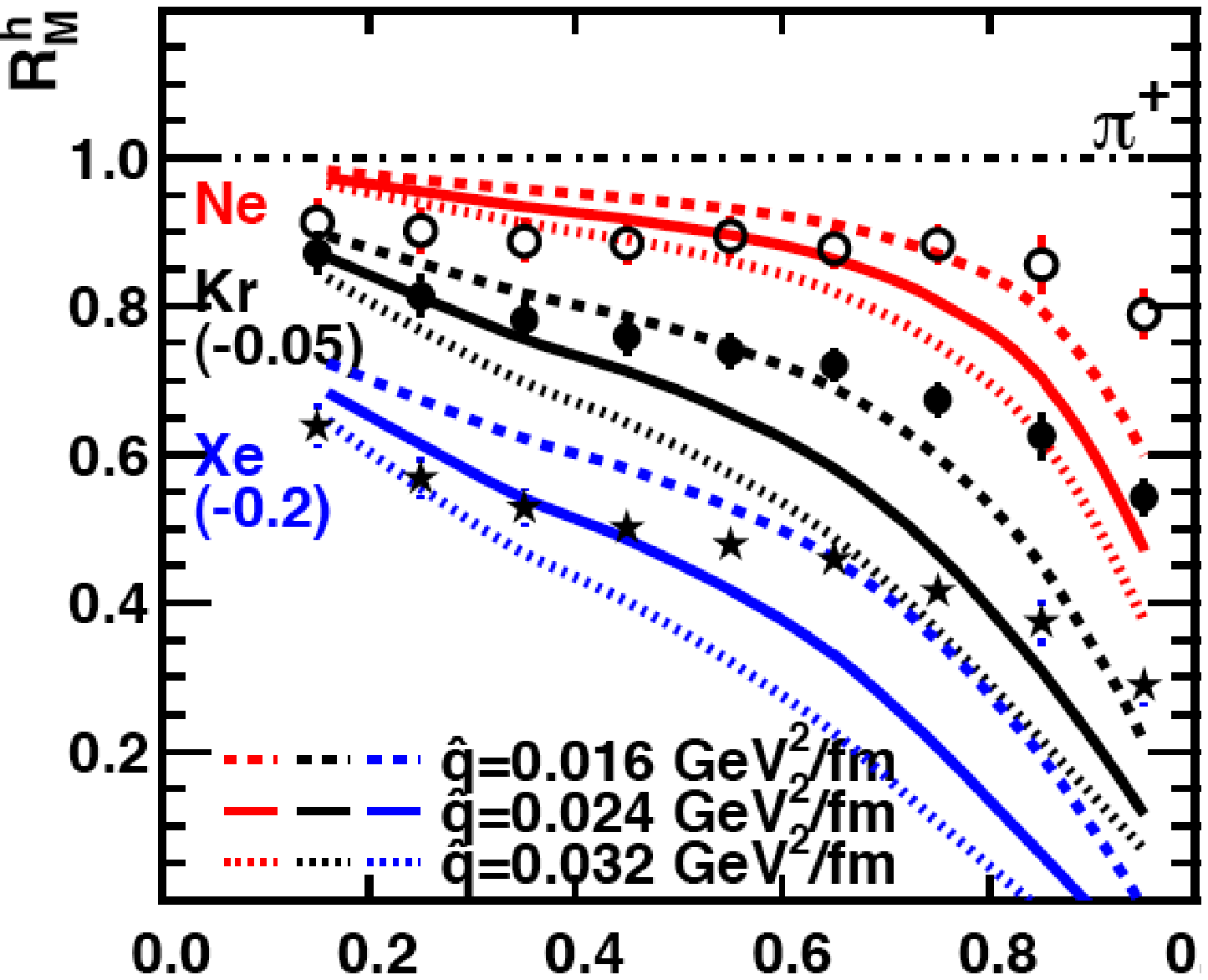}
\caption{\label{hermes-fig} {\it Left:} HERMES data \cite{hermes-first} for 
the multiplicity ratio as function 
of $z_h$ for all charged pions (open circles) and all 
charged hadrons including pions (closed squares). The 
full curve represents the prediction of the gluon-bremsstrahlung model \cite{knp}.
Other curves correspond to different fits to the data. 
{\it Right:}  The modified multiplicity ratios as function of $z_h$ with different values of the jet transport parameter $\hat q_0$ \cite{wang-last}
compared with the HERMES \cite{hermes-last} data for Ne, Kr and Xe targets. For clear presentation the modification factors for different 
targets are shifted vertically.}
 \ec
 \end{figure}

The shrinkage of the production length towards $z_h=1$, Eq.~(\ref{600}), and corresponding increase of nuclear suppression was predicted long time ago \cite{k-nied}, but only the Hermes experiment  performed the first measurements confirming such a behavior of nuclear ratios \cite{hermes-first}.
Moreover, the parameter-free quantitative predictions for the magnitude of nuclear attenuation, were well confirmed by later measurements \cite{hermes-first}, as is demonstrated in the left panel of Fig.~\ref{hermes-fig} from \cite{hermes-first} for a nitrogen target.
The main source of attenuation in the model \cite{knp} is absorption of the pre-hadron, which is a $\bar qq$ dipole. Once produced inside the nucleus, it attenuates with a cross section, which is fluctuating during the formation of the hadron wave function. The fluctuations are summed up with the path integral technique. 

Because the production length Eq.~(\ref{600}) linearly rises with energy (at a fixed virtuality), eventually the mean production length becomes so long, that any absorption corrections will be excluded, and the nuclear ratio should approach one. The energy dependence of nuclear attenuation predicted in \cite{knp} was well confirmed by the HERMES measurements. At the energies of the EMC experiment
$\nu>100\GeV$ no nuclear effects have been observed within the experimental errors. This is why
the energy range of HERMES is most interesting and why this measurement were proposed in \cite{knp} for HERMES.

In the energy loss scenario, the whole effect of nuclear suppression at large $z_h$ is attributed to medium induced energy loss, which modifies the splitting functions in the evolution equation and eventually the fragmentation function for hadronization in a medium. The authors admitted that the model failed to explain data at  large $z_h>0.5$ even by adjusting the transport coefficient $\hat q_0$, as is demonstrated
in the right panel of Fig.~\ref{hermes-fig} from \cite{wang-last}.

\subsubsection{Test \#2: Heavy flavor production}

Heavy flavors have been predicted \cite{yura} to lose much less energy for in-medium induced radiation
than light quarks, because of the dead-cone effect. Therefore they were expected to be less suppressed in heavy ion collisions.
However in Sect.~\ref{t-dep} we demonstrated that because of another, stronger dead cone, related to the very intensive vacuum radiation, charm quarks experience a reduced energy loss only on long path lengths, which are not important for quarks produced out of a dense medium. Otherwise, charm and light quarks radiate gluons with similar rates. This would be sufficient to explain the large
suppression observed at RHIC high-$p_T$ electrons produced in central $Au$-$Au$ collisions \cite{phenix-heavy,star-heavy}. However, recent measurements \cite{quark-composition} found that an essential fraction of the electrons originates from decays of $b$-quarks. This fact leads to a serious problem for the energy loss scenario, which has failed so far to explain why beauty is suppressed as much as charm and light quarks \cite{miklos2}.

On the other hand, we have just demonstrated in Eq.~(\ref{750}) that the production length for beauty hadrons is very short, $l_p\sim 1-2\fm$, so the energy loss scenario is quite irrelevant. In this case absorption resulting from in-medium production is more important and is able to explain the observed strong suppression of $b$-quarks as is discussed further in Sect.~\ref{alternative}.

\subsubsection{Test \#3: Alternative probes for the dense medium}

Another problem of the energy loss scenario is too large density of the medium, which one needs to explain the observed jet quenching. For instance, the analysis performed in \cite{phenix-theor}
led to the value of transport coefficient $\hat q_0$, which is more than order of magnitude larger than the conventional expectation \cite{bdmps}. This may indicate missed mechanisms of suppression, which were absorbed in the fit and may lead to wrong values of fitting parameters. A good test would be an alternative probe, which independently measures the same parameters. 

$\J$ production in nuclear collisions has been debated  for a long time as a probe for the produced matter. It was predicted to dissolve at high temperatures as a result of Debye screening \cite{satz}.
However, no clear signal of this phenomenon has been observed so far, and eventually it was suggested \cite{dima-satz} that it does not dissolve at all in the medium produced in heavy ion collisions in all previous measurements including RHIC.

Nevertheless, a sizable suppression by break-up of $\J$ propagating through the created medium was found in \cite{kps-psi}. The key point of this consideration is the relation between the dipole break-up cross section and $p_T$ broadening of a quark propagating through the medium \cite{jkt,dhk}, which provide direct access to the transport coefficient.

\begin{figure}[htb]
\bc
 \includegraphics[height=7cm]{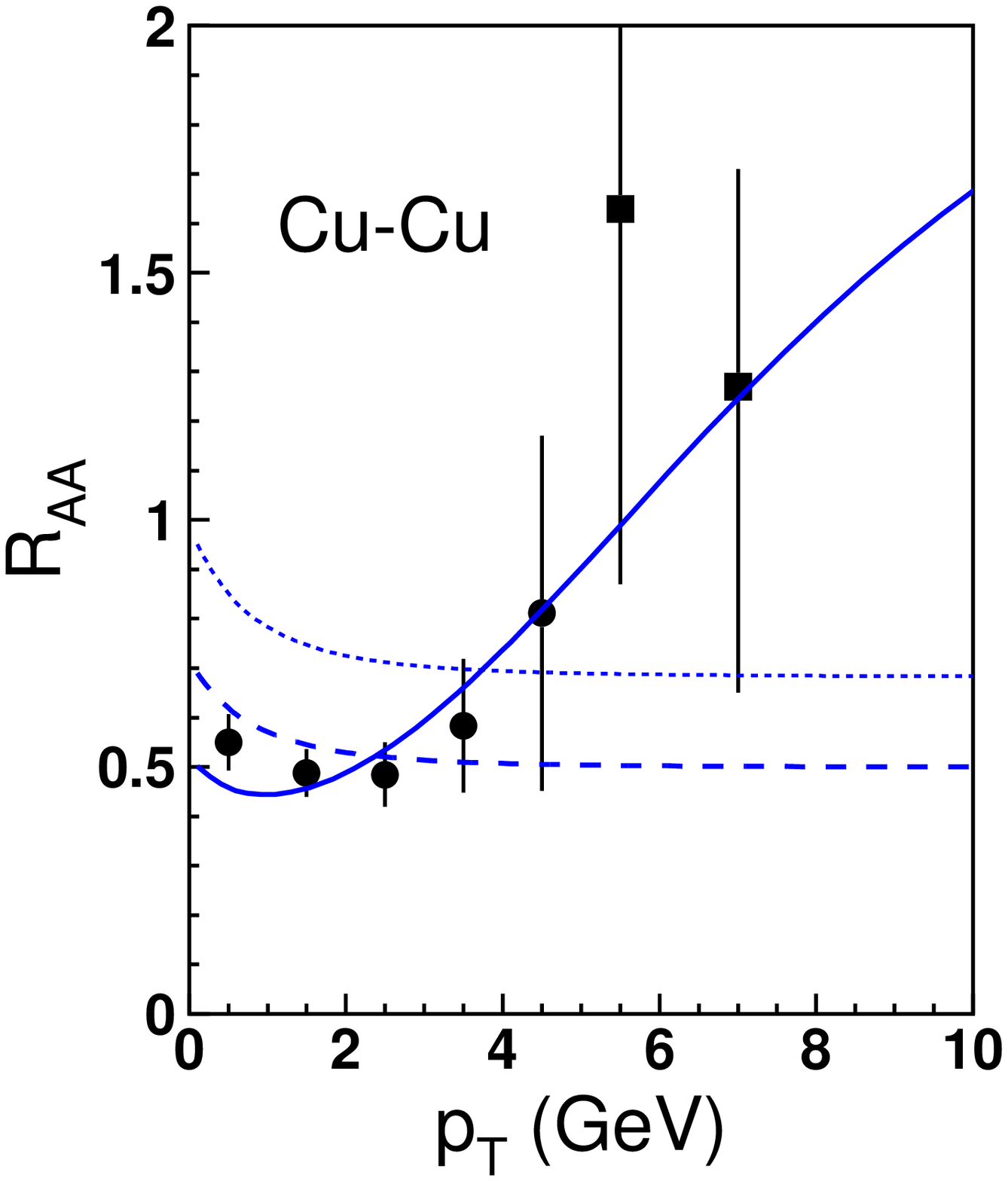}\hspace{20mm}
 \includegraphics[height=7cm]{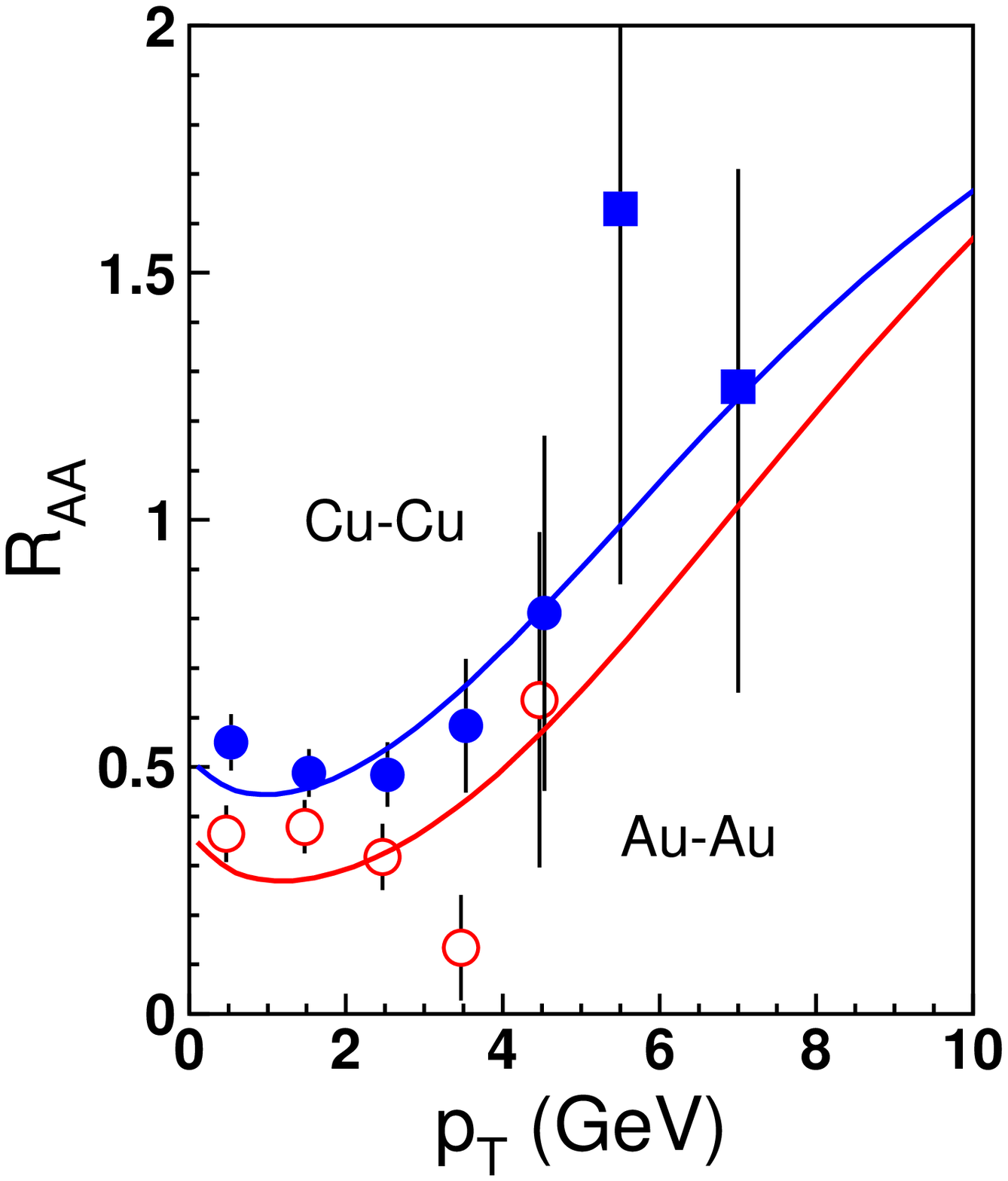}
\caption{ \label{psi} {\it Left:} Nuclear ratio $R_{AA}$ for central (0-20\%) copper-copper collisions as function of $\J$ transverse momentum  from the PHENIX \cite{phenix2} (circles) and STAR \cite{star1} (squares) measurements at $\sqrt{s}=200\GeV$.
The dotted curve shows the FSI attenuation of $\J$ in the produced dense medium.
The dashed curve also includes the ISI effects, charm shadowing and absorption.
The final solid curve is also corrected for the Cronin effect.
{\it Right:}  Nuclear ratio $R_{AA}$ for central copper-copper (full circles and squares, upper curve) and gold-gold (empty circles, bottom curve).}
 \ec
 \end{figure}

In the left panel of Fig.~\ref{psi} we present the contributions of three different sources of nuclear effects for $\J$ production in $Cu$-$Cu$ collisions. at $\sqrt{s}=200\GeV$.
The curves from top to bottom show: (i) the contribution of $\J$ absorption in the created medium,
(ii) the nuclear suppression including initial state interactions (charm shadowing and break-up),
and (iii) the final curve corrected for the Cronin effect. All the calculations are parameter free, except
the upper (dotted) curve controlled by the transport coefficient, which was fitted and found
$\hat q_0=0.3-0.5\GeV^2/\fm$. This value well agrees with the pQCD estimates \cite{bdmps},
but is significantly smaller than what results from the jet quenching analyses based on the energy loss scenario.
The right panel of Fig.~\ref{psi} demonstrates the $A$ dependence of $\J$ suppression.

It is worth mentioning that the predicted steep rise of $R_{AA}$ at $p_T>5GeV$ is not reliable, since it is based on the parametrization  of the $p_T$ dependence in $pp$ collisions, which is extrapolated beyond the measured interval of $p_T$. An alternative method of calculation of the Cronin effect, which was tried in \cite{kps-psi} led to the same results at $p_T<5\GeV$, but a substantially weaker enhancement at larger $p_T$.

Notice, that the above analyses might miss some important mechanisms contributing to the observed nuclear suppression. One of them is the mutual boosting of the saturation scales in nuclei \cite{boosting}, which results in an increased break-up cross section of $\J$.
Another mechanism is related to the observation \cite{hk1,hk2,hk3} that the so called "cold nuclear matter" participating in the initial state interactions, is not really cold, as is discussed in \cite{nontrivial}.
Inclusion of these effects can only lead to a further decrease of the transport coefficient extracted from data. Therefore the above value of $\hat q_0$ found from RHIC data, should be considered as an upper limit.

\section{Quenching of high-$p_T$ hadrons by absorption in a dense medium}\label{alternative}

Saying that a pre-hadron produced inside the medium is absorbed, one should
make it clear what it means, and why absorption suppresses the hadron production rate.
Soft interactions cannot stop or absorb a high energy hadron or a parton, which only can change their color, while the longitudinal momentum remains unchanged.  
"Absorption" is a kind of jargon, which means that the interaction in the medium excludes or strongly suppresses the chance to detect the final state hadron in the given kinematic domain.
This is not obvious for every interaction, e.g. in the case of break-up of a dipole. 
After a pre-hadron, i.e. a colorless dipole is produced it propagates without radiative energy loss,
and the collision energy loss also is tiny, because the elastic cross section is very small.
However, the dipole has a large color-exchange (inelastic) cross section. In the case of, say, charmonium dipole $\bar cc$, a color exchange results in a break up and creation of two
open charm hadrons. The chance to detect a charmonium becomes vanishingly small.
However if the final hadron to be detected is a heavy-light quark dipole, say $D$-meson,
originated from an initial charm quark, the break up is not so harmful, since the charm quark 
will always escape from the nucleus and produce a $D$-meson with a high probability. 
Important, however, is to produce a leading $D$-meson with a certain and  large fractional momentum $z_h$. This is actually why the production length shrinks at large $z_h$, Eq.~(\ref{600}),
as is depicted in Fig.~\ref{dd-lp}. The only way to stop the dissipation of energy is to produce a colorless pre-hadron. Once the latter breaks-up, the energy loss is again initiated and either a hadron with the given large $z_h$ cannot be produced any more, or the Sudakov factor suppresses its production.

\subsection{How large is the pre-hadron?}

The pre-hadron propagating in a medium attenuates  with a dipole cross section proportional to $r_T^2$, where the mean transverse dipole separation $r_T$ is rising with time.
A low energy dipole quickly expands to the hadronic size, but at high energies, Lorentz time dilation freezes the initial small size of the dipole for the time of propagation through the medium. So  with rising energy of the dipole $E$, the medium becomes more transparent.

The transverse  expansion of a $\bar qq$ dipole moving can be described as,
\beq
\frac{dr_T}{dt}=\frac{k_T(t)}{\alpha(1-\alpha)\,E},
\label{760}
\eeq
where $\alpha$ is the fractional light-cone momentum of the quark.
Applying the uncertainty relation $k_T\sim1/r_T$, we get,
\beq
r_T^2(t)=\frac{2t}{\alpha(1-\alpha)\,E}+r_0^2,
\label{765}
\eeq
Where $r_0$ is the initial dipole separation.
Such a behavior of the mean size squared can be also obtained within the more rigorous path integral technique \cite{kz91,kst1} for the early stage of expansion.

The mean fractional momentum $\alpha$ can be estimated in the perturbative model for hadronization
\cite{berger,pert}, 
\beq
q\to q+g^*\to (q\bar q)+q,
\label{770}
\eeq
where the radiated virtual gluon decays to $\bar qq$, and the $\bar q$ merges the parent quark creating a colorless $\bar qq$ dipole, which carries fraction $z_h$ of the initial quark momentum. This process is illustrated in Fig.~\ref{interference}.
Assuming equal momentum sharing in the $g^*\to\bar qq$ decay, one gets,
\beq
\alpha={1\over z_h}-1.
\label{775}
\eeq

The expansion described by Eq.~(\ref{765}) starts from the moment of the hard reaction which initiated the
jet.  The initial transverse extension of the color field in the high-$p_T$ quark is very small, $r_0\sim1/p_T$, and is quickly forgotten, since the second term in the right hand side  of (\ref{765}) can be neglected. When the dipole pre-hadron is produced, its mean initial size is given by Eq.~(\ref{765}) at the distance from the hard interaction point $l=l_p$. Then it keeps expanding in accordance with Eq.~(\ref{765}).

The size expansion at the early stage of jet development is so fast that the pre-hadron is produced with quite a large initial size,
\beq
r_T^2(l=l_p) = 
\frac{2\,z_h^2\,l_p}{(1-z_h)(2z_h-1)\,E}.
\label{777}
\eeq
 For example, for production length $l_p=2\fm$, $z_h=0.7$ (the result is almost $z_h$ independent
 for $z_h=0.5-0.8$), and jet energy $E=p_T/z_h=15\GeV$,  the dipole is produced with separation $r_T\approx 0.5\fm$. This is a rather large size, which should cause a strong absorption in a dense medium. 
 
 \subsection{Attenuation of high-$p_T$ pions}

\begin{figure}[htb]
\bc
 \includegraphics[height=4.5cm]{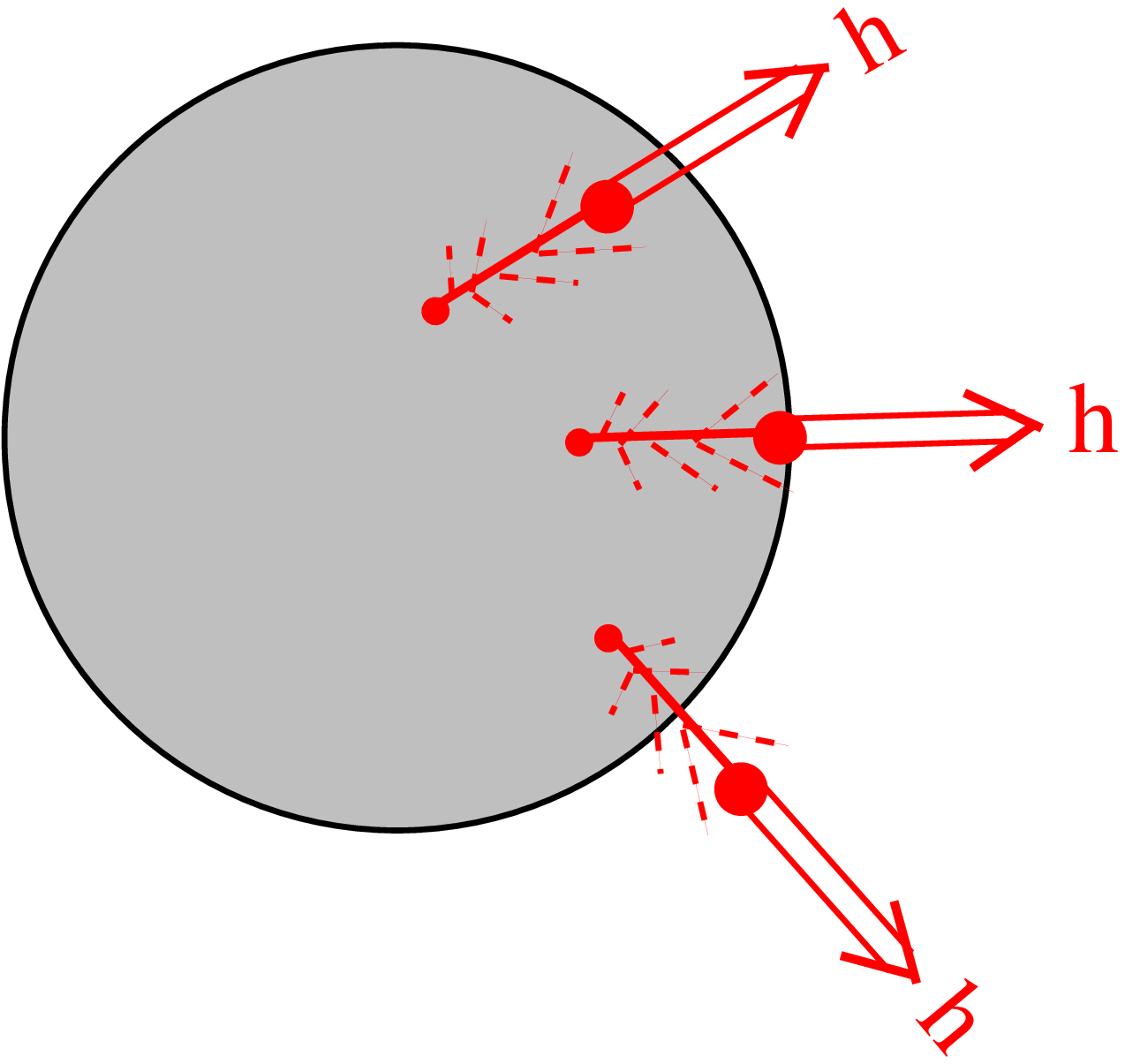}\hspace{15mm}
 \includegraphics[height=5cm]{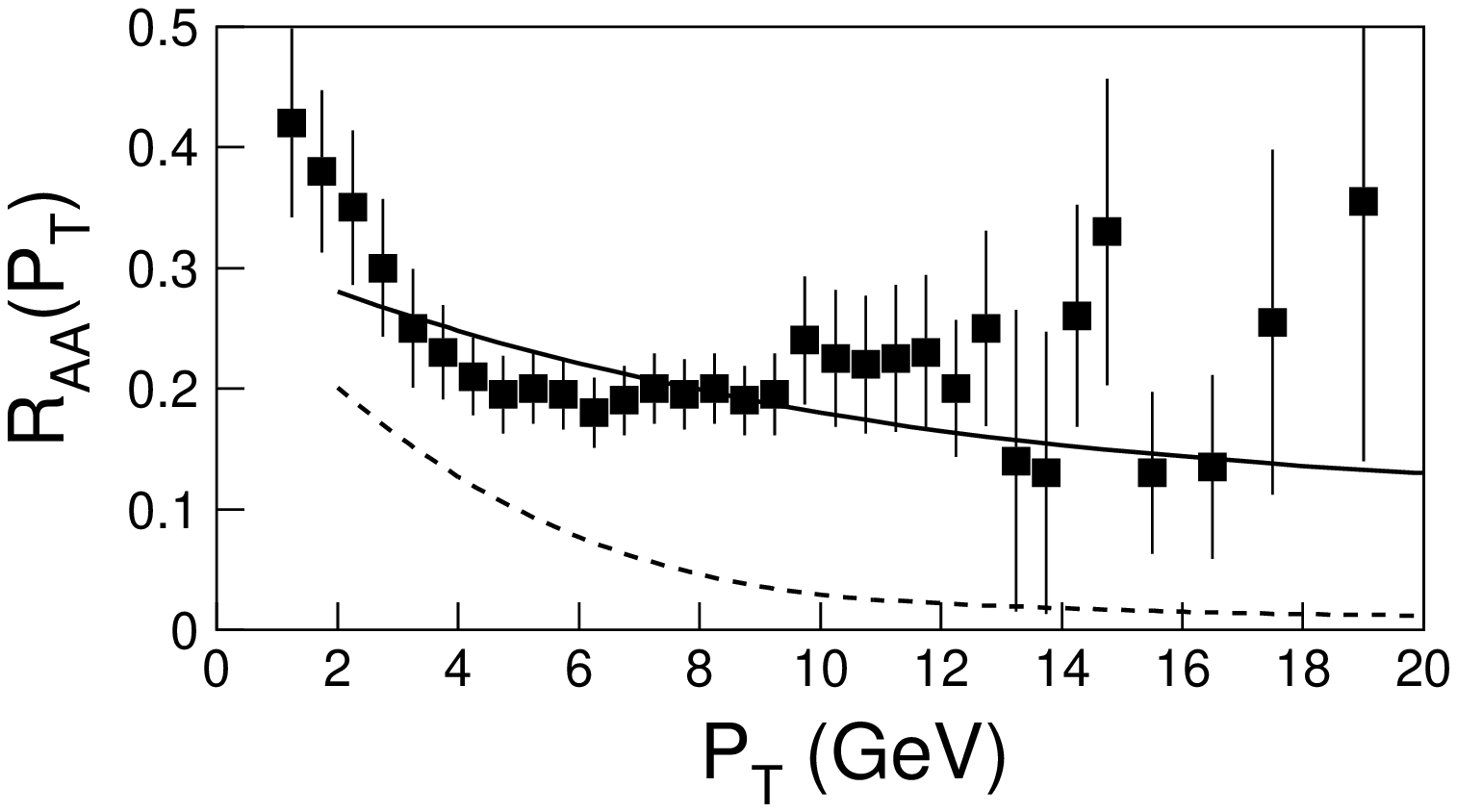}
\caption{\label{alternative-fig} {\it Left:} Pictorial explanation of the space-time development
of high-$p_T$ hadron production in a dense medium created in a central nuclear collision. 
{\it Right:}  Pion suppression in central $AA$ collisions  
($A\sim 200$) at $\sqrt{s}=200\GeV$ (solid) and $\sqrt{s}=5500\GeV$
(dashed). Data are from the PHENIX experiment \cite{phenix-pt}.}
 \ec
  \end{figure}

Let us  consider a central, $b=0$, collision of identical heavy nuclei with nuclear density
$\rho_A(r)=\rho_A\Theta(R_A-r)$, where $\rho_A=0.16\fm^{-3}$. After the nuclei pass through each other the leave behind a quark-gluon medium, which is probed by produced high-$p_T$ hadrons, as is illustrated in Fig.~\ref{alternative-fig}.
Partons are produced in a hard collisions at impact parameter $\tau$ with a rate proportional
to $T_A^2(\tau)=4(R_A^2-\tau^2)\rho_A$.
Then the parton radiates gluons and propagates transversely through the expanding medium, whose density is diluting inversely with time. Besides vacuum radiation,  the parton experiences multiple interactions in the medium which induce an additional radiation. At some point inside the medium, or outside, the parton picks up an anti-parton and produces a dipole, pre-hadron, which does not radiate any more.
 
Integrating over the location of  the hard collision and over the production time $l_p$ weighted with the distribution depicted in Fig.~\ref{dd-lp}, we arrive at the nuclear suppression factor \cite{last-call,kps-cern},
 \beq
R_{AA}=\frac{\la l_p^2\ra}{R_A^2}\left[1-
\alpha\,\frac{L}{\la l_p\ra} +
\beta\,\frac{L^2}{\la l_p^2\ra}
\right]\,.
\label{780}
 \eeq
 Here the numerical factors $\alpha$, $\beta$ are independent of $p_T$ and the medium properties and are not important for further evaluations.  The effective mean free path $L$ has the form,
 \beq
L^3=\frac{3p_T}{8\,\rho_A^2\,R_A\,X},
\label{800}
\eeq
where the factor $X$ besides other kinematic factors also includes 
the unknown initial density of the medium, which should be considered as a 
fitting  parameter adjusted to data on $R_{AA}$. However, there is a unique possibility
to predict $R_{AA}$ without knowing the medium density. Assume 
that the medium density is so high, and the absorption of pre-hadron dipoles is so strong that
their mean free path is very short, $L\ll\la l_p\ra$. In this case the survived pre-hadrons should be born either near the surface of the medium, or outside.
Then, the further details of the medium features are not important, since the jet can be produced only inside the outer slice of the medium of thickness $\la l_p\ra$, and this geometry defines the value of $R_{AA}$. Indeed, in this case the last two terms in (\ref{800}) can be neglected, and $R_{AA}$ much simplifies,
 \beq
R^h_{AA}=\frac{\la l_p^2\ra}{R_A^2}.
\label{820}
 \eeq
This regime of hadron production assuming a very short mean free path $L$,  looks quite plausible. As was demonstrated above, the starting transverse separation of the produced pre-hadron dipole given by Eq.~(\ref{777}), is quite large and keeps expanding, what leads to a strong absorption in a dense medium. 
Indeed, the comparison of Eq.~(\ref{820}) shown by the solid curve in the right panel of Fig.~\ref{alternative-fig}, with RHIC data for pion production \cite{phenix-pt} demonstrates good agreement, both in magnitude and $p_T$ dependence. This success encourages to believe that the high-density
regime indeed occurs in the central collisions at RHIC. In this case it should be valid at the energies of LHC as well. Although a further increase of the medium density expected for LHC, should not affect $R_{AA}$, Eq.~(\ref{820}), the jet structure changes. While at RHIC the high $p_T$ pion production
is dominated by valence quarks, at LHC gluons are expected to be the main source of jets.
In this case the rate of perturbative vacuum radiation should be larger by the Casimir factor $9/4$,
and correspondingly $\la l_p\ra$ shorter. With the production length distribution calculated for gluon jets \cite{last-call,kps-cern} the prediction for $R_{AA}$ at the energies of LHC is shown in Fig.~\ref{alternative-fig} by the dashed curve. Notice that as far as the medium is sufficiently dense for the regime of strong absorption, the dependence on collision energy should be very mild, most important is whether the hadron originates from quark or gluon jets.

The validity of the assumption that the mean free path Eq.~(\ref{800}) is short
depends not only on the medium density embedded in the parameter $X$, but also on $p_T$. Because $L$ is rising with $p_T$, eventually the approximation Eq.~(\ref{820}) will break down and
$R_{AA}$ will start rising tending to restore the expectation based on QCD factorization, $R_{AA}\to1$.
This hardly may happen at RHIC, because of the predicted suppression at large $x_T$ caused by the kinematic restrictions discussed in Sect.~\ref{xT}. Nevertheless this should be expected for LHC.
The position of the turn-over point in the $p_T$ scale should be sensitive to the medium density and deserves further study.

Notice that in the case of production of two high-$p_T$ hadrons absorption does not double, as one could naively expect. $R_{AA}$ will be similar to the case of single hadron detection. Indeed, in both cases the jet from which the one or two detected hadrons originate, must be within the outer medium slice of thickness equal to the shortest production length. Since $l_p$ does not vary much with the jet energy (see Fig.~\ref{dd-lp}), it does not make difference whether one, or two correlated hadrons are detected.

\subsection{Heavy flavors}

As was already mentioned, a high-$p_T$ heavy quark always escapes from the medium
and produces an open flavor hadron. Such a process should have no suppression. 
Therefore a break-up of a light-heavy dipole in a medium does not lead to a suppression,
unless the fractional momentum $z_h$ of the detected hadron is fixed at a large value. In such a case break up of the dipole ignites continuation of vacuum energy loss, which slows down the quark to smaller values of $z_h$. This is why a quark should stop radiating at a distance $l\sim l_p$ and produce a colorless dipole, which then survives through the medium. 

At first glance, since heavy quarks radiate less because of the dead-cone effect, their production length $\la l_p\ra$ should be longer compared with  light quarks.
However, as was pointed out in Sect.~\ref{lc}, a charm and light quarks radiate similarly at 
short distances $l\sim \la l_p\ra$, therefore they should have similar production lengths.

The main puzzle for the energy loss scenario, which remains to be a challenge, is the same strong suppression observed for bottom quarks. Although $b$-quarks radiate much less than light ones \cite{yura}, according to the time-dependence of energy loss shown in the left panel of Fig.~\ref{l-dep}, this occurs mainly because of prompt regeneration by $b$-quarks of their color field. After that the $b$-quark stops radiating, but forms a color flux tube \cite{jet-lag}, which becomes the main source of energy loss. This leads to a quite short production length Eq.~(\ref{750}). 

Interesting that the produced heavy-light, $c-q$ or $b-q$ dipoles expand their sizes faster than  a light $\bar qq$ dipole. This happens because of a very asymmetric sharing of the longitudinal momentum in such dipoles. Minimizing the energy denominator one gets the fractional momentum carried by the light quark,
\beq
\alpha\sim\frac{m_q}{m_Q},
\label{840}
\eeq
which indeed is very small, about $0.1$ for charm and $0.03$ for bottom.
Then according to Eq.~(\ref{765}) a $c-q$ dipole is expanding similar to $\bar qq$, but a $b-q$ dipole does it much faster.

We conclude that hadronization of charm and bottom quarks ends up at a short distance $l_p$ with production of a colorless dipole. These dipoles are expanding similar, or even faster than a light $\bar qq$ dipole,  therefore they are strongly absorbed by the dense medium. This justifies
the application of the strong absorption scenario and Eq.~(\ref{820}) to heavy charm and bottom quarks, and explains why both of them are strongly suppressed in $AA$ collisions.

\section{Summary}

This paper presents an attempt at a critical overview of the current status of our understanding of the dynamics of high-$p_T$ processes on nuclei. Here some of the important observations are highlighted.

\begin{itemize}

\item
The effects of coherence for gluon radiation are expected to be insignificant, since diffractive gluon radiation is suppressed in data. This means that gluon shadowing should be weak, what was confirmed by less biased global analyses of data for DIS on nuclei. Having a deeper insight at the contemporary global analyses, we concluded that the results for gluon nPDFs in some of them are not really constrained by data, but based on ad hoc assumptions, or otherwise rely on incorrect models for high-$p_T$ hadronic reactions.

\item
The same mechanisms which make gluon shadowing weak, also suppress the CGC effects. The saturation scale in nuclei, which is frequently overestimated, can be directly accessed by measuring $p_T$ broadening of heavy flavors produced on nuclei. The dipole phenomenology fitted to DIS data correctly reproduces the data on broadening.

\item   
The bound nucleons in colliding nuclei considerably change their  properties drifting to higher Fock components due to mutual multiple interactions. This boosts the saturation scale in $AA$ compared with $pA$ collisions as is described by the reciprocity equations.
As a result, the nuclear medium becomes more opaque for colorless dipoles.
A direct way to observe this effect in data is to compare the magnitudes of broadening observed in $pA$ and $AA$ collisions. 

\item
Trying to enhance the effects of coherence by decreasing Bjorken {$ x$}, one should be cautious going to forward rapidities.
Energy conservation becomes an issue towards the kinematic limits and may cause a strong suppression. The same effect leads to a similar suppression of particles produced with large $p_T$ ($x_T$) in $pA$ and $AA$ collisions, even if $x_L$ is small. Indeed, data from RHIC for production of high-$p_T$ pions and direct photons in $dA$ and $AA$ collisions provide evidence for such a suppression. 

\item
A highly virtual parton produced with high $ p_T$ in a hard collision starts regenerating its color field and dissipating energy via gluon radiation.
The energy loss is very intensive at the early stage, the parton radiates almost half of the radiated energy during the first {$ 1\fm$}. Medium induced radiation speeds up this process, though usually it is a small correction.  In order to respect energy conservation a leading hadron carrying a large fractional momentum $z_h$ has to be produced at a short time scale after the hard reaction. This puts in doubt the main (unjustified) assumption of the energy loss scenario for jet quenching, that the hadronization is lasting a long time and ends up with production of a leading hadron always outside the medium.

\item
The vacuum radiation of a highly virtual parton, which is lacking its color field, is subject to the dead-cone effect, making the radiation flavor independent during the initial stage of fragmentation. As a result, high-$p_T$ charm and light quarks should be suppressed similarly in heavy ion collisions.
Bottom quarks behave differently, they promptly restore the color field and stop radiating, on a distance of about $1\fm$.  The subsequent medium induced energy loss is too weak to explain the observed strong suppression of bottom produced in $AA$ collisions. This remains a serious challenge for the energy loss scenario. 

\item
Neglecting the possibility of pre-hardon production and strong attenuation  inside the medium, one should overestimate the medium density in order to reproduce the experimentally observed suppression of high-$p_T$ hadrons. Indeed, this seems to be the case, the transport coefficient 
fitted to data on jet quenching within the energy loss scenario, substantially exceeds the conventional expectations. It is also much larger than the result of the alternative probe for the dense medium, suppression of $J/\Psi$ produced in $AA$ collisions.

\item
Once a pre-hadron is produced inside the medium, its subsequent fate depends
on the initial dipole size and the speed of its expansion. Since the mean transverse momentum of the gluons radiated by a highly virtual parton steeply decreases with time, the initial transverse separation of the produced dipole
rises as function of production length. For the mean production length we estimated that the produced pre-hadron dipole is rather large $r_T\sim 0.5\fm$, and it keeps rising. Dipoles containing a heavy quark evolve even faster than light ones. As a result, the mean free path of such dipoles is very short, provided that the medium is substantially denser than the cold nuclear matter. 

This observation leads to a specific scenario for hadron production in a dense medium created in $AA$ collisions. The hard interaction point must be located not deeper than at production length $l_p$ from the surface, otherwise the produced pre-hadron will be promptly destroyed by the medium.
As far as the model-dependent mean production length is known, this scenario predicts in a parameter-free way the suppression factor $R_{AA}$, which well agrees with RHIC data.
The suppression is similar for light and heavy quarks, including beauty.

\end{itemize}

\begin{acknowledgments}

B.K. is grateful to Lanny Ray for inviting to the Workshop and encouraging to write this paper.
This work was supported in part by Fondecyt (Chile) grant 1090291, by DFG
(Germany) grant PI182/3-1, by Conicyt-DFG grant No. 084-2009, by the
Slovak Funding Agency, Grant 2/0092/10 and by Grants VZ M\v SMT
6840770039 and LC 07048 (Ministry of Education of the Czech Republic).

\end{acknowledgments}

\end{document}